\input harvmac
\newcount\tocnew\tocnew=1\newcount\tocopage
\newbox\tocbox\newdimen\tocsize
\def\tocstrut{{\vrule height8.5pt depth3.5pt width0pt}} 
\def\tocref#1#2{\tocbanner#1               
    \global\setbox\tocbox=\vbox{
    \box\tocbox\vbox{
    \line{\tocstrut{#2#1}~\dotfill~\folio} 
    }}\tocsuffix}
\def\tocline#1{\tocbanner
    \global\setbox\tocbox=\vbox{
    \box\tocbox\vbox{
    \line{\tocstrut{#1}}
    }}\tocsuffix}
\def\tocbanner{\ifnum\tocnew=1\tocstart\fi
    \ifnum\tocnew=3\tocont\fi}
\def\tocsuffix{\ifdim\tocsize<\ht\tocbox\tocgen
    \global\tocnew=3\fi}
\def\tocstart{\global\tocsize=.95\vsize   
    \global\setbox\tocbox=\vbox{
    \centerline{\tocstrut\bf Table Of Contents} 
    \line{\tocstrut\hfil}                 
    }\global\tocnew=2}
\def\tocont{\global\setbox\tocbox=\vbox{
    \centerline{\tocstrut Table Of Contents (Continued)} 
    }\global\tocnew=2}
\def\tocgen{\ifnum\tocnew=1
    \message{No TOC entries found.}\else
    \tocopage=\pageno\pageno=0\message{(TOC}
    \shipout\box\tocbox\message{)}
    \pageno=\tocopage\global\tocnew=1\fi}

%
\message{S-Tables Macro v1.0, ACS, TAMU (RANHELP@VENUS.TAMU.EDU)}
%
%
\newhelp\stablestylehelp{You must choose a style between 0 and 3.}%
\newhelp\stablelinehelp{You should not use special hrules when
stretching
a table.}%
\newhelp\stablesmultiplehelp{You have tried to place an S-Table
inside another S-Table.  I would recommend not going on.}%
%
%
\newdimen\stablesthinline
\stablesthinline=0.4pt
\newdimen\stablesthickline
\stablesthickline=1pt
%
%
\newif\ifstablesborderthin
\stablesborderthinfalse
\newif\ifstablesinternalthin
\stablesinternalthintrue
\newif\ifstablesomit
\newif\ifstablemode
\newif\ifstablesright
\stablesrightfalse
%
%
\newdimen\stablesbaselineskip
\newdimen\stableslineskip
\newdimen\stableslineskiplimit
%
%
\newcount\stablesmode
\newcount\stableslines
\newcount\stablestemp
\stablestemp=3
\newcount\stablescount
\stablescount=0
\newcount\stableslinet
\stableslinet=0
%
%
%
\newcount\stablestyle
\stablestyle=0
%
%
\def\stablesleft{\quad\hfil}%
\def\stablesright{\hfil\quad}%
%
%
\catcode`\|=\active%
%
%
\newcount\stablestrutsize
\newbox\stablestrutbox
\setbox\stablestrutbox=\hbox{\vrule height10pt depth5pt width0pt}
\def\stablestrut{\relax\ifmmode%
                         \copy\stablestrutbox%
                       \else%
                         \unhcopy\stablestrutbox%
                       \fi}%
%
%
\newdimen\stablesborderwidth
\newdimen\stablesinternalwidth
\newdimen\stablesdummy
\newcount\stablesdummyc
\newif\ifstablesin
\stablesinfalse
%
%
%
%
%
\def\stablesadj{%
  \ifcase\stablestyle%
    \hbox to \hsize\bgroup\hss\vbox\bgroup%
  \or%
    \hbox to \hsize\bgroup\vbox\bgroup%
  \or%
    \hbox to \hsize\bgroup\hss\vbox\bgroup%
  \or%
    \hbox\bgroup\vbox\bgroup%
  \else%
    \errhelp=\stablestylehelp%
    \errmessage{Invalid style selected, using default}%
    \hbox to \hsize\bgroup\hss\vbox\bgroup%
  \fi}%
\def\stablesend{\egroup%
  \ifcase\stablestyle%
    \hss\egroup%
  \or%
    \hss\egroup%
  \or%
    \egroup%
  \or%
    \egroup%
  \else%
    \hss\egroup%
  \fi}%
\def\stablestart{%
  \ifstablesin%
    \errhelp=\stablesmultiplehelp%
    \errmessage{An S-Table cannot be placed within an S-Table!}%
  \fi
  \global\stablesintrue%
  \global\advance\stablescount by 1%
  \message{<S-Tables Generating Table \number\stablescount}%
  \begingroup%
  \stablestrutsize=\ht\stablestrutbox%
  \advance\stablestrutsize by \dp\stablestrutbox%
  \ifstablesborderthin%
    \stablesborderwidth=\stablesthinline%
  \else%
    \stablesborderwidth=\stablesthickline%
  \fi%
  \ifstablesinternalthin%
    \stablesinternalwidth=\stablesthinline%
  \else%
    \stablesinternalwidth=\stablesthickline%
  \fi%
  \tabskip=0pt%
  \stablesbaselineskip=\baselineskip%
  \stableslineskip=\lineskip%
  \stableslineskiplimit=\lineskiplimit%
  \offinterlineskip%
  \def\borderrule{\vrule width \stablesborderwidth}%
  \def\internalrule{\vrule width \stablesinternalwidth}%
  \def\thinline{\noalign{\hrule height \stablesthinline}}%
  \def\thickline{\noalign{\hrule height \stablesthickline}}%
  \def\trule{\omit\leaders\hrule height \stablesthinline\hfill}%
  \def\ttrule{\omit\leaders\hrule height \stablesthickline\hfill}%
  \def\tttrule##1{\omit\leaders\hrule height ##1\hfill}%
  \def\stablesel{&\omit\global\stablesmode=0%
    \global\advance\stableslines by 1\borderrule\hfil\cr}%
  \def\el{\stablesel&}%
  \def\elt{\stablesel\thinline&}%
  \def\eltt{\stablesel\thickline&}%
  \def\elttt##1{\stablesel\noalign{\hrule height ##1}&}%
  \def\elspec{&\omit\hfil\borderrule\cr\omit\borderrule&%
              \ifstablemode%
              \else%
                \errhelp=\stablelinehelp%
                \errmessage{Special ruling will not display properly}%
              \fi}%
  \def\stmultispan##1{\mscount=##1 \loop\ifnum\mscount>3
\stspan\repeat}%
  \def\stspan{\span\omit \advance\mscount by -1}%
  \def\multicolumn##1{\omit\multiply\stablestemp by ##1%
     \stmultispan{\stablestemp}%
     \advance\stablesmode by ##1%
     \advance\stablesmode by -1%
     \stablestemp=3}%
  \def\multirow##1{\stablesdummyc=##1\parindent=0pt\setbox0\hbox\bgroup%
    \aftergroup\emultirow\let\temp=}
  \def\emultirow{\setbox1\vbox to\stablesdummyc\stablestrutsize%
    {\hsize\wd0\vfil\box0\vfil}%
    \ht1=\ht\stablestrutbox%
    \dp1=\dp\stablestrutbox%
    \box1}%
%
  \def\stpar##1{\vtop\bgroup\hsize ##1%
     \baselineskip=\stablesbaselineskip%
     \lineskip=\stableslineskip%

\lineskiplimit=\stableslineskiplimit\bgroup\aftergroup\estpar\let\temp=}%
  \def\estpar{\vskip 6pt\egroup}%
  \def\stparrow##1##2{\stablesdummy=##2%
     \setbox0=\vtop to ##1\stablestrutsize\bgroup%
     \hsize\stablesdummy%
     \baselineskip=\stablesbaselineskip%
     \lineskip=\stableslineskip%
     \lineskiplimit=\stableslineskiplimit%
     \bgroup\vfil\aftergroup\estparrow%
     \let\temp=}%
  \def\estparrow{\vfil\egroup%
     \ht0=\ht\stablestrutbox%
     \dp0=\dp\stablestrutbox%
     \wd0=\stablesdummy%
     \box0}%
  \def|{\global\advance\stablesmode by 1&&&}%
  \def\|{\global\advance\stablesmode by 1&\omit\vrule width 0pt%
         \hfil&&}%
\def\vt{\global\advance\stablesmode
by 1&\omit\vrule width \stablesthinline%
          \hfil&&}%
  \def\vtt{\global\advance\stablesmode by 1&\omit\vrule width
\stablesthickline%
          \hfil&&}%
  \def\vttt##1{\global\advance\stablesmode by 1&\omit\vrule width ##1%
          \hfil&&}%
  \def\vtr{\global\advance\stablesmode by 1&\omit\hfil\vrule width%
           \stablesthinline&&}%
  \def\vttr{\global\advance\stablesmode by 1&\omit\hfil\vrule width%
            \stablesthickline&&}%
\def\vtttr##1{\global\advance\stablesmode
 by 1&\omit\hfil\vrule width ##1&&}%
  \stableslines=0%
  \stablesomitfalse}
\def\stablesdef{\bgroup\stablestrut\borderrule##\tabskip=0pt plus 1fil%
  &\stablesleft##\stablesright%
  &##\ifstablesright\hfill\fi\internalrule\ifstablesright\else\hfill\fi%
  \tabskip 0pt&&##\hfil\tabskip=0pt plus 1fil%
  &\stablesleft##\stablesright%
  &##\ifstablesright\hfill\fi\internalrule\ifstablesright\else\hfill\fi%
  \tabskip=0pt\cr%
  \ifstablesborderthin%
    \thinline%
  \else%
    \thickline%
  \fi&%
}%
\def\endtable{\advance\stableslines by 1\advance\stablesmode by 1%
   \message{- Rows: \number\stableslines, Columns:
\number\stablesmode>}%
   \stablesel%
   \ifstablesborderthin%
     \thinline%
   \else%
     \thickline%
   \fi%
   \egroup\stablesend%
\endgroup%
\global\stablesinfalse}
%

\overfullrule=0pt \abovedisplayskip=12pt plus 3pt minus 3pt
\belowdisplayskip=12pt plus 3pt minus 3pt

\noblackbox
\input epsf
\newcount\figno
\figno=0
\def\fig#1#2#3{
\par\begingroup\parindent=0pt\leftskip=1cm\rightskip=1cm\parindent=0pt
\baselineskip=11pt \global\advance\figno by 1 \midinsert
\epsfxsize=#3 \centerline{\epsfbox{#2}} \vskip 12pt
\centerline{{\bf Figure \the\figno:} #1}\par
\endinsert\endgroup\par}
\def\figlabel#1{\xdef#1{\the\figno}}

\def\IR{\relax{\rm I\kern-.18em R}}


\font\cmss=cmss10 \font\cmsss=cmss10 at 7pt
\def\rlx{\relax\leavevmode}
\def\inbar{\vrule height1.5ex width.4pt depth0pt}
\def\IC{\relax\,\hbox{$\inbar\kern-.3em{\rm C}$}}
\def\IN{\relax{\rm I\kern-.18em N}}
\def\IP{\relax{\rm I\kern-.18em P}}
\def\IR{\relax{\rm I\kern-.18em R}}
\def\ZZ{\rlx\leavevmode\ifmmode\mathchoice{\hbox{\cmss Z\kern-.4em Z}}
 {\hbox{\cmss Z\kern-.4em Z}}{\lower.9pt\hbox{\cmsss Z\kern-.36em Z}}
 {\lower1.2pt\hbox{\cmsss Z\kern-.36em Z}}\else{\cmss Z\kern-.4em
 Z}\fi}
\def\IZ{\relax\ifmmode\mathchoice
{\hbox{\cmss Z\kern-.4em Z}}{\hbox{\cmss Z\kern-.4em Z}}
{\lower.9pt\hbox{\cmsss Z\kern-.4em Z}} {\lower1.2pt\hbox{\cmsss
Z\kern-.4em Z}}\else{\cmss Z\kern-.4em Z}\fi}

\def\narrowplus{\kern -.04truein + \kern -.03truein}
\def\narrowminus{- \kern -.04truein}
\def\narrowminussub{\kern -.02truein - \kern -.01truein}

\def\O{{\cal O}}

\def\frac#1#2{{#1\over #2}}

\def\IZ{\relax\ifmmode\mathchoice
{\hbox{\cmss Z\kern-.4em Z}}{\hbox{\cmss Z\kern-.4em Z}}
{\lower.9pt\hbox{\cmsss Z\kern-.4em Z}} {\lower1.2pt\hbox{\cmsss
Z\kern-.4em Z}}\else{\cmss Z\kern-.4em Z}\fi}
\def\IB{\relax{\rm I\kern-.18em B}}
\def\IC{{\relax\hbox{$\inbar\kern-.3em{\rm C}$}}}
\def\ID{\relax{\rm I\kern-.18em D}}
\def\IE{\relax{\rm I\kern-.18em E}}
\def\IF{\relax{\rm I\kern-.18em F}}
\def\IG{\relax\hbox{$\inbar\kern-.3em{\rm G}$}}
\def\IGa{\relax\hbox{${\rm I}\kern-.18em\Gamma$}}
\def\IH{\relax{\rm I\kern-.18em H}}
\def\II{\relax{\rm I\kern-.18em I}}
\def\IK{\relax{\rm I\kern-.18em K}}
\def\IP{\relax{\rm I\kern-.18em P}}

\font\cmss=cmss10 \font\cmsss=cmss10 at 7pt
\def\IR{\relax{\rm I\kern-.18em R}}

\def\1{{\bf 1}}
\def\3{{\bf 3}}
\def\7{{\bf 7}}
\def\2{{\bf 2}}
\def\8{{\bf 8}}

\def\hat{\widehat}
\def\quabla{{\sqcap}\!\!\!\!{\sqcup}}

\def\o{\over}
\def\IP{\relax{\rm I\kern-.18em P}}

\def\cE{{\cal E}}
\def\cF{{\cal F}}
\def\cI{{\cal I}}
\def\O{{\cal O}}

\def\det{{\rm det}}
\def\Ext{{\rm Ext}}
\def\uExt{\underline{\rm Ext}}
\def\Hom{{\rm Hom}}
\def\uHom{\underline{{\rm Hom}}}

%

%
%
\def\eqnn#1{\xdef #1{(\secsym\the\meqno)}\writedef{#1\leftbracket#1}%
\global\advance\meqno by1\wrlabeL#1}
\def\eqna#1{\xdef #1##1{\hbox{$(\secsym\the\meqno##1)$}}
\writedef{#1\numbersign1\leftbracket#1{\numbersign1}}%
\global\advance\meqno by1\wrlabeL{#1$\{\}$}}
\def\eqn#1#2{\xdef #1{(\secsym\the\meqno)}\writedef{#1\leftbracket#1}%
\global\advance\meqno by1$$#2\eqno#1\eqlabeL#1$$}



\lref\vagudi{
 R.~Dijkgraaf, S.~Gukov, A.~Neitzke and C.~Vafa,
  ``Topological M-theory as unification of form theories of gravity,'' hep-th/0411073.
}

\lref\micuram{J.~Louis and A.~Micu,
  ``Heterotic-type IIA duality with fluxes,'' hep-th/0608171.}

\lref\vafai{C.~Vafa,
``Superstrings and topological strings at large N,''
J.\ Math.\ Phys.\  {\bf 42}, 2798 (2001), hep-th/0008142.}

\lref\land{K.~Landsteiner and S.~Montero,
  ``KK-masses in dipole deformed field theories,'' hep-th/0602035.
}

\lref\rustse{J.~G.~Russo and A.~A.~Tseytlin,
  ``Exactly solvable string models of curved space-time backgrounds,''
  Nucl.\ Phys.\ B {\bf 449}, 91 (1995), hep-th/9502038;
A.~A.~Tseytlin,
  ``Exact solutions of closed string theory,''
  Class.\ Quant.\ Grav.\  {\bf 12}, 2365 (1995), hep-th/9505052.}

\lref\civ{F.~Cachazo, K.~A.~Intriligator and C.~Vafa,
``A large N duality via a geometric transition,''
Nucl.\ Phys.\ B {\bf 603}, 3 (2001), hep-th/0103067.}

\lref\nekrasovtop{L.~Baulieu, A.~S.~Losev and N.~A.~Nekrasov,
  ``Target space symmetries in topological theories. I,''
  JHEP {\bf 0202}, 021 (2002), hep-th/0106042.}

\lref\swncg{N.~Seiberg and E.~Witten,
  ``String theory and noncommutative geometry,''
  JHEP {\bf 9909}, 032 (1999), hep-th/9908142.}

\lref\cveticone{M.~Cvetic, G.~W.~Gibbons, H.~Lu and C.~N.~Pope,
 ``Cohomogeneity one manifolds of Spin(7) and G(2) holonomy,''
 Phys.\ Rev.\ D {\bf 65}, 106004 (2002), hep-th/0108245;
``M-theory conifolds,''
 Phys.\ Rev.\ Lett.\  {\bf 88}, 121602 (2002), hep-th/0112098;
``A G(2) unification of the deformed and resolved conifolds,''
Phys.\ Lett.\ B {\bf 534}, 172 (2002), hep-th/0112138.}

\lref\cvetictwo{M.~Cvetic, G.~W.~Gibbons, H.~Lu and C.~N.~Pope,
``New complete non-compact Spin(7) manifolds,''
 Nucl.\ Phys.\ B {\bf 620}, 29 (2002), hep-th/0103155.}

\lref\brand{A.~Brandhuber, J.~Gomis, S.~S.~Gubser and S.~Gukov,
``Gauge theory at large N and new G(2) holonomy metrics,''
Nucl.\ Phys.\ B {\bf 611}, 179 (2001), hep-th/0106034.}

\lref\gkp{
  S.~B.~Giddings, S.~Kachru and J.~Polchinski,
  ``Hierarchies from fluxes in string compactifications,''
  Phys.\ Rev.\ D {\bf 66}, 106006 (2002), hep-th/0105097.}

\lref\pisin{P.~Chen, K.~Dasgupta, K.~Narayan, M.~Shmakova and M.~Zagermann,
  ``Brane inflation, solitons and cosmological solutions: I,''
  JHEP {\bf 0509}, 009 (2005), hep-th/0501185.}

\lref\Kgukov{S.~Gukov, S.~Kachru, X.~Liu and L.~McAllister,
  ``Heterotic moduli stabilization with fractional Chern-Simons invariants,''
  Phys.\ Rev.\ D {\bf 69}, 086008 (2004)
  hep-th/0310159.}

\lref\fawad{S.~F.~Hassan,
``T-duality, space-time spinors and R-R fields in curved backgrounds,''
Nucl.\ Phys.\ B {\bf 568}, 145 (2000), hep-th/9907152.}

\lref\brandtwo{A.~Brandhuber,
``G(2) holonomy spaces from invariant three-forms,''
Nucl.\ Phys.\ B {\bf 629}, 393 (2002), hep-th/0112113.}

\lref\salamontwo{R. ~Bryant, S.~Salamon,
``On the construction of some complete metrics with exceptional
holonomy,'' Duke Math. J. {\bf 58} (1989) 829;
G.~W.~Gibbons, D.~N.~Page and C.~N.~Pope,
``Einstein Metrics On S**3 R**3 And R**4 Bundles,''
Commun.\ Math.\ Phys.\  {\bf 127}, 529 (1990).}

\lref\kovalev{A. Kovalev,
``Twisted connected sums and special Riemannian holonomy,''
math-DG/0012189.}

\lref\joyce{D. ~Joyce,
``Compact Riemannian 7 manifolds with holonomy $G_2$ I, J. Diff. Geom.
{\bf 43} (1996) 291;
II: J. Diff. Geom.{\bf 43} (1996) 329.}

\lref\grayone{A. Gray, L. Hervella,
``The sixteen classes of almost Hermitian manifolds and their linear
invariants,'' Ann. Mat. Pura Appl.(4) {\bf 123} (1980) 35.}

\lref\salamon{ S. Chiossi, S. Salamon,
``The intrinsic torsion of $SU(3)$ and $G_2$ structures,''
 Proc. conf. Differential Geometry Valencia 2001 [math.DG/0202282].}

\lref\gvw{S.~Gukov, C.~Vafa and E.~Witten,
   ``CFT's from Calabi-Yau four-folds,''
  Nucl.\ Phys.\ B {\bf 584}, 69 (2000)
  [Erratum-ibid.\ B {\bf 608}, 477 (2001)] hep-th/9906070.}

\lref\gukov{S.~Gukov,
``Solitons, superpotentials and calibrations,''
Nucl.\ Phys.\ B {\bf 574}, 169 (2000), hep-th/9911011.}

\lref\ach{B.~S.~Acharya and B.~Spence,
``Flux, supersymmetry and M theory on 7-manifolds,''
arXiv:hep-th/0007213.}

\lref\bw{C.~Beasley and E.~Witten,
 ``A note on fluxes and superpotentials in M-theory compactifications on
manifolds of G(2) holonomy,'' JHEP {\bf 0207}, 046 (2002),
hep-th/0203061.}

\lref\behr{K.~Behrndt and C.~Jeschek,
``Fluxes in M-theory on 7-manifolds and G structures,''
JHEP {\bf 0304}, 002 (2003), hep-th/0302047;
``Fluxes in M-theory on 7-manifolds: G-structures and
superpotential,'' hep-th/0311119.}

\lref\bertwo{K.~Behrndt and C.~Jeschek,
``Superpotentials from flux compactifications of M-theory,'', hep-th/0401019.}

\lref\gray{M. Fernandez and A. Gray, ``Riemannian manifolds with
structure group $G_2$,'' Ann. Mat. Pura. Appl. {\bf 32} (1982),
19-45.}

\lref\graytwo{M. Fernandez and L. Ugarte,
``Dolbeault cohomology for $G_2$ manifolds,''
Geom. Dedicata, {\bf 70} (1998) 57.}

\lref\ivan{T.~Friedrich and S.~Ivanov,
 ``Parallel spinors and connections with skew-symmetric torsion
  in string theory,'' math.dg/0102142;
T.~Friedrich and S.~Ivanov,
``Killing spinor equations in dimension 7 and geometry of integrable
 $G_2$-manifolds,'' math.dg/0112201.
P.~Ivanov and S.~Ivanov,
``SU(3)-instantons and $G_2$, Spin(7)-heterotic string solitons,'' math.dg/0312094.}

\lref\tseytii{A.~A.~Tseytlin,
  ``Harmonic superpositions of M-branes,''
  Nucl.\ Phys.\ B {\bf 475}, 149 (1996), hep-th/9604035;
  ``Composite BPS configurations of p-branes in 10 and 11 dimensions,''
  Class.\ Quant.\ Grav.\  {\bf 14}, 2085 (1997), hep-th/9702163.}

\lref\tp{G.~Papadopoulos and A.~A.~Tseytlin, ``Complex geometry of conifolds
and 5-brane wrapped on 2-sphere,''
Class.\ Quant.\ Grav.\  {\bf 18}, 1333 (2001).hep-th/0012034.}

\lref\papad{S.~Ivanov and G.~Papadopoulos,
  ``A no-go theorem for string warped compactifications,''
  Phys.\ Lett.\ B {\bf 497}, 309 (2001).}

\lref\smit{B.~de Wit, D.~J.~Smit and N.~D.~Hari Dass,
  ``Residual Supersymmetry Of Compactified D = 10 Supergravity,''
  Nucl.\ Phys.\ B {\bf 283}, 165 (1987).}

\lref\lust{G.~L.~Cardoso, G.~Curio, G.~Dall'Agata, D.~Lust, P.~Manousselis and G.~Zoupanos,
``Non-Kaehler string backgrounds and their five torsion classes,''
Nucl.\ Phys.\ B {\bf 652}, 5 (2003), hep-th/0211118.}

\lref\louis{S.~Gurrieri, J.~Louis, A.~Micu and D.~Waldram,
``Mirror symmetry in generalized Calabi-Yau compactifications,''
Nucl.\ Phys.\ B {\bf 654}, 61 (2003), hep-th/0211102.}

\lref\intril{
  K.~A.~Intriligator,
  ``'Integrating in' and exact superpotentials in 4-d,''
  Phys.\ Lett.\ B {\bf 336}, 409 (1994), hep-th/9407106.}

\lref\rstrom{A.~Strominger, ``Superstrings with torsion,'' Nucl.\
Phys.\ B {\bf 274}, 253 (1986).}

\lref\mal{J.~M.~Maldacena,
``The large N limit of superconformal field theories and supergravity,''
Adv.\ Theor.\ Math.\ Phys.\  {\bf 2}, 231 (1998)
[Int.\ J.\ Theor.\ Phys.\  {\bf 38}, 1113 (1999), hep-th/9711200.}

\lref\ks{I.~R.~Klebanov and M.~J.~Strassler,
``Supergravity and a confining gauge theory: Duality cascades and
chiSB-resolution of naked singularities,'' JHEP {\bf 0008}, 052 (2000), hep-th/0007191.}

\lref\klebts{I.~R.~Klebanov and A.~A.~Tseytlin,
  ``Gravity duals of supersymmetric SU(N) x SU(N+M) gauge theories,''
  Nucl.\ Phys.\ B {\bf 578}, 123 (2000), hep-th/0002159.}

\lref\mn{J.~M.~Maldacena and C.~Nunez,
``Towards the large N limit of pure N = 1 super Yang Mills,''
Phys.\ Rev.\ Lett.\  {\bf 86}, 588 (2001), hep-th/0008001.}

\lref\gcon{S.~Ferrara, L.~Girardello and H.~P.~Nilles,
  ``Breakdown Of Local Supersymmetry Through Gauge Fermion Condensates,''
  Phys.\ Lett.\ B {\bf 125}, 457 (1983);
M.~Dine, R.~Rohm, N.~Seiberg and E.~Witten,
  ``Gluino Condensation In Superstring Models,''
  Phys.\ Lett.\ B {\bf 156}, 55 (1985);
J.~P.~Derendinger, L.~E.~Ibanez and H.~P.~Nilles,
  ``On The Low-Energy D = 4, ${\cal N}=1$ Supergravity Theory
Extracted From The D = 10,
  ${\cal N}=1$ Superstring,''
  Phys.\ Lett.\ B {\bf 155}, 65 (1985).}

\lref\dabbie{A.~Sen,
  ``Orbifolds of M-Theory and String Theory,''
  Mod.\ Phys.\ Lett.\ A {\bf 11}, 1339 (1996), hep-th/9603113;
``Duality and Orbifolds,''

  Nucl.\ Phys.\ B {\bf 474}, 361 (1996), hep-th/9604070;
A.~Dabholkar,
  ``Lectures on orientifolds and duality,''
{\it Trieste 1997, High energy physics and cosmology 128-191},
 hep-th/9804208.}

\lref\dvo{
  R.~Dijkgraaf and C.~Vafa,
  ``A perturbative window into non-perturbative physics,''
  hep-th/0208048.}

\lref\dvt{
  R.~Dijkgraaf and C.~Vafa,
  ``Matrix models, topological strings, and supersymmetric gauge theories,''
  Nucl.\ Phys.\ B {\bf 644}, 3 (2002),hep-th/0206255.}

\lref\vafai{C.~Vafa,
``Superstrings and topological strings at large N,''
J.\ Math.\ Phys.\  {\bf 42}, 2798 (2001), hep-th/0008142.}

\lref\civ{F.~Cachazo, K.~A.~Intriligator and C.~Vafa,
``A large N duality via a geometric transition,''
Nucl.\ Phys.\ B {\bf 603}, 3 (2001), hep-th/0103067;
J.~D.~Edelstein, K.~Oh and R.~Tatar,
  ``Orientifold, geometric transition and large N duality for SO/Sp gauge
  theories,''
  JHEP {\bf 0105}, 009 (2001), hep-th/0104037.}

\lref\syz{A.~Strominger, S.~T.~Yau and E.~Zaslow,
``Mirror symmetry is T-duality,''
Nucl.\ Phys.\ B {\bf 479}, 243 (1996), hep-th/9606040.}

\lref\tduality{E.~Bergshoeff, C.~M.~Hull and T.~Ortin,
``Duality in the type II superstring effective action,''
Nucl.\ Phys.\ B {\bf 451}, 547 (1995), hep-th/9504081;
P.~Meessen and T.~Ortin,
``An Sl(2,Z) multiplet of nine-dimensional type II supergravity theories,''
Nucl.\ Phys.\ B {\bf 541}, 195 (1999), hep-th/9806120.}

\lref\eot{J.~D.~Edelstein, K.~Oh and R.~Tatar,
``Orientifold,
 geometric transition and large N duality for SO/Sp gauge  theories,''
JHEP {\bf 0105}, 009 (2001), hep-th/0104037.}

\lref\dotu{K.~Dasgupta, K.~Oh and R.~Tatar, {``Geometric
transition, large N dualities and MQCD dynamics,''} Nucl.\ Phys.\
B {\bf 610}, 331 (2001), hep-th/0105066; {``Open/closed string
dualities and Seiberg duality from geometric transitions in
M-theory,''} JHEP {\bf 0208}, 026 (2002), hep-th/0106040;
K.~h.~Oh and R.~Tatar,
  ``Duality and confinement in N = 1 supersymmetric theories from geometric
  transitions,''
  Adv.\ Theor.\ Math.\ Phys.\  {\bf 6}, 141 (2003), hep-th/0112040.}

\lref\dotd{K.~Dasgupta, K.~h.~Oh, J.~Park and R.~Tatar, ``Geometric
transition versus cascading solution,'' JHEP {\bf 0201}, 031
(2002), hep-th/0110050.}

\lref\ohta{K.~Ohta and T.~Yokono,
``Deformation of conifold and intersecting branes,''
JHEP {\bf 0002}, 023 (2000), hep-th/9912266.}

\lref\dott{K.~h.~Oh and R.~Tatar,
``Duality and confinement
in N = 1 supersymmetric theories from geometric  transitions,''
Adv.\ Theor.\ Math.\ Phys.\  {\bf 6}, 141 (2003), hep-th/0112040.}

\lref\edelstein{J.~D.~Edelstein and C.~Nunez,
``D6 branes and M-theory geometrical transitions from gauged  supergravity,''
JHEP {\bf 0104}, 028 (2001), hep-th/0103167.}

\lref\chsw{P.~Candelas, G.~T.~Horowitz, A.~Strominger and E.~Witten,
  ``Vacuum Configurations For Superstrings,''
  Nucl.\ Phys.\ B {\bf 258}, 46 (1985).}

\lref\candelas{P.~Candelas and X.~C.~de la Ossa, ``Comments on
conifolds,'' Nucl.\ Phys.\ B {\bf 342}, 246 (1990).}

\lref\minakas{
   P. Kaste, Ruben Minasian, M. Petrini, A. Tomasiello
  ``Nontrivial RR two-form field strength and SU(3)-structure''
   Fortsch.Phys. {\bf 51} 764 (2003), hep-th/0412187.
}

\lref\minasianone{R.~Minasian and D.~Tsimpis,
``Hopf reductions, fluxes and branes,''
Nucl.\ Phys.\ B {\bf 613}, 127 (2001), hep-th/0106266.}

\lref\robbins{K.~Dasgupta, G.~Rajesh, D.~Robbins and S.~Sethi,
``Time-dependent warping, fluxes, and NCYM,''
JHEP {\bf 0303}, 041 (2003), hep-th/0302049;
K.~Dasgupta and M.~Shmakova,
``On branes and oriented B-fields,''
Nucl.\ Phys.\ B {\bf 675}, 205 (2003), hep-th/0306030.}

\lref\gstructure{
  M.~Falcitelli, A.~Farinola and S.~Salamon,
    ``Almost--Hermitian Geometry'', Diff. Geo {\bf 4} (1994) 259;
  T.~Friedrich and S.~Ivanov, ``Parallel spinors and connections with skew-symmetric
    torsion in string theory,'' math.dg/0102142;
  S.~Salamon, ``Almost Parallel Structures,''
    Contemp. Math. {\bf 288} (2001), 162-181, math.DG/0107146.}

\lref\gauntlett{J.~P.~Gauntlett, D.~Martelli and D.~Waldram,
    ``Superstrings with intrinsic torsion,''
    Phys.\ Rev.\ D {\bf 69}, 086002 (2004) [arXiv:hep-th/0302158];
    [the former] and S.~Pakis ``G-structures and wrapped NS5-branes,''
    Commun.\ Math.\ Phys.\  {\bf 247}, 421 (2004), hep-th/0205050.}

\lref\lust{G.~L.~Cardoso, G.~Curio, G.~Dall'Agata, D.~Lust, P.~Manousselis and G.~Zoupanos,
``Non-Kaehler string backgrounds and their five torsion classes,''
Nucl.\ Phys.\ B {\bf 652}, 5 (2003), hep-th/0211118.}

\lref\salamon{ S. Chiossi, S. Salamon,
``The intrinsic torsion of $SU(3)$ and $G_2$ structures,''
 Proc. conf. Differential Geometry Valencia 2001, math.DG/0202282.}

\lref\svw{S.~Sethi, C.~Vafa and E.~Witten,
``Constraints on low-dimensional string compactifications,''
Nucl.\ Phys.\ B {\bf 480}, 213 (1996), hep-th/9606122;
K.~Dasgupta and S.~Mukhi,
``A note on low-dimensional string compactifications,''
Phys.\ Lett.\ B {\bf 398}, 285 (1997), hep-th/9612188.}

\lref\kachruone{S.~Kachru, M.~B.~Schulz, P.~K.~Tripathy and S.~P.~Trivedi,
``New supersymmetric string compactifications,''
JHEP {\bf 0303}, 061 (2003), hep-th/0211182;
S.~Kachru, M.~B.~Schulz and S.~Trivedi,
``Moduli stabilization from fluxes in a simple IIB orientifold,''
JHEP {\bf 0310}, 007 (2003), hep-th/0201028.}

\lref\hitchin{N. ~Hitchin,
``Stable forms and special metrics'',
Contemp. Math., {\bf 288}, Amer. Math. Soc. (2000).}

\lref\giveon{S.~S.~Gubser,
 ``Supersymmetry and F-theory realization of the deformed conifold with
three-form flux,'' hep-th/0010010;
A.~Giveon, A.~Kehagias and H.~Partouche,
``Geometric transitions, brane dynamics and gauge theories,''
JHEP {\bf 0112}, 021 (2001), hep-th/0110115.}

\lref\bonan{E. Bonan,
``Sur le varietes remanniennes a groupe d'holonomie $G_2$ ou Spin(7),''
C. R. Acad. Sci. paris {\bf 262} (1966) 127.}



\lref\amv{M.~Atiyah, J.~M.~Maldacena and C.~Vafa,
``An M-theory flop as a large N duality,''
J.\ Math.\ Phys.\  {\bf 42}, 3209 (2001), hep-th/0011256.}

\lref\realm{S.~Alexander, K.~Becker, M.~Becker, K.~Dasgupta,
A.~Knauf and R.~Tatar,
  ``In the realm of the geometric transitions,''
  Nucl.\ Phys.\ B {\bf 704}, 231 (2005), hep-th/0408192.}

\lref\cvetic{M.~Cvetic, G.~W.~Gibbons, H.~Lu and C.~N.~Pope,
  ``Ricci-flat metrics, harmonic forms and brane resolutions,''
  Commun.\ Math.\ Phys.\  {\bf 232}, 457 (2003), hep-th/0012011.}

\lref\hulltown{C.~M.~Hull and P.~K.~Townsend,
  ``Finiteness And Conformal Invariance In Nonlinear Sigma Models,''
  Nucl.\ Phys.\ B {\bf 274}, 349 (1986);
``The Two Loop Beta Function For Sigma Models With Torsion,''
  Phys.\ Lett.\ B {\bf 191}, 115 (1987).}

\lref\hullwit{C.~M.~Hull and E.~Witten,
  ``Supersymmetric Sigma Models And The Heterotic String,''
  Phys.\ Lett.\ B {\bf 160}, 398 (1985).}

\lref\gross{D.~J.~Gross, J.~A.~Harvey, E.~J.~Martinec and R.~Rohm,
  ``The Heterotic String,''
  Phys.\ Rev.\ Lett.\  {\bf 54}, 502 (1985);
``Heterotic String Theory. 1. The Free Heterotic String,''
  Nucl.\ Phys.\ B {\bf 256}, 253 (1985);
``Heterotic String Theory. 2. The Interacting Heterotic String,''
  Nucl.\ Phys.\ B {\bf 267}, 75 (1986).}

\lref\syz{A.~Strominger, S.~T.~Yau and E.~Zaslow,
``Mirror symmetry is T-duality,''
Nucl.\ Phys.\ B {\bf 479}, 243 (1996), hep-th/9606040.}

\lref\tduality{E.~Bergshoeff, C.~M.~Hull and T.~Ortin,
``Duality in the type II superstring effective action,''
Nucl.\ Phys.\ B {\bf 451}, 547 (1995), hep-th/9504081;
P.~Meessen and T.~Ortin,
``An Sl(2,Z) multiplet of nine-dimensional type II supergravity theories,''
Nucl.\ Phys.\ B {\bf 541}, 195 (1999), hep-th/9806120.}

\lref\adoptone{J.~D.~Edelstein, K.~Oh and R.~Tatar,
``Orientifold, geometric transition and large N duality for SO/Sp
gauge theories,'' JHEP {\bf 0105}, 009 (2001), hep-th/0104037.}

\lref\grifhar{P.~Griffiths and J.~Harris, \underbar{Principles
of Algebraic Geometry}, Wiley, New York 1978.}

\lref\hart{R.~Hartshorne, \underbar{Algebraic Geometry}, Springer-Verlag,
Berlin, 1977.}

\lref\cortismith{A.~Corti and I.~Smith, ``Conifold transitions and Mori
theory'', math.SG/0501043.}

\lref\ohta{K.~Ohta and T.~Yokono,
``Deformation of conifold and intersecting branes,''
JHEP {\bf 0002}, 023 (2000), hep-th/9912266.}

\lref\adoptfo{K.~h.~Oh and R.~Tatar, ``Duality and confinement in
N = 1 supersymmetric theories from geometric transitions,'' Adv.\
Theor.\ Math.\ Phys.\  {\bf 6}, 141 (2003), hep-th/0112040.}

\lref\edelstein{J.~D.~Edelstein and C.~Nunez, ``D6 branes and
M-theory geometrical transitions from gauged supergravity,'' JHEP
{\bf 0104}, 028 (2001), hep-th/0103167.}

\lref\candelas{P.~Candelas and X.~C.~de la Ossa,
``Comments On Conifolds,''
Nucl.\ Phys.\ B {\bf 342}, 246 (1990).}

\lref\tsimpis{R.~Minasian and D.~Tsimpis,
  ``On the geometry of non-trivially embedded branes,''
  Nucl.\ Phys.\ B {\bf 572}, 499 (2000), hep-th/9911042.}

\lref\minasianone{R.~Minasian and D.~Tsimpis,
``Hopf reductions, fluxes and branes,''
Nucl.\ Phys.\ B {\bf 613}, 127 (2001), hep-th/0106266.}

\lref\pandoz{L.~A.~Pando Zayas and A.~A.~Tseytlin,
``3-branes on resolved conifold,''
JHEP {\bf 0011}, 028 (2000), hep-th/0010088.}

\lref\rBB{K.~Becker and M.~Becker, ``M-Theory on
eight-manifolds,'' Nucl.\ Phys.\ B {\bf 477}, 155 (1996),
hep-th/9605053.}

\lref\bbdgs{K.~Becker, M.~Becker, P.~S.~Green, K.~Dasgupta and
E.~Sharpe, ``Compactifications of heterotic strings on
non-K\"ahler complex manifolds. II,'' Nucl.\ Phys.\ B {\bf 678},
19 (2004), hep-th/0310058.}

\lref\bbdg{K.~Becker, M.~Becker, K.~Dasgupta and P.~S.~Green,
``Compactifications of heterotic theory on non-K\"ahler complex
manifolds. I,'' JHEP {\bf 0304}, 007 (2003), hep-th/0301161.}

\lref\hull{C.~M.~Hull,
  ``Compactifications Of The Heterotic Superstring,''
  Phys.\ Lett.\ B {\bf 178}, 357 (1986); C.~M.~Hull and E.~Witten,
  ``Supersymmetric Sigma Models And The Heterotic String,''
  Phys.\ Lett.\ B {\bf 160}, 398 (1985); C.~M.~Hull,
  ``Superstring Compactifications With Torsion And Space-Time Supersymmetry,''
Print-86-0251 (CAMBRIDGE), Published in Turin Superunif.1985:347.}

\lref\gates{J.~M.~Gates, C.~M.~Hull and M.~Rocek,
  ``Twisted Multiplets And New Supersymmetric Nonlinear Sigma Models,''
  Nucl.\ Phys.\ B {\bf 248}, 157 (1984).}

\lref\bbdp{K.~Becker, M.~Becker, K.~Dasgupta and S.~Prokushkin,
  ``Properties of heterotic vacua from superpotentials,''
  Nucl.\ Phys.\ B {\bf 666}, 144 (2003), hep-th/0304001.}

\lref\GP{E.~Goldstein and S.~Prokushkin,
 ``Geometric model for complex non-K\"ahler manifolds with SU(3) structure,''
  Commun.\ Math.\ Phys.\  {\bf 251}, 65 (2004), hep-th/0212307.}

\lref\townsend{P.~K.~Townsend,
``D-branes from M-branes,''
Phys.\ Lett.\ B {\bf 373}, 68 (1996), hep-th/9512062.}

\lref\sav{K.~Dasgupta, G.~Rajesh and S.~Sethi,
``M theory, orientifolds and G-flux,''
JHEP {\bf 9908}, 023 (1999), hep-th/9908088.}

\lref\beckerD{K.~Becker and K.~Dasgupta,
``Heterotic strings with torsion,''
JHEP {\bf 0211}, 006 (2002), hep-th/0209077.}

\lref\lisheng{K.~Becker and L.~S.~Tseng,
  ``Heterotic flux compactifications and their moduli,''
  Nucl.\ Phys.\ B {\bf 741}, 162 (2006), hep-th/0509131.}

\lref\ks{I.~R.~Klebanov and M.~J.~Strassler,
 ``Supergravity and a confining gauge theory: Duality cascades
  and $\chi_{SB}$-resolution of naked singularities,''
JHEP {\bf 0008}, 052 (2000), hep-th/0007191.}

\lref\radu{I.~Bena, R.~Roiban and R.~Tatar,
  ``Baryons, boundaries and matrix models,''
  Nucl.\ Phys.\ B {\bf 679}, 168 (2004), hep-th/0211271;~R.~Roiban, R.~Tatar and J.~Walcher,
``Massless flavor in geometry and matrix models,''
Nucl.\ Phys.\ B {\bf 665}, 211 (2003), hep-th/0301217;~K.~Landsteiner, C.~I.~Lazaroiu and R.~Tatar,
``(Anti)symmetric matter and superpotentials from IIB orientifolds,''
JHEP {\bf 0311}, 044 (2003), hep-th/0306236;~``Chiral field Theories, Konishi Anomalies and Matrix Models'',
JHEP {\bf 0402}, 044 (2004), hep-th/0307182,~``Chiral field theories from conifolds,''
JHEP {\bf 0311}, 057 (2003), hep-th/0310052;~K.~Landsteiner, C.~I.~Lazaroiu and R.~Tatar,
``Puzzles for matrix models of chiral field theories,'' hep-th/0311103.}

\lref\gv{R.~Gopakumar and C.~Vafa,
``On the gauge theory/geometry correspondence,''
Adv.\ Theor.\ Math.\ Phys.\  {\bf 3}, 1415 (1999), hep-th/9811131.}

\lref\dvu{R.~Dijkgraaf and C.~Vafa,
``Matrix models, topological strings, and supersymmetric gauge theories,''
Nucl.\ Phys.\ B {\bf 644}, 3 (2002), hep-th/0206255.}

\lref\ps{J.~Polchinski and M.~J.~Strassler, ``The String Dual of a
Confining Four-Dimensional Gauge Theory ,'' hep-th/0003136.}

\lref\susskind{L.~Susskind,
``The anthropic landscape of string theory,'' hep-th/0302219.}

\lref\banks{T.~Banks, M.~Dine and E.~Gorbatov,
``Is there a string theory landscape?,'' hep-th/0309170.}

\lref\hori{K.~Hori and C.~Vafa,
``Mirror symmetry,'' hep-th/0002222;
M.~Aganagic, A.~Klemm and C.~Vafa,
``Disk instantons, mirror symmetry and the duality web,''
Z.\ Naturforsch.\ A {\bf 57}, 1 (2002), hep-th/0105045;
M.~Aganagic, A.~Klemm, M.~Marino and C.~Vafa,
``Matrix model as a mirror of Chern-Simons theory,''
JHEP {\bf 0402}, 010 (2004), hep-th/0211098.}

\lref\agavafa{M.~Aganagic and C.~Vafa,
  ``G(2) manifolds, mirror symmetry and geometric engineering,''
 hep-th/0110171.}

\lref\karch{M.~Aganagic, A.~Karch, D.~Lust and A.~Miemiec,
``Mirror symmetries for brane configurations and branes at singularities,''
Nucl.\ Phys.\ B {\bf 569}, 277 (2000), hep-th/9903093.}

\lref\ksschub{S.~Katz and S.A.\ Stromme, ``Schubert: a maple package
for intersection theory'', http://www.mi.uib.no/schubert/}

\lref\dmconi{K.~Dasgupta and S.~Mukhi,
``Brane constructions, conifolds and M-theory,''
Nucl.\ Phys.\ B {\bf 551}, 204 (1999), hep-th/9811139.}

\lref\gtone{M.~Becker, K.~Dasgupta, A.~Knauf and R.~Tatar,
  ``Geometric transitions, flops and non-K\"ahler manifolds. I,''
  Nucl.\ Phys.\ B {\bf 702}, 207 (2004), hep-th/0403288.}

\lref\gttwo{M.~Becker, K.~Dasgupta, S.~Katz, A.~Knauf and R.~Tatar,
  ``Geometric transitions, flops and non-Kaehler manifolds. II,''
  Nucl.\ Phys.\ B {\bf 738}, 124 (2006), hep-th/0511099.}

\lref\toappear{S.~Alexander, K.~Becker, M.~Becker, K.~Dasgupta,
A.~Knauf, R.~Tatar,
``In the realm of the geometric transitions,'' hep-th/0408192.}

\lref\dvd{R.~Dijkgraaf and C.~Vafa,
``A perturbative window into non-perturbative physics,'' hep-th/0208048.}

\lref\civu{F.~Cachazo, S.~Katz and C.~Vafa,
``Geometric transitions and N = 1 quiver theories,'' hep-th/0108120.}

\lref\civd{F.~Cachazo, B.~Fiol, K.~A.~Intriligator, S.~Katz and C.~Vafa,
``A geometric unification of dualities,'' Nucl.\ Phys.\ B {\bf 628}, 3 (2002),
hep-th/0110028.}

\lref\radu{R.~Roiban, R.~Tatar and J.~Walcher,
``Massless flavor in geometry and matrix models,''
Nucl.\ Phys.\ B {\bf 665}, 211 (2003), hep-th/0301217.}

\lref\radd{K.~Landsteiner, C.~I.~Lazaroiu and R.~Tatar,
``(Anti)symmetric matter and superpotentials from IIB orientifolds,''
JHEP {\bf 0311}, 044 (2003), hep-th/0306236.}

\lref\radt{K.~Landsteiner, C.~I.~Lazaroiu and R.~Tatar,
``Chiral field theories from conifolds,''
JHEP {\bf 0311}, 057 (2003), hep-th/0310052.}

\lref\radp{K.~Landsteiner, C.~I.~Lazaroiu and R.~Tatar,
``Puzzles for matrix models of chiral field theories,'' hep-th/0311103.}

\lref\ber{M.~Bershadsky, S.~Cecotti, H.~Ooguri and C.~Vafa,
 ``Kodaira-Spencer theory of gravity and exact results for quantum string amplitudes,''
Commun.\ Math.\ Phys.\  {\bf 165}, 311 (1994), hep-th/9309140.}

\lref\bismut{J. M. Bismut,
``A local index theorem for non-K\"ahler manifolds,''
Math. Ann. {\bf 284} (1989) 681.}

\lref\monar{F. Cabrera, M. Monar and A. Swann,
``Classification of $G_2$ structures,''
J. London Math. Soc. {\bf 53} (1996) 98;
F. Cabrera,
``On Riemannian manifolds with $G_2$ structure,''
Bolletino UMI A {\bf 10} (1996) 98.}

\lref\kath{Th. Friedrich and I. Kath,
``7-dimensional compact Riemannian manifolds with killing spinors,''
Comm. Math. Phys. {\bf 133} (1990) 543;
Th. Friedrich, I. Kath, A. Moroianu and U. Semmelmann,
``On nearly parallel $G_2$ structures,'' J. geom. Phys. {\bf 23} (1997) 259;
S. Salamon,
``Riemannian geometry and holonomy groups,''
Pitman Res. Notes Math. Ser., {\bf 201} (1989);
V. Apostolov and S. Salamon,
``K\"ahler reduction of metrics with holonomy $G_2$,''
math-DG/0303197.}

\lref\ooguri{H.~Ooguri and C.~Vafa,
``The C-deformation of gluino and non-planar diagrams,''
Adv.\ Theor.\ Math.\ Phys.\  {\bf 7}, 53 (2003), hep-th/0302109;
``Gravity induced C-deformation,''
Adv.\ Theor.\ Math.\ Phys.\  {\bf 7}, 405 (2004), hep-th/0303063.}

\lref\tv{T.~R.~Taylor and C.~Vafa,
``RR flux on Calabi-Yau and partial supersymmetry breaking,''
Phys.\ Lett.\ B {\bf 474}, 130 (2000), hep-th/9912152.}

\lref\wittencoho{E.~Witten,
  ``Topological Sigma Models,''
  Commun.\ Math.\ Phys.\  {\bf 118}, 411 (1988);
``Mirror manifolds and topological field theory,''
 hep-th/9112056; ``Topological Quantum Field Theory,''
  Commun.\ Math.\ Phys.\  {\bf 117}, 353 (1988).}

\lref\toprev{M.~Marino,
  ``Les Houches lectures on matrix models and topological strings,''
  hep-th/0410165; A.~Neitzke and C.~Vafa,
  ``Topological strings and their physical applications,''
  hep-th/0410178; M.~Vonk,
  ``A mini-course on topological strings,'' hep-th/0504147.}

\lref\hitchin{N.~Hitchin,
  ``Generalized Calabi-Yau manifolds,''
  Quart.\ J.\ Math.\ Oxford Ser.\  {\bf 54}, 281 (2003) math.dg/0209099.}

\lref\gualtieri{M.~Gualtieri,
``Generalised Complex Geometry,''
math.DG/0401221.}

\lref\yaufu{J.~X.~Fu and S.~T.~Yau,
  ``Existence of supersymmetric Hermitian metrics with torsion on non-Kaehler
  manifolds,'' hep-th/0509028.}

\lref\yauli{J.~Li and S.~T.~Yau,
  ``The existence of supersymmetric string theory with torsion,''
  hep-th/0411136; ``Hermitian Yang-Mills Connection On Nonkahler Manifolds,''
in {\it San Diego 1986, Proceedings, Mathematical Aspects of String Theory} 560-573.}

\lref\kapuli{A.~Kapustin and Y.~Li,
  ``Topological sigma-models with H-flux and twisted generalized complex
  manifolds,''
  hep-th/0407249.}

\lref\sensethi{A.~Sen and S.~Sethi,
  ``The mirror transform of type I vacua in six dimensions,''
  Nucl.\ Phys.\ B {\bf 499}, 45 (1997)
  hep-th/9703157; J.~de Boer, R.~Dijkgraaf, K.~Hori,
A.~Keurentjes, J.~Morgan, D.~R.~Morrison and S.~Sethi,
  ``Triples, fluxes, and strings,''
  Adv.\ Theor.\ Math.\ Phys.\  {\bf 4}, 995 (2002)
 hep-th/0103170; D.~R.~Morrison and S.~Sethi,
  ``Novel type I compactifications,''
  JHEP {\bf 0201}, 032 (2002)
  hep-th/0109197.}

\lref\wittentop{E.~Witten,
  ``Mirror manifolds and topological field theory,''
  hep-th/9112056;  In Yau, S.T. (ed.): {\it Mirror symmetry I} 121.}

\lref\malikov{F. ~Malikov, V. ~Schechtman, A. ~Vaintrob,
``Chiral de Rham complex'', math.AG/9803041.}

\lref\schect{ F. ~ Malikov, V. ~Schechtman,
``Chiral de Rham complex. II,''
math.AG/9901065; ``Chiral Poincar\'e duality,'' math.AG/9905008;
V.~ Gorbounov, F. ~Malikov, V. ~Schechtman,
``Gerbes of chiral differential operators,'' math.AG/9906117.}

\lref\zerotwo{J.~Distler,
  ``Resurrecting (2,0) Compactifications,''
  Phys.\ Lett.\ B {\bf 188}, 431 (1987);
J.~Distler and B.~R.~Greene,
  ``Aspects Of (2,0) String Compactifications,''
  Nucl.\ Phys.\ B {\bf 304}, 1 (1988);
J.~Distler and S.~Kachru,
  ``(0,2) Landau-Ginzburg theory,''
  Nucl.\ Phys.\ B {\bf 413}, 213 (1994), hep-th/9309110;
J.~Distler and S.~Kachru,
  ``Duality of (0,2) string vacua,''
  Nucl.\ Phys.\ B {\bf 442}, 64 (1995), hep-th/9501111;
T.~M.~Chiang, J.~Distler and B.~R.~Greene,
  ``Some features of (0,2) moduli space,''
  Nucl.\ Phys.\ B {\bf 496}, 590 (1997), hep-th/9702030.}

\lref\gsvy{B.~R.~Greene, A.~D.~Shapere, C.~Vafa and S.~T.~Yau,
  ``Stringy Cosmic Strings And Noncompact Calabi-Yau Manifolds,''
  Nucl.\ Phys.\ B {\bf 337}, 1 (1990).}

\lref\reczt{A.~Adams, A.~Basu and S.~Sethi,
  ``(0,2) duality,''
  Adv.\ Theor.\ Math.\ Phys.\  {\bf 7}, 865 (2004), hep-th/0309226;
S.~Katz and E.~Sharpe,
  ``Notes on certain (0,2) correlation functions,''
Commun.\ Math.\ Phys.\  {\bf 262}, 611 (2006),  hep-th/0406226;
E.~Sharpe,
  ``Notes on correlation functions in (0,2) theories,'' hep-th/0502064;
``Notes on certain other (0,2) correlation functions,'' hep-th/0605005.}

\lref\wittwo{E.~Witten,
  ``Two-dimensional models with (0,2) supersymmetry: Perturbative aspects,''
  hep-th/0504078.}

\lref\zumino{B.~Zumino,
  ``Supersymmetry And Kahler Manifolds,''
  Phys.\ Lett.\ B {\bf 87}, 203 (1979).}

\lref\horwit{P.~Horava and E.~Witten,
  ``Heterotic and type I string dynamics from eleven dimensions,''
  Nucl.\ Phys.\ B {\bf 460}, 506 (1996), hep-th/9510209;
``Eleven-Dimensional Supergravity on a Manifold with Boundary,''
  Nucl.\ Phys.\ B {\bf 475}, 94 (1996), hep-th/9603142.}

\lref\dmone{K.~Dasgupta and S.~Mukhi,
  ``Orbifolds of M-theory,''
  Nucl.\ Phys.\ B {\bf 465}, 399 (1996), hep-th/9512196;
E.~Witten,
  ``Five-branes and M-theory on an orbifold,''
  Nucl.\ Phys.\ B {\bf 463}, 383 (1996), hep-th/9512219.}

\lref\kimn{J.~P.~Gauntlett, N.~Kim, D.~Martelli and D.~Waldram,
  ``Wrapped fivebranes and ${\cal N} = 2$ super Yang-Mills theory,''
  Phys.\ Rev.\ D {\bf 64}, 106008 (2001), hep-th/0106117;
F.~Bigazzi, A.~L.~Cotrone and A.~Zaffaroni,
  ``${\cal N} = 2$ gauge theories from wrapped five-branes,''
  Phys.\ Lett.\ B {\bf 519}, 269 (2001), hep-th/0106160;
R.~Apreda, F.~Bigazzi, A.~L.~Cotrone, M.~Petrini and A.~Zaffaroni,
  ``Some comments on ${\cal N} = 1$ gauge theories from wrapped branes,''
  Phys.\ Lett.\ B {\bf 536}, 161 (2002), hep-th/0112236;
P.~Di Vecchia, A.~Lerda and P.~Merlatti,
  ``${\cal N} = 1$ and ${\cal N} = 2$ super Yang-Mills theories from wrapped branes,''
  Nucl.\ Phys.\ B {\bf 646}, 43 (2002), hep-th/0205204;
F.~Bigazzi, A.~L.~Cotrone, M.~Petrini and A.~Zaffaroni,
  ``Supergravity duals of supersymmetric four dimensional gauge theories,''
  Riv.\ Nuovo Cim.\  {\bf 25N12}, 1 (2002), hep-th/0303191.}

\lref\axell{G.~Curio and A.~Krause,
  ``Four-flux and warped heterotic M-theory compactifications,''
  Nucl.\ Phys.\ B {\bf 602}, 172 (2001), hep-th/0012152.}

\lref\axelk{G.~Curio and A.~Krause,
  ``G-fluxes and non-perturbative stabilisation of heterotic M-theory,''
  Nucl.\ Phys.\ B {\bf 643}, 131 (2002), hep-th/0108220;
M.~Becker, G.~Curio and A.~Krause,
  ``De Sitter vacua from heterotic M-theory,''
  Nucl.\ Phys.\ B {\bf 693}, 223 (2004), hep-th/0403027;
G.~Curio, A.~Krause and D.~Lust,
  ``Moduli stabilization in the heterotic / IIB discretuum,'' hep-th/0502168.}

\lref\sezgin{A.~Salam and E.~Sezgin,
  ``$SO(4)$ Gauging Of N=2 Supergravity In Seven-Dimensions,''
  Phys.\ Lett.\ B {\bf 126}, 295 (1983).}

\lref\buchkov{E.~I.~Buchbinder and B.~A.~Ovrut,
  ``Vacuum stability in heterotic M-theory,''
  Phys.\ Rev.\ D {\bf 69}, 086010 (2004), hep-th/0310112.}

\lref\jefgreg{R.~Gregory, J.~A.~Harvey and G.~W.~Moore,
  ``Unwinding strings and T-duality of Kaluza-Klein and H-monopoles,''
  Adv.\ Theor.\ Math.\ Phys.\  {\bf 1}, 283 (1997), hep-th/9708086.}

\lref\senbanks{A.~Sen,
  ``F-theory and Orientifolds,''
  Nucl.\ Phys.\ B {\bf 475}, 562 (1996), hep-th/9605150;
T.~Banks, M.~R.~Douglas and N.~Seiberg,
  ``Probing F-theory with branes,''
  Phys.\ Lett.\ B {\bf 387}, 278 (1996), hep-th/9605199.}

\lref\senF{A.~Sen,
  ``Orientifold limit of F-theory vacua,''
  Phys.\ Rev.\ D {\bf 55}, 7345 (1997), hep-th/9702165;
  ``Orientifold limit of F-theory vacua,''
  Nucl.\ Phys.\ Proc.\ Suppl.\  {\bf 68}, 92 (1998), hep-th/9709159.}

\lref\ouyang{P.~Ouyang,
  ``Holomorphic D7-branes and flavored N = 1 gauge theories,''
  Nucl.\ Phys.\ B {\bf 699}, 207 (2004), hep-th/0311084.}

\lref\orione{S.~Chakravarty, K.~Dasgupta, O.~J.~Ganor and G.~Rajesh,
  ``Pinned branes and new non Lorentz invariant theories,''
  Nucl.\ Phys.\ B {\bf 587}, 228 (2000), hep-th/0002175.}

\lref\difuli{A.~Bergman and O.~J.~Ganor,
  ``Dipoles, twists and noncommutative gauge theory,''
  JHEP {\bf 0010}, 018 (2000), hep-th/0008030;
K.~Dasgupta, O.~J.~Ganor and G.~Rajesh,
  ``Vector deformations of N = 4 super-Yang-Mills theory, pinned branes,  and
  arched strings,''
  JHEP {\bf 0104}, 034 (2000), hep-th/0010072;
A.~Bergman, K.~Dasgupta, O.~J.~Ganor, J.~L.~Karczmarek and G.~Rajesh,
  ``Nonlocal field theories and their gravity duals,''
  Phys.\ Rev.\ D {\bf 65}, 066005 (2002), hep-th/0103090;
K.~Dasgupta and M.~M.~Sheikh-Jabbari,
  ``Noncommutative dipole field theories,''
  JHEP {\bf 0202}, 002 (2002), hep-th/0112064.}

\lref\ramdam{O.~J.~Ganor and U.~Varadarajan,
  ``Nonlocal effects on D-branes in plane-wave backgrounds,''
  JHEP {\bf 0211}, 051 (2002), hep-th/0210035;
M.~Alishahiha and O.~J.~Ganor,
  ``Twisted backgrounds, pp-waves and nonlocal field theories,''
  JHEP {\bf 0303}, 006 (2003), hep-th/0301080;
D.~W.~Chiou and O.~J.~Ganor,
  ``Noncommutative dipole field theories and unitarity,''
  JHEP {\bf 0403}, 050 (2004), hep-th/0310233.}

\lref\fawad{S.~F.~Hassan,
  ``T-duality, space-time spinors and R-R fields in curved backgrounds,''
  Nucl.\ Phys.\ B {\bf 568}, 145 (2000), hep-th/9907152.}

\lref\ganorha{O.~J.~Ganor and A.~Hanany,
  ``Small $E_8$ Instantons and Tensionless Non-critical Strings,''
  Nucl.\ Phys.\ B {\bf 474}, 122 (1996), hep-th/9602120.}

\lref\dall{G.~Dall'Agata and N.~Prezas,
  ``${\cal N} = 1$ geometries for M-theory and type IIA strings with fluxes,''
  Phys.\ Rev.\ D {\bf 69}, 066004 (2004), hep-th/0311146;
  ``Scherk-Schwarz reduction of M-theory on G2-manifolds with fluxes,''
  JHEP {\bf 0510}, 103 (2005), hep-th/0509052.}

\lref\lilia{L.~Anguelova and D.~Vaman,
  ``${\cal R}^4$ corrections to heterotic M-theory,'' hep-th/0506191;
P.~Manousselis, N.~Prezas and G.~Zoupanos,
  ``Supersymmetric compactifications of heterotic strings with fluxes and
  condensates,'' hep-th/0511122.}

\lref\strobl{P.~Schaller and T.~Strobl,
  ``Poisson structure induced (topological) field theories,''
  Mod.\ Phys.\ Lett.\ A {\bf 9}, 3129 (1994), hep-th/9405110.}

\lref\tscvetic{M.~Cvetic and A.~A.~Tseytlin,
  ``General class of BPS saturated dyonic black holes as exact superstring
  solutions,''
  Phys.\ Lett.\ B {\bf 366}, 95 (1996), hep-th/9510097;
``Solitonic strings and BPS saturated dyonic black holes,''
  Phys.\ Rev.\ D {\bf 53}, 5619 (1996)
  [Erratum-ibid.\ D {\bf 55}, 3907 (1997)], hep-th/9512031.}

\lref\afm{O.~Aharony, A.~Fayyazuddin and J.~M.~Maldacena,
  ``The large N limit of N = 2,1 field theories from three-branes in
  F-theory,''
  JHEP {\bf 9807}, 013 (1998), hep-th/9806159.}

\lref\witp{E.~Witten,
  ``Bound states of strings and p-branes,''
  Nucl.\ Phys.\ B {\bf 460}, 335 (1996), hep-th/9510135;
E.~Gava, K.~S.~Narain and M.~H.~Sarmadi,
  ``On the bound states of p- and (p+2)-branes,''
  Nucl.\ Phys.\ B {\bf 504}, 214 (1997), hep-th/9704006.}

\lref\nifuli{O.~Lunin and J.~Maldacena,
  ``Deforming field theories with $U(1) \times U(1)$ global symmetry and their gravity
  duals,''
  JHEP {\bf 0505}, 033 (2005), hep-th/0502086;
U.~Gursoy and C.~Nunez,
  ``Dipole deformations of N = 1 SYM and supergravity backgrounds with $U(1) \times
  U(1)$ global symmetry,''
  Nucl.\ Phys.\ B {\bf 725}, 45 (2005), hep-th/0505100;
 R.~Casero, C.~Nunez and A.~Paredes,
  ``Towards the string dual of N = 1 SQCD-like theories,''
  Phys.\ Rev.\ D {\bf 73}, 086005 (2006), hep-th/0602027.}
 
\lref\bobby{B.~S.~Acharya,
  ``On realising N = 1 super Yang-Mills in M theory,'' hep-th/0011089;
B.~Acharya and E.~Witten,
  ``Chiral fermions from manifolds of G(2) holonomy,'' hep-th/0109152.}

\lref\gwyn{R.~ Gwyn,
{Work in progress}.}

\lref\maldacena{J.~M.~Maldacena,
  ``The large N limit of superconformal field theories and supergravity,''
  Adv.\ Theor.\ Math.\ Phys.\  {\bf 2}, 231 (1998)
  [Int.\ J.\ Theor.\ Phys.\  {\bf 38}, 1113 (1999)], hep-th/9711200;
O.~Aharony, S.~S.~Gubser, J.~M.~Maldacena, H.~Ooguri and Y.~Oz,
  ``Large N field theories, string theory and gravity,''
  Phys.\ Rept.\  {\bf 323}, 183 (2000), hep-th/9905111.}

\lref\imamura{Y.~Imamura,
  ``Born-Infeld action and Chern-Simons term from Kaluza-Klein monopole in
  M-theory,''
  Phys.\ Lett.\ B {\bf 414}, 242 (1997), hep-th/9706144;
J.~P.~Gauntlett and D.~A.~Lowe,
  ``Dyons and S-Duality in N=4 Supersymmetric Gauge Theory,''
  Nucl.\ Phys.\ B {\bf 472}, 194 (1996), hep-th/9601085;
K.~M.~Lee, E.~J.~Weinberg and P.~Yi,
  ``Electromagnetic Duality and $SU(3)$ Monopoles,''
  Phys.\ Lett.\ B {\bf 376}, 97 (1996), hep-th/9601097;
A.~Sen,
  ``Dynamics of multiple Kaluza-Klein monopoles in M and string theory,''
  Adv.\ Theor.\ Math.\ Phys.\  {\bf 1}, 115 (1998), hep-th/9707042;
``A note on enhanced gauge symmetries in M- and string theory,''JHEP {\bf 9709}, 001 (1997), hep-th/9707123.}

\lref\beckeryau{K.~Becker, M.~Becker, J.~X.~Fu, L.~S.~Tseng and S.~T.~Yau,
  ``Anomaly cancellation and smooth non-Kahler solutions in heterotic string
  theory,'' hep-th/0604137;
M.~Cyrier and J.~M.~Lapan,
  ``Towards the Massless Spectrum of Non-Kahler Heterotic Compactifications,'' hep-th/0605131.}

\lref\BPV{W.~ Barth, C.~ Peters and A.~ Van de Ven,
{\bf Compact complex surfaces}, Springer Verlag, 1984.}

\lref\shiu{B.~R.~Greene, K.~Schalm and G.~Shiu,
  ``Warped compactifications in M and F theory,''
  Nucl.\ Phys.\ B {\bf 584}, 480 (2000), hep-th/0004103.}

\lref\sltwoz{J.~H.~Schwarz,
  ``An SL(2,Z) multiplet of type IIB superstrings,''
  Phys.\ Lett.\ B {\bf 360}, 13 (1995)
  [Erratum-ibid.\ B {\bf 364}, 252 (1995)] hep-th/9508143;
J.~X.~Lu and S.~Roy,
   ``Non-threshold (f,D p) bound states,''
  Nucl.\ Phys.\ B {\bf 560}, 181 (1999) hep-th/9904129.}

\lref\dgg{K.~Dasgupta, M.~Grisaru, R.~Gwyn, S.~Katz, A.~Knauf and R.~Tatar,
  ``Gauge - gravity dualities, dipoles and new non-Kaehler manifolds,''hep-th/0605201.}

\lref\lwtu{Loring~W.~Tu,
  ``Bundles over an elliptic curve''
  Adv. Math. {\bf 98}, 1-26 (2006).}

\lref\atvb{M.~F.~Atiyah, 
	``Vector bundles over an elliptic curve'',
	Proc. London Math. Soc. {\bf 7}, 414-452 (1957).}

\lref\gaud{P.~Gauduchon, ``Le th\'eor\`em de l'excentricit\'e nulle'',
	C.R.A.S. Paris {\bf 285} (1977), pp. 387-390.}

\lref\brone{Marian~Aprodu, Vasile~Br\^\i nz\u anescu and Matei~Toma, 
	``Holomorphic vector bundles on primary Kodaira surfaces,''
	Math. Z. {\bf 242} (2002), p. 63-73, math.CV/9909136.}

\lref\brtwo{Vasile~Br\^\i nz\u anescu and Ruxandra~Moraru,
   ``Holomorphic rank-2 vector bundles on non-K\"ahler elliptic surfaces
   (Fibr\'es holomorphes de rang 2 sur des surfaces elliptiques non-k\"ahleriennes),''
   Annales de l'institut Fourier, {\bf 55} no. 5 (2005), p. 1659-1683, math.AG/0306191.}

\lref\brthree{Vasile~Br\^\i nz\u anescu and Ruxandra~Moraru,
   ``Stable bundles on non-K\"ahler elliptic surfaces,'' math.AG/0306192.}

\lref\brfour{Vasile~Br\^\i nz\u anescu and Ruxandra~Moraru,
	``Twisted Fourier-Mukai transforms and bundles on non-K\"ahler elliptic surfaces,''
	Math. Res. Lett. {\bf 13} no. 4 (2006), pp. 501-514, math.AG/0309031.}

\lref\nekra{N.~Nekrasov,~H.~Ooguri and C.~Vafa, `S-duality and
Topological Strings,'' hep-th/0403167.}

\lref\lustu{G.~L.~Cardoso, G.~Curio, G.~Dall'Agata and D.~Lust,
``BPS action and superpotential for heterotic string compactifications  with
fluxes,'' JHEP {\bf 0310}, 004 (2003),hep-th/0306088.}

\lref\lustd{G.~L.~Cardoso, G.~Curio, G.~Dall'Agata and D.~Lust,
``Heterotic string theory on non-K\"ahler manifolds with H-flux
and gaugino condensate,'' hep-th/0310021.}

\lref\bd{M.~Becker and K.~Dasgupta, ``K\"ahler versus non-K\"ahler
compactifications,'' hep-th/0312221;
K.~Becker, M.~Becker, K.~Dasgupta and R.~Tatar,
  ``Geometric transitions, non-Kaehler geometries and string vacua,''
  Int.\ J.\ Mod.\ Phys.\ A {\bf 20}, 3442 (2005), hep-th/0411039;
A.~Knauf,
  ``Geometric transitions on non-Kaehler manifolds,'' hep-th/0605283.}

\lref\micuhet{S.~Gurrieri, A.~Lukas and A.~Micu, ``Heterotic on half-flat,''
  Phys.\ Rev.\ D {\bf 70}, 126009 (2004), hep-th/0408121;
A.~Micu,
  ``Heterotic compactifications and nearly-Kaehler manifolds,''
  Phys.\ Rev.\ D {\bf 70}, 126002 (2004), hep-th/0409008;
A.~R.~Frey and M.~Lippert,
  ``AdS strings with torsion: Non-complex heterotic compactifications,''
  Phys.\ Rev.\ D {\bf 72}, 126001 (2005), hep-th/0507202;
P.~Manousselis, N.~Prezas and G.~Zoupanos,
  ``Supersymmetric compactifications of heterotic strings with fluxes and
  condensates,''
  Nucl.\ Phys.\ B {\bf 739}, 85 (2006), hep-th/0511122.}

\lref\micu{S.~Gurrieri and A.~Micu,
``Type IIB theory on half-flat manifolds,''
class.\ Quant.\ Grav.\  {\bf 20}, 2181 (2003), hep-th/0212278.}

\lref\dal{G.~Dall'Agata and N.~Prezas,
``N = 1 geometries for M-theory and type IIA strings with fluxes,''
  Phys.\ Rev.\ D {\bf 69}, 066004 (2004), hep-th/0311146;
A.~Franzen, P.~Kaura, A.~Misra and R.~Ray,
  ``Uplifting the Iwasawa,''
  Fortsch.\ Phys.\  {\bf 54}, 207 (2006), hep-th/0506224;
A.~Misra,
  ``Flow equations for uplifting half-flat to Spin(7) manifolds,''
  J.\ Math.\ Phys.\  {\bf 47}, 033504 (2006), hep-th/0507147.}

\lref\douo{M.~R.~Douglas,``The statistics of string / M theory vacua,''
JHEP {\bf 0305}, 046 (2003), hep-th/0303194.}

\lref\wittenchern{E.~Witten,``Chern-Simons gauge theory as a
string theory,'' hep-th/9207094.}

\lref\ms{D.~Martelli and J.~Sparks,``G Structures, fluxes and
calibrations in M Theory,'' hep-th/0306225.}

\lref\grana{A.~Butti, M.~Grana, R.~Minasian, M.~Petrini and A.~Zaffaroni,
    ``The baryonic branch of Klebanov-Strassler solution: A supersymmetric
family
    of SU(3) structure backgrounds,''
    JHEP {\bf 0503} (2005) 069, hep-th/0412187.}

\lref\fg{A.~F.~Frey and A.~Grana,``Type IIB solutions with
interpolating supersymetries ,'' hep-th/0307142.}

\lref\bbs{K.~Becker, M.~Becker and R.~Sriharsha,``PP-waves,
M-theory and fluxes,'' hep-th/0308014.}

\lref\minu{P. Kaste,~R. Minasian,~A. Tomasiello,''Supersymmetric M theory
Compactifications with
Fluxes on Seven-Manifolds and G Structures'', JHEP {\bf 0307} 004, 2003,
hep-th/0303127.}

\lref\minasianone{
  A.~Butti, M.~Grana, R.~Minasian, M.~Petrini and A.~Zaffaroni,
  ``The baryonic branch of Klebanov-Strassler solution: A supersymmetric family
  of SU(3) structure backgrounds,''
  JHEP {\bf 0503}, 069 (2005), hep-th/0412187.
}
\lref\seiberg{S.~S.~Gubser, C.~P.~Herzog and I.~R.~Klebanov,
  ``Symmetry breaking and axionic strings in the warped deformed conifold,''
  JHEP {\bf 0409}, 036 (2004) hep-th/0405282;
A.~Dymarsky, I.~R.~Klebanov and N.~Seiberg,
  ``On the moduli space of the cascading SU(M+p) x SU(p) gauge theory,''
  JHEP {\bf 0601}, 155 (2006) hep-th/0511254.}

\lref\minasian{S.~Fidanza, R.~Minasian and A.~Tomasiello,
  ``Mirror symmetric SU(3)-structure manifolds with NS fluxes,''
  Commun.\ Math.\ Phys.\  {\bf 254}, 401 (2005), hep-th/0311122;
M.~Grana, R.~Minasian, M.~Petrini and A.~Tomasiello,
  ``Supersymmetric backgrounds from generalized Calabi-Yau manifolds,''
  JHEP {\bf 0408}, 046 (2004), hep-th/0406137;
``Type II strings and generalized Calabi-Yau manifolds,''
  Comptes Rendus Physique {\bf 5}, 979 (2004), hep-th/0409176;
`Generalized structures of N = 1 vacua,'' hep-th/0505212.}

\lref\lindstrom{U.~Lindstrom,
  ``Generalized N = (2,2) supersymmetric non-linear sigma models,''
  Phys.\ Lett.\ B {\bf 587}, 216 (2004), hep-th/0401100.}

\lref\minalind{
U.~Lindstrom, R.~Minasian, A.~Tomasiello and M.~Zabzine,
  ``Generalized complex manifolds and supersymmetry,''
  Commun.\ Math.\ Phys.\  {\bf 257}, 235 (2005), hep-th/0405085;
U.~Lindstrom,
  ``Generalized complex geometry and supersymmetric non-linear sigma models,''
  hep-th/0409250;
U.~Lindstrom, M.~Rocek, R.~von Unge and M.~Zabzine,
  ``Generalized Kaehler geometry and manifest N = (2,2) supersymmetric
  nonlinear sigma-models,'' hep-th/0411186.}

\lref\urangapark{J.~Park, R.~Rabadan and A.~M.~Uranga,
  ``Orientifolding the conifold,''
  Nucl.\ Phys.\ B {\bf 570}, 38 (2000), hep-th/9907086.}

\lref\uranga{A.~M.~Uranga,
  ``Brane configurations for branes at conifolds,''
  JHEP {\bf 9901}, 022 (1999), hep-th/9811004.}

\lref\dm{K.~Dasgupta and S.~Mukhi,
  ``Brane constructions, conifolds and M-theory,''
  Nucl.\ Phys.\ B {\bf 551}, 204 (1999), hep-th/9811139;
``Brane constructions, fractional branes and anti-de Sitter domain walls,''
  JHEP {\bf 9907}, 008 (1999), hep-th/9904131.}

\lref\beru{K.~Behrndt and M.~Cvetic, ``General N=1 Supersymmetric Flux
Vacua of
(Massive) Type IIA String Theory'', hep-th/0403049.}

\lref\dmftheory{K.~Dasgupta and S.~Mukhi,
  ``F-theory at constant coupling,''
  Phys.\ Lett.\ B {\bf 385}, 125 (1996), hep-th/9606044.}

\lref\dalu{G.~Dall'Agata, ``On Supersymmetric Solutions of Type IIB
Supergravity with General Fluxes'', hep-th/0403220.}

\lref\bercve{K.~Behrndt, M.~Cvetic and T.~Liu,
  ``Classification of supersymmetric flux vacua in M theory,'' hep-th/0512032;
K.~Behrndt and C.~Jeschek,
 ``Fluxes in M-theory on 7-manifolds and G structures,''
  JHEP {\bf 0304}, 002 (2003), hep-th/0302047;
``Fluxes in M-theory on 7-manifolds: G-structures and superpotential,''
  Nucl.\ Phys.\ B {\bf 694}, 99 (2004), hep-th/0311119.}

\Title{\vbox{\hbox{hep-th/0610001}}}
{\vbox{ \vskip-10in
\hbox{\centerline{Dipole-Deformed Bound States and}}
\hbox{\centerline{Heterotic Kodaira Surfaces}}}}

\vskip.02in \centerline{\bf Keshav
Dasgupta$^1$, ~Josh Guffin$^2$,~Rhiannon Gwyn$^1$, ~Sheldon Katz$^3$}
\vskip.18in
\centerline{\it ${}^1$ Rutherford Physics Building, McGill University,
Montreal, QC H3A 2T8, Canada}
\centerline{\tt keshav,gwynr@hep.physics.mcgill.ca}
\centerline{\it ${}^2$ Department of Physics, University of Illinois at Urbana-Champaign, IL 61801, USA}
\centerline{\tt guffin@uiuc.edu}
\centerline{\it ${}^3$ Depts. of Math \& Phys, University of Illinois at Urbana-Champaign, IL 61801, USA}
\centerline{\tt katz@math.uiuc.edu}

\vskip.5in

\centerline{\bf Abstract}

\noindent We study a particular ${\cal N} = 1$ confining gauge theory with
fundamental flavors realised as seven branes in the background of wrapped
five branes on a rigid two-cycle of a non-trivial global geometry.
In parts of the moduli space, the five branes form bound states with the
seven branes.  We show that in this regime the local supergravity solution
is surprisingly tractable, even though the background topology is
non-trivial. New effects such as dipole deformations may be studied in
detail, including the full backreactions.
Performing the dipole deformations in other ways leads to different warped
local geometries.  In the dual heterotic picture, which is locally given by
a ${\bf C}^\ast$ fibration over a Kodaira surface, we study details of the
geometry and the construction of bundles. We also point out the existence of 
certain exotic bundles in our framework.

\Date{September 2006}

\centerline{\bf Contents}\nobreak\medskip{\baselineskip=12pt
\parskip=0pt\catcode`\@=11

\noindent {1.} {Introduction} \leaderfill{1} \par
\noindent {2.} {${\cal N} = 1$ bound-state metric} \leaderfill{3} \par \noindent \quad{2.1.} 
{Bound states of D5 branes and D7 branes: a first look} \leaderfill{4} \par \noindent \quad{2.2.}
{Bound states in a non-trivial topology}
\leaderfill{10} \par \noindent {3.} {Dipole deformed bound states} \leaderfill{20} \par \noindent
\quad{3.1.} {First dipole deformation} \leaderfill{24} \par \noindent \quad{3.2.}
{Second dipole deformation} \leaderfill{26} \par \noindent
{4.} {Heterotic Kodaira surfaces} \leaderfill{27} \par \noindent 
\quad{4.1.}
{The Atiyah bundle} \leaderfill{29} \par \noindent
\quad{4.2.}
{The Serre construction} \leaderfill{31} \par \noindent
\quad{4.3.}
{Families of bundles} \leaderfill{32} \par
\noindent {5.} {Summary and discussion}\leaderfill{34} \par \catcode`\@=12 \bigbreak\bigskip}

\newsec{Introduction}

Recently, it has become apparent that in order to deal with
flux compactifications the target manifold has to be generically
non-K\"ahler \sav, although conformally K\"ahler manifolds may sometimes
arise. Examples of compact complex manifolds have been explicitly
constructed in heterotic theories \beckerD, \bbdg, \GP, \bbdgs, \beckeryau,
\yauli, \yaufu\ and the classification of torsional manifolds \gstructure,
\salamon\ has appeared in \lust, \louis, \gauntlett. These manifolds 
form the foundation on which string compactifications to lower
dimension can  be performed.  In the absence of fluxes, and for 
heterotic theories, when the spin-connection is embedded in the gauge
connection (via the so-called standard embedding) the original Calabi-Yau
compactifications are valid \chsw.  However, when spin-connection is {\it
not} embedded in the gauge connection, the compactification manifolds are no
longer Calabi-Yau but are the generic non-K\"ahler manifolds \bd,
\gttwo, \bbdgs\foot{Even when the fluxes are absent!}.

On the other hand, non-compact non-K\"ahler manifolds are interesting from
a different point of view. They sometimes appear as gravity duals to
certain confining ${\cal N} =1$ gauge theories in type IIA \gtone, \gttwo\
and heterotic theories \gttwo, \dgg, but may or may not be
complex\foot{Examples of compact non-complex non-K\"ahler manifolds are
constructed in \micuhet, \micu.}.  In type IIB, the known gravity duals for
confining gauge theories are generically K\"ahler (more appropriately
conformally K\"ahler), which could be further dual to non-K\"ahler manifolds
with both kinds of background three-forms.  

The conformally K\"ahler geometry we studied in our earlier works
\gtone, \realm, \gttwo, \dgg\ exhibits a globally integrable complex
structure derived from an F-theory picture. As such, both bi--fundamental
and fundamental matter appeared in the construction due to the presence of
three, five and seven-branes.  Unfortunately, the global metric was too
involved to derive, mainly because the construction involved a patch by
patch description. However, with some effort the local geometry near the
origin (i.e the far IR in the dual gauge theory) was derived by keeping
only the D5 branes and the seven branes \gtone, \gttwo, \dgg\ in the local
neighborhood. The resulting metric turned out to be rather simple in the
far IR, at least when the seven branes are kept away and the D5 branes
wrap local patches of a ${\bf T}^2$ rather than a two-sphere.  

An underlying F-theory picture implies that special points in the moduli
space of the solutions may exist.  There are two possible consequences 
of being at such a point:

$\bullet$ The system could be governed by an orientifold model, and

$\bullet$ Bound states of D5 branes with the underlying seven branes could appear.

\noindent The second possibility does not imply the first, although once we
are in an orientifold picture the existence of bound states is automatic:
there will always be a point in the same moduli space where such states can
appear. Elaborating on this will be the topic of sec. (2.1) and sec. (2.2). 

Imagine we are away from the orientifold point. This is possible by
considering any small perturbation to the orientifold picture.
One may ask whether bound states can appear in this scenario.
It turns out that bound states are very generic, and appear whether we
are at the orientifold point or not.  The question then is how to
distinguish between the two scenarios?  Herein lies the subtlety. When we
are {\it at} the orientifold point in the moduli space, the bound-state
configurations undergo a {\bf dipole} deformation in addition to any other
possible deformations. On the other hand, once we are away from the
orientifold point there seems to be no constraint on the system to undergo
a dipole deformation, and other deformations may take over.

Thus the crucial point seems to be the dipole deformation of bound states.
Things become complicated however, because the underlying topology becomes
non-trivial. Details on this will be presented in sec. 3. We will calculate
the local backreactions due to the dipole deformations on the background
geometry. We will show in sec. (3.1) and (3.2) that there are multiple ways
to perform dipole deformations. In both cases, the final results
nicely confirm our earlier predictions regarding the local geometry. 

On the heterotic side, we analyse some more aspects of the local geometry
given by a ${\bf C}^\ast$ fibration over a Kodaira surface. Recall that the
heterotic geometry is {\it not} U-dual to the corresponding type IIB
picture. In sec. 4 we study details of the bundle structure on the
local geometry. We provide three different ways to construct a bundle 
on our manifold. In sec (4.1) we give an intrinsic construction on the 
local geometry. In sec. (4.2) we use the method of pulling back a bundle 
constructed on the base torus, and in sec. (4.3) we use the method of pulling
back a bundle constructed on the primary Kodaira surface.

\noindent We conclude with a short summary of our results.

\newsec{${\cal N} =1$ bound state metric}

\noindent In our earlier paper \dgg, we discussed a set-up in type IIB
where $N$ D5 branes and some local and non-local seven branes wrap a
two-cycle of some non-trivial global geometry, giving rise to ${\cal N} =
1$ gauge theory with fundamental flavors.  This is basically the
gauge-gravity scenario, where the gravity dual is given by another geometry
with at least one topological non-trivial three-cycle on which we allow
three-form $H_{RR}$ fluxes and one non-compact three-cycle on which we have
three-form $H_{NS}$ fluxes. 

F-theory provides the full global geometry both before and after the
geometric transition\gtone, \gttwo, \realm, \dgg.  The geometries are
conformally K\"ahler on both sides, but full global metrics are not known.
However, the precise local geometries on any given patch are determined in
\gtone, \realm, \gttwo\ and \dgg\ up to possible subtleties mentioned therein.

We shall encounter several subtleties in our construction.  Firstly, due to
an inherent orientifold action only certain components of $B_{NS}$ can
survive. These $B_{NS}$ fields give rise to the dipole deformation \difuli,
\ramdam\ in our theory (see \dgg\ for more details). Secondly the metric
involves non-trivial background topology, fluxes and branes (both D7 and
D5). Thirdly, due to the existence of both kind of branes, there will be a
point in the moduli space where the D5 branes form a bound state with the
D7 branes.
The dipole deformation backreacts on the ${\bf P}^1$-wrapped bound state,
and in \dgg\ we computed a precise local solution taking it into account
for a D7 brane wrapping a ${\bf P}^1$.  We argued in \dgg\ that bound
states of D5 branes can be constructed by
allowing a non-trivial first Chern class on the D7 brane.  However, we did
not compute the {\it backreactions} from allowing $c_1 \not = 0$.
Our conclusion then was that the backreactions would be small, and
therefore the local geometry would be no different from the ones we
examined earlier in \gtone, \realm, and \gttwo.
In this section, we aim to compute precisely the backreaction and see how
far we are from the predicted local geometry of \gtone, \realm, \gttwo,
which was determined for a {\it separated} system of D5 and D7 branes in a
non-trivial geometry\foot{Supersymmetry for this system was discussed in
detail in \gtone, \gttwo, \dgg\ using an F-theory construction with
primitive G-fluxes.  Notice that the global geometry is {\it not} a
Calabi-Yau resolved conifold, but is a K\"ahler geometry with at least one
${\bf P}^1$. Locally near $r \to 0$ the geometry somewhat resembles a
resolved conifold with vanishing ${\bf P}^1$. Clearly this kind of geometry
is expected for the IR description of our gauge theory to make sense.}.  We
will show that at this point in the moduli space, the local metric is
surprisingly simple to determine even after we take into account all
possible backreactions. However before we delve into the determination of the local
metric, we first need some details on bound states of branes in a {\it
flat} background.

\subsec{Bound state of D5 branes and D7 branes: a first look}

\noindent Following our earlier works \gtone, \realm, \gttwo, \dgg\ we keep
a D5 brane oriented along the $x^{0,1,2,3,8,9}$ directions and a D7
parallel to the D5 brane at some point in the $x^4, x^5$ directions.  Using
the harmonic function ansatz, it is easy to write down the metric for a
system of parallel Dp-branes.  We expect the metric to be given by
\eqn\toymetric{ ds^2  =  f_1^{-1}f_2^{-1} ds_{012389}^2 + f_1
f_2^{-1}ds_{67}^2 + f_1f_2 ~ds_{45}^2.}
where $f_1$ and $f_2$ are related respectively to the harmonic functions of
the D5 brane and D7 brane.  However, this configuration is clearly not
supersymmetric and is therefore not stable.  Strings stretching between
these branes have a mass given by \eqn\mass{m^2 = -{1\o 2} + {d^2\o
(\pi\alpha')^2,}} with $d$ being the distance in the $x^{4,5}$ plane.  For $d <
{\pi\alpha'\o \sqrt{2}}$ the string becomes tachyonic and therefore reduces
the total energy of the system to its bound-state value (see for example
\witp\ for more details)\foot{The above results are strictly valid in the
flat-space limit \witp. For a curved background simple mode expansions are
not possible and the result can change as we saw earlier \gtone,\gttwo,
\dgg.}.

One way to construct the metric of a bound state of D5 and D7 branes is to
start directly with the metric \toymetric\ and calculate the change due to
tachyon condensation from the D5-D7 strings. Alternatively, we can start
with the metric of a D7 brane \eqn\metseven{ds^2 = f_2^{-1} ds^2_{01236789}
+ f_2~ds^2_{45},} and switch on gauge fluxes along the $x^{6,7}$ directions.
Finally, once we determine the backreactions of the fluxes on the geometry
we will obtain the desired metric. Our ansatz for the final bound-state
metric in a flat background can therefore be presented as
\eqn\bound{ds^2_{\rm bound} = {\tilde f}_2^{-1} ds^2_{012389} + f_3
~ds^2_{67} + {\tilde f}_2~ds^2_{45},} where we have a new warp factor $f_3$
along the directions $x^{6,7}$ to take into account the gauge fluxes
$F_{67}$ on the D7 brane. The ${\tilde f}_2$ factor signals any possible changes
to the original warp factor $f_2$. In the following we will give a simple
way to determine these warp factors.  

To begin, first observe that in addition to the metric, we expect all the
other type IIB background fields to have non-zero expectation values.  For
example, we will now have both axion and dilaton, respectively ($\chi,
\phi$), and also $H_{NS}$ and $H_{RR}$.  To describe $SL(2, Z)$ invariant
quantities in type IIB, we must define the following factor that depends on
the asymptotic values of the axion-dilaton ($\chi_0, \phi_0$):
\eqn\deltaa{\Delta = e^{-\phi_0} + e^{\tilde\phi_0}} with $\tilde\phi_0 =
\phi_0 + 2~ {\rm ln}~(1-\chi_0)$. The above relation \deltaa\ is valid only
for a single D5 and a single D7 (which is our present interest). For a
system with $m$ D5s and $n$ D7s we will need the following replacement:
\eqn\repla{\phi_0 ~\to~ \phi_0 - 2~{\rm ln}~n, ~~~~ \tilde\phi_0
~\to~\tilde\phi_0 + 2 ~{\rm ln}~\Big\vert{m-n\chi_0 \o 1-\chi_0}\Big\vert.}
Using \deltaa, one can fix the seven brane central charge $Q_7$ uniquely as 
\eqn\charge{ Q_7~=~ {\sqrt{\Delta} \o 2\pi},}
where the effect of the bound D5 appears from the definition of \deltaa.
Fixing the D7 brane charge also implies that all subsequent charges in the
$SL(2, Z)$ multiplets are completely fixed. This becomes important when we
have to define non-local seven branes in our scenario (which we will
encounter later). 
 
The bound state of $m$ D5s and $n$ D7s can now be determined using the
techniques elaborated in \sltwoz. The simplest way is to take $n$
coincident D3 branes with $m$ units of electric flux and U-dualise the
resulting background to get the D5/D7 bound state. The metric for this
configuration is exactly as predicted in \bound\ with $f_3$ and ${\tilde
f}_2$ given as \eqn\fdef{\eqalign{&{\tilde f}_2 ~\equiv~ \sqrt{h} ~=~
\sqrt{1-Q_7~{\rm ln}~r} \cr & f_3 = {e^{-\phi_0}\sqrt{h}~\Delta\o \Delta -
n^2 e^{-\phi_0}(1-h)}}} where $\Delta$ is now defined for an ($m,n$) bound
state, and $h$ is the modified harmonic function (notice the relative minus
sign). Thus, we see that switching on $F_{67}$ fluxes on $n$ coincident D7
branes changes the metric along the $x^{6,7}$ directions from ${1\o
\sqrt{h}}$ to $a\sqrt{\Delta\o h}$, with $a$ given by \eqn\adef{a ~ = ~
{h\sqrt{\Delta} \o \Delta e^{\phi_0} - n^2 (1 - h)},} and so tells us the
backreaction of the gauge flux on the bulk geometry. In a non-trivial
topology, such backreactions would be useful to evaluate the full IR
geometry of the corresponding ${\cal N} =1$ gauge theory. 

S-dualising this background will give rise to coincident NS5 branes with
{\it magnetic} seven branes. The behavior of string coupling near NS5
branes has been studied earlier in various papers. The magnetic seven brane
could also contribute to the coupling constant. Therefore, in the original
bound-state configuration we expect non-trivial behavior of the string
coupling as we approach the region near the core of the bound state. The
behavior is \eqn\gstring{g_s^2 ~ = ~ {g_0^2 \o (1- Q_7~{\rm ln}~r)(1-
Q_8~{\rm ln}~r)},} with $Q_8 \equiv {n^2 e^{-\phi_0}\o 2\pi\sqrt{\Delta}}$ and $Q_7$ as
defined in \charge. As we approach the core of the bound state, the theory
becomes weakly coupled. The constant coupling $g^2_0$ is defined at a point
where ${\rm ln}~r = 0$, and is in fact given by $g_0^2 = {4\pi^2 Q_7 Q_8\o n^2}$.
We also observe that the description is valid only in the regime $r
< e^{1/Q_7}$ or $r > e^{1/Q_8}$, while in the regime \eqn\regime{e^{1/Q_7} ~
< ~r ~ < ~ e^{1/Q_8}} we must use a different description\foot{Note that
$Q_7 ~ > ~ Q_8$ for our case.}.  That is also one of the reasons why our
local description is particularly good when dealing with seven branes.  

One can show that $n$ coincident D3 branes with $m$ units of electric
flux will exhibit exactly the same behavior for the bound-state metric.
This is expected since the fluxes contribute equally. What would be
different in this case is the spatial dependence of the warp factors.  That
is, the string coupling near the bound state will be different from the one
that we found earlier in \gstring. The precise dependence is easy to work
out, and is given by \eqn\gstnow{g_s ~ = ~ g_0 \sqrt{Q+r^4 \o Q + {\tilde
r}^4},} where $Q$ is the charge and ${\tilde r}$ is the scaled radius with
the scale factor given by $\Big({\Delta e^{\phi_0}\o n^2}\Big)^{1\o 4}$.
We see that near $r = 0$, the string coupling is not necessarily small, and
is given by $g_s = g_0 = {\Delta^{1\o 2}\o n}$.  Indeed the coupling
constant lies in the range \eqn\ccons{e^{-{\phi_0\o 2}}
~ \le ~ g_s ~ \le ~ {\Delta^{1\o 2}\o n}} defined for $0 \le r \le\infty $.
$g_s$ is nowhere divergent, and the metric therefore serves as
a good description in the whole region.  It turns out that the generic
behavior of the dilaton for {\it any} ($m, n$) bound state can always be
put in the form \eqn\genbound{\phi ~ = ~ \alpha~ {\rm ln}~\Delta
+ \beta~ {\rm ln}~h + \gamma~ {\rm ln}~a,} where $a$ is defined in \adef\ 
and ($\alpha,\beta,\gamma$) are constants\foot{One might be concerned by
the fact that \genbound\ has no apparent $\phi_0$ term. This is not an
issue because $a$ has the required powers of $\phi_0$ (see \adef).}.  As an
example, one can check that in type IIB, the dilatons $\phi_p$ of 
fundamental string\slash Dp-brane bound states are described in terms of 
$(\alpha, \beta, \gamma)$ as \eqn\dil{\phi_1~=~\Big(-{1\o 2}, {1\o 2}, -1\Big), ~~~
\phi_3~=~\Big(-{1\o 4}, 0, -{1\o 2}\Big), ~~~ \phi_5~=~\Big(0, -{1\o
2}, 0\Big), ~~~ \phi_7~=~\Big({1\o 4}, -{1}, {1\o 2}\Big),} with the
values of ($a, h$) changing accordingly for each bound state. One can now
see that for our type IIB D5/D7 bound state, the dilaton is given by
\eqn\dilfive{\phi~ = ~ \Big({1\o 4}, -{1}, {1\o 2}\Big),} which reduces to
\gstring\ and is therefore not globally defined. As we observed above, this
is not a matter of concern because we will only have a description on a
given patch once we go to a more involved scenario \gtone, \gttwo, \dgg. 

Existence of this bound state can also be inferred from M-theory with a
Taub-NUT background. Consider the $n=1$ case with $m$ arbitrary. Then the
harmonic form changes from its standard value to the one given in \robbins\
(see for example equations (33) and (51) in the first reference of
\robbins), due to the backreactions of the G-fluxes
\foot{These G-fluxes have two legs in the internal TN space and two legs
along seven-dimensional spacetime.}.  The $m$ five branes can be thought of
as wrapping the degenerating cycle of our Taub-NUT space. Clearly the
G-fluxes source these five-branes, so that we have a bound-state
configuration.

Having determined the dilaton, our next goal is to find the axion $\chi$.
The axion is expected because we have seven branes. For our bound-state
configuration, the axion can easily be determined, and is given by
\eqn\axion{d\chi ~ = ~ {1\o 9!}{1\o 1 + b(h-1)}~\big[f(x_4)~dx_5 -
f(x_5)~dx_4\big].} One can integrate this equation\foot{One should note
that $dh ~ = ~ -\pi^{-1} r^{-2} \sqrt{\Delta}(x_4~dx_4 + x_5 ~dx_5)$ so
$\chi$ is not linearly dependent on $h$.} to determine $\chi$. The
functions $f(x_i)$ are determined by using the relation ${\del h\o \del
x_i} = - {\sqrt{\Delta} x_i \o \pi r^2}$ to get \eqn\fxdef{f(x_i) ~ = ~
{x_i \o \pi r^2}\Big[(m-n\chi_0) ~\big(6 h^{-1} g(h) - \chi_0 h^{-2}\big) +
n e^{-2\phi_0} h^{-2}\Big],} where we have denoted the asymptotic value of
the axion as $\chi_0$, and $b$ in \axion\ above is given by $b = {n^2\o
e^{\phi_0}\Delta} \equiv {\tilde h -1 \o h -1},$ with the latter equality
serving as a definition of ${\tilde h}$. One can use this definition to
write the function $g(h)$ in a compact form as 
\eqn\ghdef{g(h)~ = ~ {\tilde
h}^{-2}\Delta^{-1} e^{-\phi_0}\Big\{\tilde h h^{-1} \big[mn(h-1)+ \chi_0 \Delta e^{\phi_0}\big]
-n(m-n\chi_0)\Big\}.}
A non-trivial axion also
exists for other bound states. For example, an electric flux on a D-string
induces a non-trivial axion given by\foot{Although electric fluxes on D3 branes
do not switch on any axion.} \eqn\chinow{\chi_1 ~ = ~ {\alpha + \gamma_1
h\o \tilde h},} where $\alpha = \chi_0 - \gamma_1$ and $\gamma_1 = {mn
e^{-\phi_0}\o \Delta}$ with $h$ defined accordingly for the bound state.
In fact, this will have important consequences when we embed our system in a
non-trivial topology and perform a dipole deformation. From the above
analysis we can now define a complex coupling $\tau \equiv \chi + i
e^{-\phi}$ for our background.

The next step is to find the $H_{RR}$ that forms the source of the bound D5
branes.  There are various ways to do this. One simple way is to compute
the $C_6$ sources from the D5 branes and then Hodge-dualise. This procedure
yields  \eqn\hrr{H_{RR}~ = ~ \Bigg({m-n\chi_0 \o 7!~3!~\pi
r^2}\Bigg)~h^{-{3\o 2}} {\tilde h}^{-2} e^{-\phi}~ \Big[x_4~ dx_5 \wedge
dx_6 \wedge dx_7 - x_5~dx_4 \wedge dx_6 \wedge dx_7\Big].} Given  the RR
field, supersymmetry requires us to have $H_{NS}$ fields also. An easy way
to see this is to construct a three-form $G = H_{RR} + \tau ~H_{NS}$ and
then consider the supersymmetry constraint $G = \pm  \ast iG$ for a given
non-trivial background, as in  \gvw, \gkp. Anticipating later
generalisations, we see that this requires non-zero $H_{NS}$ as well.  It
turns out that the NS three-form is given by \eqn\nsform{\eqalign{H_{NS}~&
= ~ d\chi_1~\delta_{h, 1-Q_7~{\rm ln}~r}\cr &= ~ {n(m-n\chi_0)\o \Delta
\tilde h^2 e^{\phi_0}}~ dh \wedge dx_6 \wedge dx_7,}} where the functional
form of $\chi_1$ is as in \chinow. We see that the NS-form is switched
on precisely because of the axion in the D-string--with--flux case. Once we
have such a $B_{NS}$ field, the dipole deformation becomes particularly
involved, as we shall soon see. 

Since $H_{NS}$ is non-trivial, $B_{NS}$ cannot be gauged away. $B_{NS}$ has
legs along the directions $x^{6,7}$.  Recall that the bound D5 branes are
oriented along the $x^{0,1,2,3,8,9}$ directions, and therefore the B-field
measures the charge of these five-branes. Incidentally the $B_{RR}$ fields
are also along the same directions. This makes sense precisely because of
the orientations of the D5 branes. 

One should note at this stage that the NS B-field \nsform\ is {not} {\it a
priori} related to the dipole deformation.
It has both of
its legs orthogonal to the D5 branes and parallel to the D7 brane. This is
a unique case, where from the D5 point of view one would expect some kind
of pinning effect as in \orione\foot{Of course not in a flat background,
but something like in a Taub-NUT space \orione.}, and from the D7 point of
view a non-commutative effect \swncg. We will discuss these possible
generalisations elsewhere once we switch on the dipole deformation. It is also
interesting to note that in lower--dimensional branes, for example for a D3
brane, once we switch on a {\it magnetic} flux the four-form charge of the
D3 brane changes to a new value given by \eqn\ffor{Q_4~ = ~ \vert
b_2\vert\Bigg(\chi_0 - 6\vert b_1\vert -{n e^{-2\phi_0}\o m - n\chi_0}\Bigg).}
Here, $\vert b_1\vert$ and $\vert b_2\vert$ are respectively the magnitudes
of background NS and RR two-forms, which satisfy the constraint that $b_1
\wedge b_2$ generates a Chern-Simons term on the D3 brane. 

\subsec{Bound states in a non-trivial topology}

So far our analysis has concentrated on determining the metric of a bound state
of $m$ D5 and $n$ D7 branes in a flat background. In the following we want
to compute the backreactions induced by allowing a non-trivial background
geometry with additional non-trivial topology.  

The reason this is important has already been emphasised. Our earlier
F-theory picture gave a supersymmetric configuration of D5 and D7 branes in
a background that locally resembled a resolved conifold \gtone, \gttwo,
\dgg.  By moving the five or seven branes, we can reach a point in the
moduli space where bound states exist. These branes would then wrap a
non-trivial two-cycle in the local geometry, whose explicit form (given
earlier in \dgg) can be written as:
\eqn\locgeom{ds^2 ~=~ {\cal A}~dr_1^2 + {\cal B}~(dz + f_1 ~dx + f_2 ~dy)^2
+ ({\cal C}~d\theta_1^2 + {\cal D}~dx^2) + ({\cal E}~d\theta_2^2 + {\cal
F}~dy^2).}
Here the warp factors are functions of the radial coordinate $r_1$ and $f_i
= f_i(\theta_i)$.
Recall that our coordinates ($r_1, z, x, y, \theta_i$) are local, and thus
the metric \locgeom\ is only for a local patch (see discussions in \dgg).  

Using the notations of our local metric, we can see that the D7 branes are
oriented along the ($z, r_1, y, \theta_2$) directions and are located at a
point on the torus described by the coordinates ($x, \theta_1$). The D5
branes are located at a different point on the ($x, \theta_1$) torus,
although they wrap the other torus ($y, \theta_2$) exactly as the D7 branes
do (see figure 2 in \dgg). 

We should now relate the local coordinates used here to the coordinates
used in the previous section. The ($x, \theta_1$) torus is related to the
($x_4, x_5$) cycle, and the ($y, \theta_2$) torus is related to the ($x_8,
x_9$) cycle. The radial coordinate used here (i.e. $r_1$) is proportional to
the $x_7$ coordinate used before. Finally, $z$ is related to the compact
$x_6$ coordinate. Thus 
\eqn\coord{(r_1, z, x, \theta_1, y, \theta_2) ~ \propto~ (x_7, x_6, x_4,
x_5, x_8, x_9)}
which means that the radial coordinate we defined earlier, namely $r =
\sqrt{x_4^2 + x_5^2}$, is {\it not} the radial coordinate $r_1$ used in the
local geometry, but is related to the distances along the ($x, \theta_1$)
torus.  In terms of the NS and RR fields, this means that both the B-fields
have components parallel to the D7 branes in the $(z, r_1)$ directions but,
as discussed before, this doesn't lead to any pinning effects for the D5
branes.  

How do we now embed a bound set-up of $m$ D5 and $n$ D7 branes in the local
geometry \locgeom? Our first observation is that the D5/D7 bound state
cannot change the topology of the manifold. This is easy to
understand, as local metric deformations do not change the topology of the
underlying manifold. What could then change? There are two possibilities:
(a) The warp factors will change, but no additional terms will appear in
the metric, or (b) The warp factors will change, and additional terms will
appear in the metric. These additional terms deform the metric
without changing the topology. 

To verify one of these cases, we make the following observations:  

\noindent (a) If we begin with the metric \locgeom\  and remove the
bound--state configuration, the resulting metric should resemble the local
solution given by 
\eqn\locsol{{\cal A}(r_1) ~ = ~ {\cal C}(r_1)^2, ~~ {\cal B}(r_1) ~ = ~
{\cal C}(r_1)^{-2}, ~~ {\cal D}(r_1) ~ = ~ {\cal C}(r_1), ~~ {\cal E}(r_1)
~ = ~ {\cal F}(r_1) ~ = ~ {\cal C}(r_1),} 
with the warp factor ${\cal C}(r_1)$ defined as
\eqn\calcnow{{\cal C} ~ = ~ 1 + \Bigg({1 \o {\cal F}_3(r_0) \sqrt{{\cal
F}_1(r_0)}} {\del {\cal F}_3 \o \del r_1}\Big\vert_{r_1 = r_0}\Bigg)~r_1 ~
\equiv ~ 1 + Q~r_1.}
These terms are explained in \dgg\ (see section 2 therein). 

%
\noindent (b) If we begin with the metric \locgeom\  and remove the bound
D5 branes, i.e.~make $(m,n)=(0,1)$, then the solution \locsol\
changes according to 
\eqn\locsolchange{{\cal A}~\to~ k^{-1} {\cal A}, ~~ {\cal B}~\to~ k^{-1}
{\cal B}, ~~ ({\cal C}, {\cal D}) ~\to ~ (k~{\cal C}, k~{\cal D}), ~~
({\cal E}, {\cal F}) ~\to~ (k^{-1}{\cal E}, k^{-1}{\cal F}),}
without generating an extra term in the metric \dgg. Here $k = k(r_1)$ is
the harmonic function whose value was left undetermined in \dgg.  On the
other hand, $(m, n) \ge (1, 0)$, the original case studied by \pandoz,
may not be supersymmetric \cvetic, \gtone, \gttwo, \realm, \dgg. 

The above set of observations might naively imply that embedding a bound
state of D5/D7 in the local geometry \locsol\ should generate no additional
terms in the metric, other than the one that we already have; only the warp
factors should change. Therefore our first ansatz for the metric of a bound
state of $m$ D5 and $n$ D7 branes in the local geometry \locsol\ is given
by
\eqn\locsoltwo{\eqalign{&{\cal A} ~\to~ {\sqrt{h}\o p+q~h}~{\cal A},
~~~~~  {\cal B} ~\to~ {\sqrt{h}\o p+q~h}~{\cal B}\cr & ({\cal C}, {\cal D})
~ \to ~ (\sqrt{h} {\cal C}, \sqrt{h} {\cal D}), ~~~~~ ({\cal E}, {\cal F})
~ \to ~ \Big({{\cal E}\o \sqrt{h}}, {{\cal F}\o \sqrt{h}}\Big),}}
where ($p, q$) are integers defined in terms of the $\Delta$ given in the
previous section:
\eqn\pq{p ~ = ~ e^{\phi_0} - q, ~~~~~~~~ q ~ = ~ {n^2 \o \Delta}.}

A little thought will tell us that this cannot be the complete answer,
since our method for constructing the bound state implies that the metric
depends non-trivially on the ($x, \theta_1$) directions instead of the
$r_1$ direction. Therefore, a simple linear superposition like \locsoltwo\
may not provide the full scenario, and we require corrections to the above
ansatz. 

To entertain possible corrections,  we first observe that we can make the
coefficients \locsol\ constant if we take $Q$ in \calcnow\ to be
vanishingly small. This is the regime where the fibration described by
\locgeom\ becomes trivial (at least to first order) \gtone, \gttwo, \dgg.
Second, we observe that a $U(n)$ gauge theory in $2+1$ dimensions with $m$
units of magnetic flux exhibits somewhat similar properties to those of the
bound state that we are looking for, as long as in a certain regime the
$2+1$ dimensional theory can be described as the dual configuration of a
$U(m)$ gauge theory with $n$ units of magnetic flux. This is nothing but
a type IIA D2 brane configuration with magnetic fluxes allowing the
{\it bulk} $U(1)$ electric flux
\eqn\uone{A_0 ~ = ~ -{1\o \kappa}\cdot{m- \gamma n\o h\sqrt{\Delta}}.}
Here $\kappa$ and $\gamma$ are constants that may be fixed by going to
a type IIB theory, where $\tau(r\to\infty)=\gamma + i \kappa$ implies
that ($\phi_0, \chi_0$) $\equiv$ ($-{\rm log}~\kappa, \gamma$) are the
respective asymptotic values of the dilaton-axion we defined earlier. We
have kept $h$ as the 2d harmonic function.

This bulk electric field is affected by the worldvolume magnetic fluxes,
as can be seen from \uone. To determine the worldvolume fluxes, we need a
D2 brane oriented along $x^{0,6,7}$ and fluxes satisfying $\int F_{67} =
m$.
Such a configuration affects the string coupling, and therefore the
behavior of the dilaton.  We compute that in terms of the parameters in
\genbound. The dilaton $\phi_2$ becomes
\eqn\phit{\phi_2~=~\Big({1\o 4}, {1\o 4}, {1\o 2}\Big),}
so that the full description of the near--core region can only be
captured by M-theory, while a type IIA description is valid away from the
core. 

Once we lift the configuration to M-theory, there will be G-fluxes and
globally--defined $C$-fields. For the present case, it is not too difficult
to work out the three-form field. It is given by 
\eqn\thfo{C~=~ {1\o h}\Big[\alpha_1 - \alpha_2 \Big({h+\beta_1\o
h+\beta_2}\Big)\Big]~dx_0 \wedge dx_6 \wedge dx_7 + {m\o n}
\Big({h+\beta_1\o h+\beta_2}\Big)~ dx_{11} \wedge dx_6 \wedge dx_7}
with $G = dC$ as the G-flux. The constants $\alpha_i$ and $\beta_i$ are
defined as:
\eqn\albe{\eqalign{&\alpha_1~=~ {\gamma(m-\gamma n) - n\kappa^{2} \o
\kappa \sqrt{\Delta}}, ~~~~ \beta_1 ~ = ~ {\gamma \Delta \o mn\kappa} -1 ,\cr &
\alpha_2 ~ = ~ {6m(m-\gamma n)\o n\kappa \sqrt{\Delta}}, ~~~~~~~~~~~
\beta_2 ~ = ~ {\Delta \o n^2 \kappa} -1.}}
We see that both the G-fluxes and the C-fields are not necessarily sourced
only by the M2 branes. Extra fluxes appear in the theory from the
backreactions of the worldvolume $F$ fluxes.

The backreaction of the worldvolume $F$ fluxes on the geometry can also be
worked out. If we are away from the core, then a type IIA description
suffices. The backreaction is only felt along the directions of the $F$
fluxes, i.e along $x^{6,7}$, and is given by\foot{We are absorbing the
unimportant constant $\kappa$ in the definition of the coordinates.}
\eqn\gsise{\delta g_{66} ~ = ~ \delta g_{77} ~=~ {1\o \sqrt{h}}\cdot
{1\o h+\beta_2}\bigg[h\Big({\Delta \o n^2}-1\Big)-\beta_2\bigg], ~~~~~ \delta g_{mn} ~ = ~ 0; ~~~ m, n \ne 6,7,}
where $\beta_2$ is defined in \albe. We should also note that in M-theory
all the metric components will change, and therefore the near--core
description will be a little more complicated. The metric will involve the
usual fibration structure due to the presence of the electric flux \uone,
and the warp factors will change due to the explicit dependence of the
string coupling on internal magnetic fluxes. 

Once we know the complete background of D2 branes with magnetic fluxes, we
can use the following set of duality arguments to determine a candidate
metric for our D5/D7 bound state on a non-trivial background. Clearly we
will {\it not} be able to simulate completely the local geometry on which
we want to have our bound state, but we can come very close. The following
are a set of steps that can potentially lead us to the required answer:

\noindent $\bullet$ The D2 brane is oriented along $x^{6,7}$ and is
orthogonal to the other directions. Consider the  $x^4$ direction.  If we
compactify the ($x^6, x^4$)  directions on a torus ${\bf T}^2$, then
generically the torus can have arbitrary complex structure $\tau_1$. We can
parametrise the complex structure by a real coordinate $\sigma_1$ such that
${\rm Im}~\tau_1 = 0$.  Such parametrisation is exactly of the form given
earlier in \orione, \difuli, \robbins. 

\noindent $\bullet$ As we discussed earlier, the near--core region of this
configuration is at strong coupling. The complete picture can therefore
only be given via M-theory. Lifting this configuration to M-theory gives
rise to a three--torus ${\bf T}^3$, where one of the toroidal directions is
the eleventh direction $x^{11}$ with radius $R_{11}$. We can pick a torus
along ($x^6, x^{11}$) and shrink it to zero size, while  keeping the $x^4$
cycle in ${\bf T}^3$ invariant.  This will take us to type IIB theory with
an orthogonal set of D1 branes. Since we kept the $x^4$ cycle inside ${\bf
T}^3$ unchanged, the original non-trivial complex structure will generate
an additional $B_{NS} = B_{64}$ field along with the bound state. 

\noindent $\bullet$ We can keep the system at a point on a ${\bf T}^4$
along $x^{1,2,3,9}$. Let the volume of the ${\bf T}^4$ be $V = 16 \pi^4~R_1
R_2 R_3 R_9$, where $R_i$ are the radii of the cycles as measured in the
warped geometry. Shrink the volume of the torus to zero size. Observe that
this doesn't affect the $B_{64}$ field, as it is orthogonal to the torus. 

\noindent $\bullet$ One can easily show that this configuration is {\it
dual}\foot{This duality, although not quite like AdS/CFT, is in the same
spirit as gauge/gravity dualities, in that a gauge theory on an
intersecting D-string configuration is mapped to a theory of gravity
without branes.} to a configuration of two intersecting Taub-NUT spaces in
M-theory where one of the TNs is along $x^{4,5,7}$ and the eleventh
direction $x^{11}$ and the other TN is along $x^{4,5,6}$ and $x^{11}$ along
with some G-fluxes. The existence of non-zero G-fluxes can be accounted
for from the fact that the corresponding three--forms thread through the
degenerating cycle of the two Taub-NUT spaces. For the present case there
are at least {\it two} non-zero components of the C-field ($C_{4,6,11}$ and
$C_{8,6,11}$) through which this duality can be explicitly analysed.

\noindent $\bullet$ It is easy to go to type IIB from M-theory by shrinking
some two-torus to zero size. For the case in question, not all two-tori
would give the kind of answer that we are looking for. In fact there is one
non-trivial two-torus along the directions ($x^6, x^{11}$) that is
particularly suited for our purpose.  In the limit where this torus shrinks
to zero volume,  we have an exact duality to a bound-state configuration of
D5/D7 branes! More interestingly however, some components of the $C$-fields
will dissolve completely in the geometry to give us a non-trivial local
solution resembling our required local solution \locgeom, at least
in certain limits. The other surviving components will provide the
necessary dipole deformation to the bound-state configuration.

Through this set of duality transformations, we hope to get a handle on our
background so that we can analyse all the field components that constitute
the supergravity solution for the system. We will begin by analysing our
duality chain from an intermediate stage that involves Taub-NUT spaces. The
duality arguments leading to that configuration are easy, though tedious,
to reproduce: we will leave them for the reader to derive.

Our starting point is to observe that the $B_{NS}$ field we obtain from the
non-trivial complex structure (parametrised by $\sigma_1$) in the first
step of our duality chain is a little involved. If we denote the asymptotic
value of $B_{64}$ as $b_\infty$, then for ${\rm tan}~\sigma_1 \equiv {\rm
x}$ we have 
\eqn\ntcs{{\rm x}^3 -c b_\infty {\rm x}^2 + {\rm x} - \kappa c b_\infty ~ = ~ 0,}
where $\kappa$ is as defined earlier and $c$ is a multiplicative constant.
The constant $c$ will in general be identity, but because of our duality to
Taub-NUT spaces, it turns out to be a non-trivial constant which has
interesting physical consequences. For the time being we will parametrise
$c$ by another angular coordinate $\sigma_2$ as 
\eqn\pbajwa{c~\equiv~ {\rm sec}~\sigma_2.}
With this description of $c$, one can show that the dual description of a
system of orthogonal D-strings at a point on ${\bf T}^4$ when vol(${\bf
T}^4$) $\to 0$ is an intersecting {\it warped} Taub-NUT background with the
following three-form fields:
\eqn\tnC{\eqalign{C~=~& {1\o \kappa_1} \Big[{{\rm tan}~\sigma_2\o \kappa_2}~dx_6 \wedge dx_8 - {\chi_\alpha ~{\rm cos}~
\sigma_1~{\rm sin}~\sigma_2\o \kappa_2}~dx_7 \wedge dx_8 + {\sqrt{h}~{\rm tan}~\sigma_1~Q_- \o {\rm cos}~\sigma_2}~
dx_4 \wedge dx_6 ~ + \cr
&~~~~~~~~~~~~ - \sqrt{h}~{\rm sin}~\sigma_1~\chi_\alpha~Q_-~dx_4 \wedge dx_7 \Big] \wedge dx_{11}}}
with the corresponding G-fluxes given by $G = dC$. In the above,  ${\tilde
h}$ is as defined earlier.  The other coefficients are defined in the
following way:  
\eqn\code{\eqalign{& \kappa_1 ~ = ~ {\tilde h}^{-1} \sqrt{h}(\kappa~{\rm cos}^2~\sigma_1 + 
{\tilde h}~{\rm sin}^2~\sigma_1) 
~\equiv~ {\tilde h}^{-1} \sqrt{h}~{\bar \kappa}_1, 
~~~~ \chi_\alpha ~ = ~ {\gamma \o \tilde h} 
+ {mn\kappa (h-1)\o \tilde h \Delta},\cr
& \kappa_2 ~ = ~ \sqrt{h}~{\rm cos}^2~\sigma_2 + \kappa_1^{-1}~{\rm sin}^2~\sigma_2 ~\equiv~ \sqrt{h}~\bar\kappa_2, 
~~~~ Q_\pm ~ = ~ 
1 \pm \kappa_1^{-1} \kappa_2^{-1}~{\rm sin}^2~\sigma_2,}}
where $h$ is the corresponding warp factor in the intersecting Taub-NUT
metric.  This warp factor's behavior can be traced through the duality chain
explicitly\foot{The value of $\chi_\alpha$ above is same as the one we
computed for D-strings, i.e $\chi_1$ in \chinow. It would be an interesting
exercise to see if this follows from our duality chain.}.  We should also
note that the coordinates $x^i$ are the {\it dual} coordinates defined for
this space, and are related to the original coordinates (i.e the
coordinates of the intersecting system of D-strings) by coordinate
transformations. 

The G-fluxes we constructed from \tnC\ each have one of their
components along the Taub-NUT circle $x^{11}$. Such a choice of G-fluxes
cannot completely specify the dual picture. We need more components of
G-fluxes that are orthogonal to the Taub-NUT circle. In other words, if we
specify the additional G-flux as $G_2$, then we require $G_2 \wedge dC =
0$.  For our specific case, $G_2$ turns out to be
\eqn\gtwo{G_2 ~ = ~{(m-\gamma n)~{\rm sin}~\sigma_2 \o \kappa
~h^2~\sqrt{\Delta}} ~ \ast \big( dh \wedge dV_0 \wedge dx_{11}\big),}
with $*$ the Hodge star for the warped metric (to be determined
below) and $dV_0 \equiv dx_0 \wedge dx_1 \wedge dx_2 \wedge dx_3 \wedge
dx_9$ a constant form. 

Looking at the three-form that we get in \tnC, we see that in some limits
it can have constant pieces. Does this mean that we can gauge them away?
For a flat background such components can be gauged away, but not for the
present case.
Since the constant pieces have one leg along the Taub-NUT circle $dx^{11}$
and a normalisable harmonic (1,1) form lives on the Taub-NUT, gauging away
such components switches on other components (see 
\imamura for more details). Therefore such constant pieces survive. 

All we now need is to specify the metric for our case. Since we expect an
intersecting Taub-NUT solution, the metric will typically be a
five-dimensional warped metric. Both the TNs share one degenerating cycle
along the eleventh direction, and therefore the fibration structure will be
non-trivial in the $x^{11}$ cycle. The precise metric turns out to be
\eqn\tnmet{\eqalign{ds^2~=~& \Bigg({\kappa_1 \kappa_2 \tilde h \o \kappa h}\Bigg)^{1\o 3}~\bigg[ ds^2_{01239} + 
 {\tilde h Q_-\o \bar\kappa_1}({\rm sec}~\sigma_2~dx_6 - \chi_\alpha~{\rm cos}~\sigma_1~dx_7)^2 ~ + \cr
& + {1\o \bar\kappa_2} \Big(dx_8  + 
 {{\tilde h}~{\rm tan}~\sigma_1~{\rm sin}~\sigma_2 \o \bar\kappa_1}~dx_4\Big)^2\bigg]
+ \Bigg({\kappa_1 \kappa_2  \tilde h  h^2 \o \kappa}\Bigg)^{1\o 3} ~\bigg[dx_5^2 + {\kappa \o \tilde h}~dx_7^2~  + \cr 
& + \Big({\rm sec}^2~\sigma_1 - {\tilde h \o \bar\kappa_1}~{\rm tan}^2~\sigma_1\Big) dx_4^2 \bigg] +
\Bigg({\kappa \o \kappa_1 \kappa_2 \tilde h \sqrt{h}}\Bigg)^{2\o 3}~\big(dx_{11} 
+ \omega \cdot dx\big)^2.}}
Here we see that the non-trivial fibration is via the one-form fields
$\omega_\mu$, the precise form of which will be derived below. The two TN
spaces can now be seen to be along $x^{4, 5, 6, 11}$ and $x^{4, 5, 7, 11}$
and therefore span a five-dimensional surface. 

To determine the one-form $\omega_\mu$ we start by choosing a hypersurface
in our space {\it orthogonal} to the two intersecting TNs. Let $dV_1$ be a
constant form on the surface. Using this we can define the following form
on the manifold:
\eqn\eform{\eqalign{&dS_1 ~ \equiv ~ -{m-\gamma n\o \kappa \sqrt{\Delta}~{\rm sec}~\sigma_1 }\Big[ 
\gamma dh^{-1} - 5~ d\Big({\chi_\alpha\o h}\Big) - 
{\rm sin}^2~\sigma_2 ~d\Big({\chi_\alpha\o h \kappa_1 \kappa_2} \Big) 
 \Big] \wedge dx_7 \wedge dV_1 \wedge dx_{11}~ + \cr 
&~ + {n\kappa\o \sqrt{\Delta} {\rm sec}~\sigma_1}~dh^{-1}\wedge dx_7 \wedge dV_1 \wedge dx_{11}
 + {m-\gamma n\o \kappa \sqrt{\Delta}}
\Big[{\rm sec}~\sigma_2~ d(h^{-1}Q_-) \wedge dx_6 \Big] \wedge dV_1 \wedge dx_{11}.}}
This is a nine-form that cannot be gauged away. It turns out that this is
not the only nine-form we can define for our case. As in the case of the
original nine-form \eform, we can use another hypersurface along $x^4$ that
intersects the original hypersurface used to define \eform\ in a
five-dimensional space and intersects the $x_8$ line at a point.  Let
$dV_2$ be a constant form on this hypersurface; we can then construct
another nine-form
\eqn\niform{\eqalign{dS_2 ~\equiv ~ & - {m-\gamma n\o \kappa \sqrt{\Delta}}\Bigg\{
~{\rm sin}~\sigma_1~{\rm sin}~\sigma_2
\bigg[\gamma d(\kappa_1^{-1} h^{-1/2}) - 5~d\bigg({\chi_\alpha \o \kappa_1\sqrt{h}}\bigg) + d\bigg({\chi_\alpha \o 
\kappa^2_1 \kappa_2 \sqrt{h}}\bigg) {\rm sin}^2~\sigma_2\bigg] ~+ \cr
& ~~~~- {n\kappa^2 \o m - \gamma n}~{\rm sin}~\sigma_1 ~{\rm sin}~\sigma_2~ d(\kappa^{-1}h^{-{1\o 2}}) \Bigg\}
\wedge dx_7 \wedge dV_2 \wedge dx_{11} ~ + \cr
& ~~~~~~ + {m-\gamma n\o \kappa \sqrt{\Delta}} ~
{\rm tan}~\sigma_1~{\rm tan}~\sigma_2 ~d\bigg({Q_+\o \kappa_1 \sqrt{h}}\bigg) 
\wedge dx_6 \wedge dV_2 \wedge dx_{11}.}}
Using the above equations \eform\ and \niform, the one form $\omega_\mu$
determining the fibrations for both the TN spaces is defined as
\eqn\oform{d\omega ~=~\lim_{R_{11} \to 0}~ \ast (dS_1 + dS_2),}
%
where the Hodge dual is on the warped metric in the above limit. The reason
we have to take a limit is that in M-theory there are no fundamental
one-forms, which only exist in the type IIA limit.  Therefore, plugging in
the value of \oform\ into \tnmet\ yields the complete fibration. Along
with the three-form \tnC\ and G-fluxes \gtwo\ this specifies the full M-theory
background. Thus, this configuration is dual to the intersecting D-string
configuration in the limit when vol(${\bf T}^4$) $= 0$.

Our last step is to shrink the ($x^{11}, x^6$) torus to zero volume in
order to reach type IIB theory. To see this explicitly, we have to put
holomorphic coordinates on a small patch of the torus.  Let $dz = dx_{11} +
\tau_m~dx_6$, where $\tau_m$ is the complex structure. If the one-form
$\omega_6 \ne 0$, one can show that
\eqn\taum{{\rm Re}~\tau_m ~ = ~ \omega_6, ~~~ {\rm Im}~\tau_m ~ = ~ 
\sqrt{{\kappa_2 {\tilde h}\sqrt{h} Q_- {\rm sec}^2~\sigma_2\o \kappa},}}
where $\kappa_2$ was defined in \code. Using this, the metric on the small patch can be written as 
\eqn\mepa{ds^2_{\rm patch}~ = ~ \bigg({\kappa\o \kappa_1 \kappa_2 {\tilde h} \sqrt{h}}\bigg)^{2\o 3} 
\vert dz\vert^2.}
The above choice of holomorphic coordinate does not imply the existence
of an integrable complex structure for our case. Nevertheless, the full metric
can be written down using real coordinates as we saw above. Shrinking
the volume of the ${\bf T}^2$ to zero size implies that there will be a
non-trivial dilaton given by
\eqn\dila{\phi_b ~ = ~ {1\o 4}~{\rm ln}~\Delta - {\rm ln}~h + {1\o 2}~{\rm ln}~a,}
where $a$ was defined in \adef. We see that \dila\ is {\it exactly} the
flat-space limit of the dilaton given in \genbound.  In fact, as we
mentioned briefly before, the ansatz \genbound\ -- although suitable for
bound states in a flat background -- works also for the curved background
with non-trivial topology as in our case.

Once we have the dilaton, we should look for the corresponding axion.  Our
earlier analysis for a flat background yielded the value given in \axion.
To determine the precise correction to our earlier result \axion, we can
use the constant form $dV_1$ defined in \eform.  With this one can show
that the axionic field undergoes a simple modification given by
\eqn\axionnow{d\chi_b ~ = ~ d\chi~{\rm cos}~\sigma_1.}
%
The axion so obtained sources the D7 branes.   
{}From the bound-state analysis, we should then expect a source
for the D5 branes also. This is indeed the case, and is given by
\eqn\dfivesou{H_{RR} ~ = ~ {m-\gamma n \o \kappa \sqrt{\Delta}}~ {\rm sec}~\sigma_2 \ast \big\{dh^{-1} \wedge 
dV_1 \big\}.}
The above two fields \axionnow\ and \dfivesou\ are not the only sources of
the axion and RR three-forms. There are in fact sources of these fields
that {\it do not} exist in the flat-space limit. For example there could be
an axion source from $dV_2$ defined in \eform\ as:
\eqn\dchi{d\chi_b^{(2)} ~ =  -{m-\gamma n \o \kappa \sqrt{\Delta}}~{\rm sin}~\sigma_1~
\ast \bigg\{ 
d\bigg({\gamma - 6 \chi_\alpha \o \kappa_1 \sqrt{h}}\bigg) \wedge {\cal D}M \bigg\} +  
{n\kappa\o \sqrt\Delta} ~
{\rm sin}~\sigma_1~\ast~\bigg\{d\bigg({1\o \kappa_1\sqrt{h}}\bigg) \wedge {\cal D}M \bigg\}}
where we have defined ${\cal D}M$ as 
\eqn\dmdef{{\cal D}M ~ = ~ dV_2 \wedge dx_7\wedge
\big({\rm sec}~\sigma_2~dx_8-{\rm sin}~\sigma_2~dx_6\big).}
It is interesting to note that our background also has additional sources
of $H_{RR}$, much like the additional sources of axion computed above in
\dchi. These sources, as one might expect, do not come from the D5 branes
only. They are given by
\eqn\dhrr{\eqalign{H_{RR}^{(2)} ~ = ~ &{m-\gamma n\o \kappa\sqrt{\Delta}}
\bigg[2{\rm tan}~\sigma_1~{\rm tan}~\sigma_2~
{\rm sin}^2~\sigma_2~\ast \bigg\{d\big(\kappa_1^{-2}\kappa_2^{-1}h^{-1/2}\big)\wedge dV_2\bigg\}~ + \cr 
& ~~~~~~~~~~ + {\rm sin}~\sigma_1~\ast \big(dh^{-1} \wedge dV_0 \wedge dx_6\big)\bigg],}} 
with $dV_0$ defined in \gtwo. We see that an equivalent term like $\ast
(dh^{-1} \wedge dV_0)$ for \dfivesou\ contributes only to ${\cal
O}(\sigma_1)$ to the $H_{RR}$ (similarly the other contribution goes like
${\cal O}(\sigma_1\sigma^3_2)$) and therefore in the limit 
\eqn\sadno{\sigma_{1,2} ~\to~ 0}
the above contributions are suppressed. It turns out that all the new
contributions to the axion and the RR three-form, other than 
 \axionnow\ and \dfivesou, are suppressed in the limit of small
$\sigma_{1,2}$. 

Such a limit will provide us with tremendous simplifications. Therefore, in
our notation, the flat-space results can be determined simply by setting
$\sigma_1 = \sigma_2 = 0$. In the flat-space limit we expect a $B_{NS}$
field that provides the Chern class of the D7 brane gauge bundle. The
corresponding $H_{NS}$ is therefore orthogonal to the D5 branes, and is
given by
\eqn\hnsn{H_{NS} ~ = ~ {\rm cos}~\sigma_1~{\rm cos}~\sigma_2~ d\chi_\alpha \wedge dx_6 \wedge dx_7.}
Thus it is related to the RR form \dfivesou\ by supersymmetry. As before
there could be new sources of $H_{NS}$ fields that are suppressed in the
limit \sadno\ as
\eqn\hnssup{H_{NS}^{(2)} ~ = ~ - {\rm sin}~\sigma_1~ d\bigg({\chi_\alpha Q_-\o \kappa_1\sqrt{h}}\bigg)\wedge dx_4 
\wedge dx_7} 
With the knowledge of NS and RR three-forms as well as the axion-dilaton,
we can construct $G = H_{RR} + \tau_b H_{NS}$ with $\tau_b = \chi_b + i
e^{-\phi_b}$. With the help of $G$, a superpotential for our background can
then be easily constructed. 

Our final venture is to determine the metric for the D5/D7 bound state.
From the intersecting Taub-NUT solution \tnmet\ the result follows from
shrinking the torus \mepa\ to zero size. This gives us:
\eqn\iibmet{\eqalign{ds^2_{\rm IIB} ~ = ~ & {ds^2_{0123}\o \sqrt{h}} + {1\o \sqrt{h}}\bigg(dx_9^2 + 
{Dx_8^2\o \bar\kappa_2}\bigg) + \sqrt{h}\bigg( {\kappa dx_4^2\o \bar\kappa_1}
 + dx_5^2\bigg) + {\kappa \sqrt{h}\o \tilde h}~dx_7^2 ~ + \cr
& ~~~~~~~~~~~~~~ +~ {\sqrt{h} \bar\kappa_1 \o \tilde h Q_-}~{\rm cos}^2~\sigma_2 
(dx_6 + {f_1}~dx_4 + {f_2}~Dx_8)^2,}}
where the various coefficients are defined as:
\eqn\defco{\eqalign{& {f_1} = \sqrt{h} \kappa_1^{-1} ~{\rm tan}~\sigma_1~{\rm sec}~\sigma_2, 
~~~~ {f_2} = -\kappa_1^{-1} \kappa_2^{-1}~{\rm tan}~\sigma_2, \cr
&Dx_8 = dx_8 + \sqrt{h}\kappa_1^{-1}~{\rm tan}~\sigma_1~{\rm sin}~\sigma_2~dx_4.}}
At this stage we can impose the coordinate redefinitions \coord\ on our
metric \iibmet. The resulting background looks almost like \locgeom\ if we
replace $dx_8$ in \locgeom\ by $Dx_8$. Therefore the metric of D5/D7 bound
states has almost the predicted form of \locgeom\ as we have been
expecting. The metric \iibmet\ however looks {\it exactly} like \locgeom\
only in the limit \sadno. In the following we will discuss this limit, and
also determine the changes to the metric after we make a dipole
deformation.

\newsec{Dipole-deformed bound states}

To study the effect of the limit \sadno\ on the metric \iibmet\ we have to make a small expansion about the
angular terms $\sigma_{1,2}$. This way we will also be able to compare our result with our earlier proposal 
\locsoltwo. Our claim is that the metric \iibmet\ resembles \locgeom. To verify this for the two tori ${\bf T}^2$ 
we find:
\eqn\twoto{\eqalign{&{\cal D} ~=~ \sqrt{h}\big(1-s \sigma_1^2\big), ~~~~~~~ {\cal C} ~ = ~ \sqrt{h}, \cr
& {\cal F} ~=~ {1\o \sqrt{h}} \bigg\{1+\sigma_2^2\bigg[{1} - {\tilde h (1- s \sigma_1^2)\o \kappa 
h}\bigg]\bigg\}, ~~~~~ {\cal E} ~ = ~ {1\o \sqrt{h}},}}
where $s \equiv {\tilde h \o \kappa} - {1}$. Thus we see that the corrections to \locsoltwo\ go like
${\cal O}(\sigma_{i}^2)$ and are therefore suppressed in the limit \sadno. This continues to hold for the
other coefficients of the metric \iibmet\ because
\eqn\calacalb{{\cal A} ~ = ~ {\sqrt{h} \o p+qh}, ~~~~ {\cal B} ~ = ~{\sqrt{h} \o p+qh} \bigg(1+ s\sigma_1^2 + 
{\tilde h\o h\kappa}\sigma_2^2\bigg)\big(1-{\sigma_2^2}\big),} 
which are again of the form \locsoltwo. Finally the fibrations in the metric \iibmet\ take the following form:
\eqn\fibk{f_1~=~ {\sigma_1 \tilde h \o \kappa }\bigg[1-s\sigma_1^2 + {\sigma^2_2\o 2} 
\bigg], ~~~~ f_2~=~ -{\sigma_2 \tilde h\o {h}\kappa}\bigg[1 - s\sigma_1^2 +{\sigma_2^2}\bigg
(1-{\tilde h\o \kappa {h}}\bigg)\bigg],}
which implies that the $f_i$s are determined by the angular coordinates $\sigma_i$ linearly as
$f_i \propto \sigma_i$. This will be crucial later.

The upshot of the above discussion is that the metric \iibmet\ follows the ans\"atze \locsoltwo\ in the limit 
$Dx_8 \approx dx_8$ and \sadno\ with the fibration terms $f_i$ given as \fibk. On the other hand, the NS three-form
field is of the form
\eqn\bnso{\eqalign{B_{NS}&~=~ \bigg(\chi_\alpha~{\rm cos}~\sigma_1~{\rm cos}~\sigma_2~dx_6 - 
{\chi_\alpha~{\rm sin}~\sigma_1~Q_- \o \kappa_1 \sqrt{h}}~dx_4\bigg) \wedge dx_7\cr
& ~\approx ~ \chi_\alpha \bigg(1-{\sigma_1^2\o 2}\bigg) \bigg(1-{\sigma_2^2\o 2}\bigg) dx_6 \wedge dx_7 - 
{\sigma_1 \chi_\alpha \tilde h \o \kappa h} \bigg(1- s \sigma_1^2 -{\tilde h \sigma_2^2\o 
h\kappa} \bigg)dx_4 \wedge dx_7}}
as can be extracted from \hnsn\ and \hnssup. We see that the second term is suppressed by $\sigma_1$. 

Among the RR fields we will have the axion and the RR three-form $dB_{RR}$. All of them will have pieces
that remain finite in the limit \sadno. This means that we have both $B_{NS}$ and $B_{RR}$ along 
$x^{6,7}$ that remain finite, with additional components that are of ${\cal O}(\sigma_i)$. Therefore our 
approximate background will be
\eqn\approxbg{\eqalign{&ds^2~ \approx~ {\sqrt h\o p+qh}~dr_1^2 + {\sqrt h\o p+qh}~(dz+f_1 dx+f_2 dy)^2 + 
\sqrt{h} \vert dz_1\vert^2 + {\vert dz_2\vert^2\o \sqrt{h}};\cr
&B_{NS} ~\approx~b_{67}, ~~~~~~ B_{RR}~ \approx~ \tilde b_{67}, ~~~~~~ h ~=~h(r, \sigma_i), ~~~~~~ r^2 ~ = ~ 
x^2 + \theta_1^2,}}
with $z_i$ defined exactly as in \gtone, \realm, \gttwo, \dgg, namely $z_1 = x+i\theta_1 = x_4 + i x_5$ and
$z_2 = y + i \theta_2 = x_8 + i x_9$; and $r_1 = x_7$ as before. There is also an axion-dilaton ($\chi_b, \phi_b$) 
that resembles the flat-space result. 

\noindent We would like to make the following comments:

\noindent $\bullet$ The metric \approxbg\ and the original metric \iibmet\ resemble the metric of  
wrapped D5 branes when $p >> q$ in \approxbg. This is the expected case when $m >> n$. On the other hand, the 
metric \iibmet\ or \approxbg\ resembles the metric of D7 branes when $n >> m$ as one might expect. However in
both the above limits the metrics {\it do not} resemble the metric of D3 branes at a point on our local geometry. 
That this is not inconsistent with our previous analysis on \gtone, \gttwo, \realm\ and \dgg\ is because we haven't
put in the required fluxes. Once the necessary fluxes are put in, the metric will resemble the metric of  
D3 branes at a point in our local geometry.

\noindent $\bullet$ The limit \sadno\ that we used above to write the type IIB solution \iibmet\ in the form  
\approxbg\ may not always hold. There could be some regime (or a different patch) where \sadno\ 
cannot be applied consistently. That would mean that \approxbg\ will not always capture the dynamics in certain
patches of the full global geometry. On the other hand, if we demand 
\eqn\demand{h~ \equiv~ h({\rm \bf Re}~z_1)}
then there are certain definite advantages over \sadno:

\noindent (1) $Dx_8$ defined in \defco\ becomes a total derivative such that $dD = D^2 =  0$ and therefore forms a 
cohomology. In fact $Dx_8 = dx_8^+ \equiv d(x_8 + x_4^+)$ where 
\eqn\xfour{x_4^+ ~ = ~ \int^{{\rm \bf Re}~z_1}~d({{\rm \bf Re}~z_1})~{{\tilde h}~{\rm tan}~\sigma_1~
{\rm sin}~\sigma_2 \o \kappa~{\cos}^2~\sigma_1 + \tilde h~{\sin}^2~\sigma_1}}
\noindent (2) The harmonic function will become linear in ${\rm \bf Re}~z_1$ and so can be approximated as 
$h = 1 + c~{\rm \bf Re}~z_1$ where $c$ is a constant. This means that $V_0$ and $V_2$ are not independent 
forms, but are related as
\eqn\vtwovo{dV_2~=~ c~dh \wedge dV_0}
\noindent (3) The following unnecessary components will vanish: 
\eqn\unreqd{dS_2 ~=~ d \chi_b^{(2)} ~=~ H_{NS}^{(2)} ~=~ H_{RR}^{(2)+} ~=~ 0}
where these components are defined above and the superscript + implies that there are a few surviving 
components\foot{In our notation, $H_{RR}^{(2)}$ defined in \dhrr\ can be written as 
$H_{RR}^{(2)} ~ = ~ H_{RR}^{(2)+} + H_{RR}^{(2)-}$ with $\pm$ forming the first and the second components of 
\dhrr\ respectively.}. 
The fact that $H_{NS}$ and $H_{RR}^{(2)+}$ vanish also means that the 
corresponding two--form fields can be gauged away. The ungauged components are precisely the ones that appear 
in \approxbg\ above along with an additional one
\eqn\survive{H_{RR}^{(2)-} ~ = ~ - {c(m-\gamma n)~ {\rm sin}~\sigma_1 \o \kappa h^2 \sqrt{\Delta}} ~ \ast 
\big(dV_2 \wedge dx_6\big)}
\noindent (4) The final metric becomes
\eqn\fimett{\eqalign{ds^2_{\rm IIB} ~ = ~ & {ds^2_{0123}\o \sqrt{h}} + {1\o \sqrt{h}}\big[dx_9^2 + 
\bar\kappa_2^{-1}(dx^+_8)^2\big] + \sqrt{h}\big[ (dx^-_4)^2
 + dx_5^2\big] + {\kappa \sqrt{h}\o \tilde h}~dx_7^2 ~ + \cr
& ~~~~~~~~~~~~~~ +~ {\sqrt{h} \bar\kappa_1 \o \tilde h Q_-}~{\rm cos}^2~\sigma_2 
(dx_6 + {f_1}~dx_4 + {f_2}~dx^+_8)^2,}}
which is the closest we get in realising the precise local geometry of our earlier papers. It is indeed remarkable 
to see that our earlier predictions fit perfectly with the above derivation; \fimett\ should then be regarded
as a derivation of our local geometry. With $h$ defined as above and 
\eqn\fours{x_4 ~=~\int dx_4^- ~{\sqrt{\kappa}~{\rm cos}^2~\sigma_1\o \sqrt{\bar\kappa_1}} + 
x_4^+~{{\rm sin}~2\sigma_1 \o 2 {\rm sin}~\sigma_2}}
we can easily argue for the form \locgeom\ with ${\cal C}(r_1)$ inserted in the {\rm local} limit.

\noindent The final background therefore consists of the metric \fimett\ with $f_i = f_i(\theta_i)$ as derived in \dgg. 
The other fields are the NS fields: 

\vskip.1in

$\bullet$ Dilaton $\phi$ and two-form $b_{67}$                   

\vskip.1in

\noindent which appear from \dila\ and \hnsn\ respectively. The RR fields are the three-forms coming from
\dfivesou\ and \dhrr. However some of these components are gauged away. Similar things happen with the 
axions \axionnow\ and \dchi. One can show that the ungauged component here is only \axionnow. We can dualise
these forms and write everything in terms of the six-forms and eight-form as

\vskip.1in

$\bullet$ ${\cal C}_1 = C^{(6)}_{012389}, ~~~~~{\cal C}_2 = C^{(6)}_{012369}, ~~~~~d{\cal C}_3 = \ast d\chi_b$

\vskip.1in

\noindent with ${\cal C}_3$ being the required eight-form. With this configuration at hand we can now 
make dipole deformations to our background. Due to the existence of $b_{67}$ there could be multiple ways to 
perform dipole deformations here. In the following we will analyse these aspects in detail.

\subsec{First dipole deformation}       

The multiple ways of doing dipole deformations that we alluded to above are related to the fact that we can either consider
the $b_{67}$ field while performing the dipole deformation or not. The deformation without an intervening 
$b_{67}$ field is predictably much easier to apply. In this section we will try this approach. 

The dipole deformation -- which we will call the first dipole deformation -- can be parametrised by an 
angular coordinate $\sigma_3$. Our starting point is the metric \fimett\ along with the NS and RR backgrounds 
discussed above. We define the following variables:
\eqn\defvari{\alpha_0 = {\sqrt{h}\bar\kappa_1\o \tilde h Q_-}~{\rm cos}^2~\sigma_2, ~~~~~ 
\alpha_2 = {1\o \bar\kappa_2} + \alpha_0 f_2^2 \sqrt{h}, ~~~~~ j_0 = {{\rm cos}^2~\sigma_3 \o \sqrt{h}} 
\big(1+\alpha_2 {\rm tan}^2~\sigma_3\big)}
where all the parameters appearing above have been described earlier. We should 
also remember that under a dipole deformation {\it both} the metric and the coordinates describing the 
underlying space change. Let the new coordinates be $y_i, i = 4, ....., 9$. The dipole deformation then
changes the metric \fimett\ to
\eqn\fimettone{\eqalign{& ds^2_{\rm I} ~ = ~  {ds^2_{0123}\o \sqrt{h}} + \bigg[\sqrt{h} (dy_4^-)^2 + 
{dy_5^2\o j_0}\bigg] + \bigg[{dy_9^2\o \sqrt{h}} + {{\rm sec}^2~\sigma_3 \o \sqrt{h} \bar\kappa_2}~dy_8^2\bigg]
+ {\kappa \sqrt{h}\o \tilde h}~dy_7^2 ~ + \cr
& + \alpha_0 \big(dy_6 + f_1~dy_4 + f_2~{\rm sec}~\sigma_3~dy_8)^2 - 
{\alpha_0^2 f_2^2 {\rm sin}^2~\sigma_3 \o j_0} \Big[dy_6 + f_1~dy_4 + \big(f_2~{\rm sec}~\sigma_3
 + {\rm x}\big)dy_8\Big]^2.}}
Looking at the above metric we see that this is almost of the form of the initial starting metric \fimett\ 
except for the $y_6$ fibration part because of the presence of an extra term. This term is defined as:
\eqn\xdefi{{\rm x} ~ = ~ {1\o f_2}\cdot {Q_- \o \kappa_1 \kappa_2} \cdot {\rm sec}^2~\sigma_2~{\rm sec}~\sigma_3.}
Can this term be made smaller? To see this we first need to figure out whether the NS two-form performing the 
dipole deformation is affected or not. It turns out that the $B_{NS}$ field is also affected in the following way
\eqn\bdipole{B_{NS} ~=~{\alpha_0 f_2~ {\rm sin}~\sigma_3 \o j_0} \Big[dy_6 + f_1~dy_4 + \big(f_2~{\rm sec}~\sigma_3
 + {\rm x}\big)dy_8\Big] \wedge dy_5 + b_{67}~dy_6 \wedge dy_7.} 
This means that the $B_{NS}$ fields do not follow the standard fibration structure of the background expected from
the initial metric \fimett. This can be rectified by making ${\rm x} << 1$. 
To allow this, observe that the free adjustable parameters in 
our problem are ($\sigma_{1,2,3}$) of which $\sigma_3$ is generically small. If we are also in the limit 
\sadno\ then $\sigma_{1,2}$ will also be small. This will make ${\rm x} \sim 1$. On the other hand we could be 
be in the limit \demand\ but {\it not} in the limit \sadno. For this case we can have 
\eqn\limit{Q_- ~<<~ 1 ~~~~\Rightarrow ~~~~ {\rm x}~<<~1}
where $Q_-$ is defined in \code. 
With this choice the dipole deformation takes the following nice form
\eqn\fidip{\eqalign{& ds^2_{\rm I} ~ = ~  {ds^2_{0123}\o \sqrt{h}} + \sqrt{h}\bigg[(dy_4^-)^2 + {dy_5^2 \o 
{\rm cos}^2~\sigma_3 + \alpha_2~ {\rm sin}^2~\sigma_3}\bigg] + {1\o \sqrt{h}}\bigg(
{dy_9^2} + {{\rm sec}^2~\sigma_3 \o \bar\kappa_2}~dy_8^2\bigg)~ + \cr
& ~~~~ + {\kappa \sqrt{h}\o \tilde h}~dy_7^2 + 
\alpha_0\bigg(1- {\beta_2~{\rm tan}^2~\sigma_3  \o 1 + \alpha_2 ~{\rm tan}^2~\sigma_3}\bigg) 
\big(dy_6 + f_1~dy_4 + f_2~{\rm sec}~\sigma_3~dy_8)^2,}}
which is exactly as we had predicted in \dgg\ if we take 
$f_i = {f_i(\theta_i)\o 1 + \delta_{i2}({\rm sec}~\sigma_3 -1)}$ and ${\cal C}(r_1) \to 1$! 
Here $\beta_2 = \alpha_0 f^2_2 \sqrt{h}$ and
denoting the $y_6$ fibration 
structure as $Dy_6$  the dipole B-field is given by
\eqn\dipb{{\cal B} ~ = ~ {\beta_2 \o f_2}\bigg(
{{\rm tan}~\sigma_3~{\rm sec}~\sigma_3  \o 1 + \alpha_2 {\rm tan}^2~\sigma_3}\bigg)~
Dy_6 \wedge dy_5 + b_{67}~dy_6 \wedge dy_7}
with appropriate field strength. It is also interesting to see that a component of the B-field \dipb\ can provide a
dipole deformation to the D7-brane gauge theory although not necessarily the component that makes a dipole 
deformation to the D5-brane gauge theory.

We will not evaluate the RR fields in detail because they can be easily worked out from the dipole deformations and 
following earlier works \difuli, but go directly into comparing the volumes of the two-cycles 
$\Sigma_2$ before and after 
the deformation. 
Before dipole deformation the volume of the two-cycle
on which we have wrapped D5 branes
is given by 
\eqn\vin{V_{\rm initial} ~=~ \int_{\Sigma_2} ~{1\o \sqrt{h}}\bigg({1\o \bar\kappa_2 \sqrt{h}} + 
\alpha_0~f_2^2\bigg),}
which is the volume of a ${\bf T}^2$ in the geometry \fimett. After dipole deformation the volume of the two-cycle
changes to 
\eqn\vfin{V_{\rm final} ~ = ~ \int_{\Sigma_2} ~{{\rm sec}^2~\sigma_3\o \sqrt{h}}\bigg[{1\o \bar\kappa_2 \sqrt{h}} +
\alpha_0~f_2^2(1- {z})\bigg],}
with ${z}$ given as ${z} = {\beta_2~{\rm tan}^2~\sigma_3  \o 1 + \alpha_2 ~{\rm tan}^2~\sigma_3}$. We see that for 
small enough deformation the volume of the two cycle {\it shrinks} making the KK states heavier. This is again consistent
with our earlier conclusion \dgg. 

\subsec{Second dipole deformation}

The second kind of dipole deformation will take into account the presence of the background $b_{67}$ field. We will
again parametrise the dipole deformation by the angular coordinate $\sigma_3$. We start by defining the 
quantity 
\eqn\jone{j_1 ~ = ~ \alpha_0^{-1} {\rm cos}^2~\sigma_3 \big(1 + \alpha_0 \sqrt{h}~{\rm tan}^2~\sigma_3\big)}
which is similar to $j_0$ defined in \defvari. As before the coordinates of the dipole-deformed background will be
different from the coordinates used in \fimett. For the sake of simplicity we shall again use $y^i, i= 4,..., 9$ 
for the new background. 

The background after the dipole deformation is different from the one that we discussed in the previous subsection. 
It is now given by:
\eqn\morot{\eqalign{& ds^2_{\rm II} ~ = ~  {ds^2_{0123}\o \sqrt{h}} + \sqrt{h} \big[(dy_4^-)^2 + 
{\rm sec}^2~\sigma_3~{dy_5^2}\big] + 
{1\o \sqrt{h}}~\bigg[{dy_9^2} + {dy_8^2 \o  \bar\kappa_2}\bigg] +
 \bigg({\kappa \sqrt{h}\o \tilde h} + {b_{67}^2\o \alpha_0}\bigg)~dy_7^2 ~ + \cr
& + j_1^{-1} \big(dy_6 + f_1~{\rm cos}~\sigma_3~dy_4 + f_2~{\rm cos}~\sigma_3~dy_8)^2 -
j_1^{-1} \bigg(\sqrt{h}~{\rm tan}~\sigma_3~dy_5 - {b_{67}~{\rm cos}~\sigma_3 \o \alpha_0}~
dy_7\bigg)^2.}}
The above metric differs from the previous one in many key respects. The $y_6$ fibration is consistent with 
expectation, whereas our earlier metric \fimettone\ had a different $y_6$ fibration structure. On the 
other hand we now have a cross term $dy_5 dy_7$, but this is suppressed by ${\rm tan}~\sigma_3$. 

In the limit where the dipole deformation is small all the ${\rm tan}~\sigma_3$-dependent terms can be dropped 
from the metric, and we get the following metric:
\eqn\kotmot{\eqalign{& ds^2_{\rm II} ~ = ~  {ds^2_{0123}\o \sqrt{h}} + \sqrt{h} \big[(dy_4^-)^2 + 
{\rm sec}^2~\sigma_3~{dy_5^2}\big] + 
{1\o \sqrt{h}}~\bigg[{dy_9^2} + {dy_8^2 \o \bar\kappa_2}\bigg] ~ + \cr
& ~~~~+ {\kappa \sqrt{h}~dy_7^2  \o \tilde h} 
+ \alpha_0 ~{\rm sec}^2~\sigma_3\big(dy_6 + f_1~{\rm cos}~\sigma_3~dy_4 + f_2~{\rm cos}~\sigma_3~dy_8)^2,}}
which again fits nicely with our ansatz. 

The dipole-deforming B field is more involved than our earlier case. If we denote $Dy_6$ as the fibration 
one-form, then we can write the B field as
\eqn\bdipp{{\cal B} ~ = ~ {{\rm cos}~\sigma_3 \o \alpha_0~j_1}~Dy_6 \wedge \big(b_{67}~dy_7 - \sqrt{h}~ \alpha_0~ 
{\rm tan}~\sigma_3~{\rm sec}~\sigma_3~dy_5\big) - b_{67}~ {\rm sec}~\sigma_3~\big(Dy_6 - dy_6\big)\wedge dy_7.} 
As before, the B field has components that perform dipole deformations to the D7-brane gauge theory also. In addition
to that the RR fields are also affected by the dipole deformations. We will not discuss them 
here as they are easy to work out. Instead we will concentrate on the volume of the two-cycle on which we have 
wrapped branes. The original volume is given by \vin. The final volume after dipole deformation will be
\eqn\vfint{V_{\rm final} ~ = ~ \int_{\Sigma_2} ~{1\o \sqrt{h}}\bigg({1\o \bar\kappa_2 \sqrt{h}} +
{\alpha_0~f_2^2 \o 1 + \alpha_0~\sqrt{h}~{\rm tan}^2~\sigma_3}\bigg),}
which is clearly smaller than \vin\ -- confirming again our earlier arguments. Notice that, compared to
the previous case, this deformation does not distinguish between the limits \sadno\ and \demand. If we are in the 
limit \sadno\ then $\sigma_i \to 0$ and the harmonic function will be logarithmic as in \ouyang.
Otherwise 
it will be a function as in \demand. 

More details of these calculations can also be worked out easily following the analysis of \nifuli. However we will 
not do so here, and instead turn to the heterotic picture where the story is equally interesting. 

\newsec{Heterotic Kodaira surfaces}  

The discussion so far about type IIB theory suggested that even when we are at a point in the moduli 
space where bound states can appear, the background metric on a local patch is of the form
\eqn\metmetnow{{ds^2_{\cal M}~\sim ~ dr_1^2 + 
\Big(dz + f_1~ dx + f_2~dy\Big)^2~
+ \vert dz_1\vert^2 + \vert dz_2\vert^2,}}
with $dz_i$ being the two tori with
complex coordinates defined above. In the limit where the D7 branes are far from the D5 branes, the $dz$ 
fibration is defined with \eqn\feye{f_i(\theta_i) ~ = ~ {\rm cot}~\theta_i, ~~~~~~~ \theta_i \ne 0} 
In the limit where we expect bound states, the functions $f_i$ are in general more complicated than \feye\ 
as described above.  For the {\it delocalised} harmonic function that we took i.e. $h(z_1)$ -- which could be
linear or logarithmic depending on the limits \sadno\ or \demand\ chosen -- $f_i$ could be brought 
in the expected form if we also make $h$ a function of $r_1$. This is like inserting correct prefactors of 
${\cal C}(r_1)$ for every term as mentioned above.  

Once such a starting point is made precise, the rest of the steps are straight-forward (up to possible subtleties 
mentioned in \gtone). The duality chain gives rise to local solutions in Type II and M-theories whose global 
metrics could possibly be constructed by joining all the local patches. 

The heterotic story, on the other hand, is equally interesting. The conjectured local metric {\it after} 
geometric transition proposed in \realm\ was shown in \gttwo\ to have a global description that resembled 
the MN type metric \mn\ in some limits! Clearly this would mean that the global metric is the gravity dual 
to the theory on wrapped 
five-branes which, here, would mean the theory on wrapped heterotic NS5 branes.    

Our next step was to look for possible metrics {\it before} geometric transition. Unfortunately there was no 
known duality chain (like the one that we had for the Type II to M-theory) that could help us here. A duality to 
F-theory via an orientifold corner of the moduli space was not very helpful to give us the complete global picture,
and for the derivation used in \gttwo\ to go to the full global metric we had to rely on various correlated ideas
and connections (see \gttwo\ for details). 
In addition to that, although 
there was no {\it a priori} reason to justify that there is a geometric transition in the
heterotic theory, 
the existence of a ``dual'' metric resembling the MN type metric gave us a hint that maybe the theory 
on wrapped NS5 branes could also be described by a dual gravity theory.  

That brought us to the next stage of determining the metric before geometric transition. This time however there 
was no known way to determine the global metric. The local metric was determined in \dgg\ to be of the form:  
\eqn\hettori{\eqalign{& ds^2 ~ = ~ {\cal H}(r)^2~dr^2 + {\cal H}(r)^{-2}~\Big(dz + F_1~ dx
+ F_2~dy\Big)^2 + 
~{\cal H}_1(r)~(1-\sigma_0)~~\vert dx ~+~ \tau_6~dy\vert^2 \cr
&+ {\cal H}_2(r)~d_6~\Big( {\rm sec}^2~\theta\Big[d\theta_1 + {\rm sin}~2\theta (a~dx - b~dy)\Big]^2 
+  \vert\tau_2\vert^2~{\rm sec}^2~\tilde{\theta}\Big[d\theta_2 - {\rm sin}~2\tilde{\theta} (\tilde{a}~dx -
\tilde{b}~dy)\Big]^2\Big)}}
where the $F_i$ are functions of $\theta_i$; and 
the new complex structure $\tau_6$ of the $(x,y)$--torus is a function of $\tau_3,~\tau_5$ and $\sigma_0$ 
and determined by\foot{All the other coefficients are defined in \dgg.}
\eqn\bigtau{\rm{Re}~\tau_6 ~ = ~ {\rm{Re}~\tau_3-\sigma_0\rm{Re}~\tau_5 \over 1-\sigma_0}~,~~~~~\vert\tau_6\vert^2 ~=~
 {\vert\tau_3\vert^2-\sigma_0\vert\tau_5\vert^2 \over 1-\sigma_0}~.}
The $(\theta_1,\theta_2)$--torus is non--trivially fibered over the $(x,y)$--torus, forming a specific family of 
Kodaira surfaces. The local manifold is therefore a ${\bf C}^\ast$ fibration over Kodaira surfaces. 

Once we determine the local geometry, the next question is to study the bundle structure. 
This is related to the fact that in the analysis of the geometry after geometric transition \gttwo, the study 
of vector bundles showed us a non-singular way to pull the bundle across a conifold transition. This transition 
would give us a two-fold result: the full global geometry and the bundle structure. However our analysis in \dgg, as 
discussed above, only provided us with a local metric. Therefore it is important to study the vector bundle on this 
local geometry. The full analysis of bundle structure with torsion 
on a non-K\"ahler manifold is particularly involved, so we will 
go to the Calabi-Yau limit of the background, and study the bundles 
there\foot{This 
is possible in special cases where {\it both} anomaly and non-K\"ahlerity are canceled 
by switching on fluxes and gluino-condensates. See sec. 5 of \gttwo\ (and also \gcon). 
This works because both the condensate term and the non-K\"ahlerity term come with the same 
powers of $\alpha'$ \gttwo. However,
this cancellation is checked 
only to some small orders in $\alpha'$, and the generic result to all orders has not been 
analysed.}.   

In the ensuing sections, we shall construct rank 2 bundles ${\cal E}$ on our
Calabi-Yau threefold $X$, utilizing several techniques.  Recall that
$X$ consists of a ${\bf C}^\ast$ fibration over a primary Kodaira surface
$\bf S$, a non-trivial holomorphic ${\bf T}^2$ fibration over a base
${\bf T}^2$.  We denote the base ${\bf T}^2$ as $B$, as in the diagram:
\vskip.17in

\centerline{\epsfbox{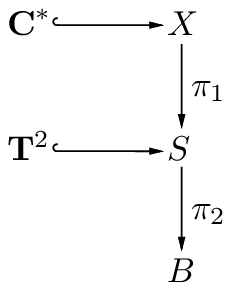}}\nobreak

\vskip.18in

We shall consider three methods for constructing a bundle on $X$; an
intrinsic construction on $X$, pulling back a bundle constructed on $B$,
and pulling back a bundle constructed on $S$.  The bundles produced by each
construction have differing Chern classes, so that the appropriate method
of construction differs from compactification to compactification.

\subsec{The Atiyah bundle}

\noindent We construct the first bundle by finding a rank 2 bundle {\cal E}
on the elliptic curve $B$, and then pulling back by $\pi_2\circ \pi_1$. We
shall require that the bundle satisfies the anomaly cancellation
requirements $c_1(X)=c_1({\cal E})$ and $c_2(X) = c_2({\cal E})$.

Since $X$ is Calabi-Yau, we must have $c_1({\cal E}) = 0$.  To deduce $c_2(X)$, we
use the multiplicative property of Chern classes under exact sequences: if
\eqn\eseq{
0\rightarrow~{\cal E}^\prime\rightarrow~{\cal E}\rightarrow~{\cal E}^{\prime\prime}\rightarrow0}
is exact and $c({\cal E})$ denotes the total Chern class of ${\cal E}$,
\eqn\totch{c({\cal E}) = c({\cal E}^\prime) \cdot c({\cal E}^{\prime\prime}).}
\noindent To compute the second
Chern class of $X$, consider the exact sequence \eqn\Xseq{ 0 \rightarrow
T_{\pi_1} \rightarrow T_X \rightarrow \pi_1^*T_S \rightarrow 0.}
%
%
%
Here $T_X$ and $T_S$
respectively denote the tangent bundles of
$X$ and $S$, 
while $T_{\pi_1}$ designates directions tangent to
the fibres of $\pi_1$, which are 
isomorphic to ${\bf
C}^*$.

Recall that in \dgg, the threefold $X$ was constructed by specifying a
torsion class $c\in H^2(S,\bf Z)$ as the image under the coboundary map
$\delta\!\!:\!\!H^1(S,{\cal O}_S^*)\rightarrow H^2(S,\bf Z)$. As a torsional
element, its image in $H^2(S,\bf C)$ vanishes and therefore
$c_1(T_{\pi_1}) = 0$.  It immediately follows from \totch\ and \Xseq\ 
that $c_1(S)=0$ and $c_2(X) =
\pi_1^* c_2(S)$.  However, $c_2(S)$ is the Euler characteristic of the Kodaira 
surface $S$, which is zero.  It follows that $c_2(X)=0$.

%
%

We turn now to the construction of a bundle on $B$ satisfying these
constraints.  For a given elliptic curve and for each natural number $n$,
Atiyah \atvb\ constructs a rank $n$ degree 0 indecomposable vector bundle
$A_n$ with $h^0(B,A_n) =1$.  The bundles are defined recursively, with
$A_n$ the unique (up to isomorphism) non-trivial extension of $O_B$ by
$A_{n-1}$:
\eqn\atiy{ 0 \rightarrow {\cal O}_B \rightarrow A_n \rightarrow A_{n-1}
\rightarrow 0.}
Recursion begins with $A_1$ defined as ${\cal O}_B$.  Indecomposable
bundles over elliptic curves are semistable \lwtu, so it follows that each
$A_n$ is semistable. Since our interest lies with rank two bundles, we
concentrate on $A_2$.  Explicitly, $A_2$ comprises the unique extension of
${\cal O}_B$ by itself:
\eqn\atseq{ 0 \rightarrow {\cal O}_B \rightarrow A_2 \rightarrow {\cal O}_B
\rightarrow 0.}
Uniqueness follows from the fact that Ext$^1({\cal O}_B,{\cal O}_B) \cong
H^1(B,{\cal O}_B) \cong \bf C$.  

To compute the Chern classes of $A_2$, we exploit the exact sequence
\atseq.  Clearly, $c_i(A_2)=0$ for $i\ge 2$ since dim${}_{\bf C} B = 1$.
Using the sequence \atseq, we compute
\eqn\catwo{c_1(A_2) = 2 c_1 ({\cal
O}_B)}
and deduce that $c_1(A_2)=0$.  

\subsec{The Serre construction}

In the previous section we considered a model with $c_2(X) = c_2({\cal
E})$, which doesn't necessarily indicate a lack of $H$-torsion. Switching
on $H$-torsion will imply that $dH \ne 0$, and that the spin connection is
not embedded in the gauge connection \smit, \papad, \bbdg, \bbdgs.

The detailed analysis of switching on a $H$ torsion and studying the bundle
structure on a non-K\"ahler manifold is particularly involved\foot{See
\yauli, \yaufu\ where bundle structure was addressed for the kind of models
studied in \beckerD, \bbdg, \bbdgs, \bbdp}. Let us therefore consider a toy
model in which $c_2(X) \ne c_2({\cal E})$. In heterotic string theory, such
a choice would generically lead to an anomaly. This anomaly could possibly
be cancelled by non-local terms contributing to $H$ at ${\cal O}(\alpha')$.
However, we haven't analysed this situation for heterotic theories, and
therefore we will only mention the following analysis as a toy 
example\foot{Observe however that, if we pull a bundle through a geometric 
transition, the chern classes of
the bundle before and after the transition will differ by the class of the
curve on which the geometric transition is based.  Before the transition,
branes wrapping the curve allow cancellation of anomalies with $c_2(E)\ne
c_2(X)$.  After the transition, we can then have $c_2(X)=c_2(E)$, as
required by the disappearance of the branes. More details on this will be presented
elsewhere. We thank Ron Donagi and Eric Sharpe for discussion.}. We
will also assume that $X$ is still the Calabi-Yau manifold described above.

In the following we would therefore like to utilize the Serre construction
as outlined in sec. (4.1) of \gttwo.  The construction requires an elliptic
curve $\cal C$ inside the Calabi-Yau threefold $X$.  In this section, the
fibre ${\bf T}^2$ shall be referred to as $E$.

Consider a point $p\in B$, and denote the fibre over $p$ as  $E_p =
\pi_2^{-1} (p)$.  We shall consider this fibre as a submanifold of $S$, so
that we may restrict the $\bf C^*$ bundle structure of
$X$ to $E_p \subset \bf S$. If $X\vert_{E_p}$ admits a
global section $s$, the image $E^\prime$ of $E_p$ under $s$ forms an
elliptic curve in $X\vert_{E_p}$, and thus one in $X$ by inclusion.  The
diagram below succinctly captures this data: \vskip.17in

\centerline{\epsfbox{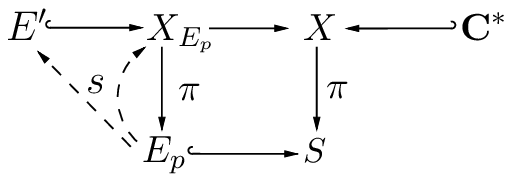}}\nobreak

\vskip.18in \noindent Because $X$ comprises a principal ${\bf C}^*$ bundle,
sections over any set exist only when the bundle restricted to that set is
trivial.  Thus, we must verify that a bundle on $S$ can restrict to a
trivial bundle on $E_p$.  We recall that a non-trivial ${\bf C}^*$ bundle
may be thought of as a trivial $\bf R$ bundle over a  non-trival $U(1)$
bundle.  Let $c_1 \in H^2(S,\bf Z)$ be the first Chern class of $X$ as a
principal $U(1)$ bundle.  Then, triviality of $X$ on $E_p$ is equivalent to
the vanishing of $c_1$ when restricted to $E_p$: 
\eqn\vanishing{ c_1\vert_{E_p}\equiv c_1 \cdot [E_p] = 0.} 
Our question as to the existence of a section is thus reduced to whether
the structure of $S$ affords enough freedom to select a $c_1$ to satisfy
\vanishing. Consider $E_p$ as map $H^2(S,{\bf
Z})~{}^{E_p}_{\longrightarrow}~H^4(S,\bf Z)$. It is known that $H^2(S,\bf
Z) \cong {\bf Z}^4 \oplus {\bf Z}_m$, and that $H^4(S,\bf Z) \cong \bf Z$.
Since $E_p$ is a group homomorphism, all the torsion elements are
automatically sent to zero.  If $S$ is torsionless, $E_p$ sends ${\bf
Z}^4\rightarrow \bf Z$, so it must have a non-trivial kernel.  Thus for any
$p$,  we can find a $c_1 \in H^2(S,\bf Z)$ which vanishes on $E_p$.  

We can then apply the Serre construction to the elliptic curve $E'$ to 
arrive at a rank 2 vector bundle $V$ on $X$ satisfying $c_1(V)=0$
and $c_2(V)=[E']\ne 0$.

\subsec{Families of bundles}

In this section, we will discuss how to obtain a family of rank-2 bundles
on a Kodaira surface.  The construction requires a preexisting
holomorphic rank-2 bundle satisfying a stability condition.  The idea stems
from a  series of papers (\brone,\brtwo,\brthree,\brfour) that explored the
existence and classification of stable rank-2 holomorphic vector bundles on
non-K\"ahler elliptic surfaces.  We will also relate a method for obtaining
a bundle of arbitrary rank $r\ge 2$ on $S$, as detailed in \brone.

%
%
%
%
We first present the method of obtaining a bundle on $S$. 
Consider a genus 2 curve $C$, and  let $f\!:\!C\rightarrow B$ be a ramified
covering of degree $r$.  Next, construct the fibre product of $S$ and $C$
over $B$; $Y\equiv S\times_B C$, and note that the projection
$\pi\!:\!Y\rightarrow S$ forms an $r$-fold cover.  If we push a line bundle
$L \rightarrow Y$ forward to a sheaf $\pi_* L$ on $S$, we actually obtain a
rank $r$ vector bundle on $S$ with Chern classes given by equation (2.1) of
\brone.

When searching for bundles with specified Chern classes, it is helpful to
know whether or not they exist.  The main result of \brone\ was to show
that on primary Kodaira surfaces, holomorphic rank 2 vector bundles exist
whenever
\eqn\discrim{\frac 1 2 \left ( c_2 - \frac 1 4 c_1^2 \right) \ge 0.}
Observe that when constructing bundles on $S$ for use in Calabi-Yau
compactifications with non-vanishing $H$-torsion, $c_2$ must be positive.

Before embarking on our construction of families, we reproduce the
definitions of degree and stability for non-K\"ahler manifolds from
\brthree.  Gauduchon\gaud\  defines a metric conformally equivalent to any
hermitian metric on a compact complex manifold $M$; a metric whose
associated (1,1) form $\omega$ satisfies $\del \bar \del \omega^{d-1} = 0$.
Using this form, we define the degree of a line bundle $\cal L$ with curvature
$\cal R$ to be 
\eqn\degdef{{\rm deg}~{\cal L} = \int_M {\cal R} \wedge \omega^{d-1}.}
The degree of any torsion-free coherent sheaf follows from the degree of
its associated determinant bundle, and the slope is defined as
its degree divided by
its rank.  A stable torsion-free coherent sheaf on $M$ is one for which
every coherent subsheaf with lesser rank has lesser slope.

Primary Kodaira surfaces possess a naturally associated surface called the
relative Jacobian of $S$: $J(S)= B \times T^\vee$, with $T^\vee$ the torus
dual to the fibre torus of the surface.  When we fix a bundle $\cal E$ on
$S$, it picks out a special divisor $S_{\cal E}$ in the Jacobian called the
spectral curve. This divisor is defined using only the bundle and intrinsic
data of the surface: see section 2.2 of \brfour\ for details.

Now, we follow the construction from section 4.2 of \brfour.
Fixing a bundle $\cal E$, we obtain the double cover
$S_{\cal E}\rightarrow B$.  Note that Kodaira surfaces are not 
multiply fibred, so the normalisation $W\cong S \times_B S_{\cal E}$,
and the maps $\bar \gamma$ and $\rho$ are just projections:

\centerline{\epsfbox{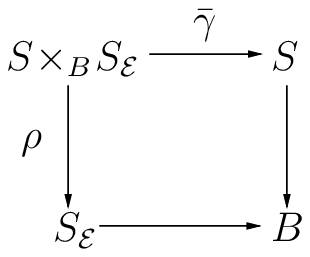}}\nobreak

The cover $S_{\cal E}\rightarrow B$ tells us that $W\rightarrow S$ is a
double cover as well.  Furthermore, $S_{\cal E}$ naturally induces a line
bundle $L$ on $W$.  If we push this line bundle forward to a sheaf on $S$,
we obtain a bundle $\delta \rightarrow S$ by taking its determinant:
$\delta = \det(\bar\gamma_* L)$.
Defining $\bar \imath$ as the involution on the Picard group of $W$ induced
by exchanging the two sheets of the covering $W\rightarrow X$, we pull the
Picard group of $S_{\cal E}$ back to $W$, $\rho^*{\rm Pic}(S_{\cal
E})\subset {\rm Pic}(W)$ and take the following subgroup:
\eqn\pzero{P := \Big \{\lambda \in \rho^*{\rm Pic}(S_{\cal E}) \enskip
\vert \enskip \imath^* \lambda \otimes \lambda = {\cal O}_W \enskip {\rm
and } \enskip \bar \gamma_* (c_1(\lambda)) = 0\enskip {\rm in} \enskip H^2(S,\bf
Z)\Big \}.}
Then, every rank 2 vector bundle on $X$ with determinant $\delta$ (that is,
fixed $c_1 = c_1(\delta)$) and spectral cover $S_{\cal E}$ is obtained as
$\bar\gamma_* (L\otimes \lambda), \enskip \lambda \in P$.
%
%
One can show that $P$ is isomorphic to the Prym variety Prym$(S_{\cal
E}\slash B)$ associated to the covering $S_{\cal E}\rightarrow B$. By
Proposition 3.2 of \brthree, if ${\cal E}$ is stable, then all bundles in
the family are stable.

\newsec{Summary and discussion}

Our main aim in this paper was to analyse the metric of D5 branes wrapped
on some two-cycle of a local geometry in the regime where the D5 branes
form a bound state with the seven branes. Our earlier study of the local
metric done in \gtone, \realm\ and \gttwo\ were always away from the D7
brane flavors. We did make some attempt in \dgg\ to determine the full
metric using an order-by-order expansion, but could only analyse the
effects {\it without} incorporating the backreactions from the full
bound-state configuration. Accordingly, the dipole deformation was also
approximate. Nevertheless we predicted in \dgg\ that the local metric with
all possible backreactions and dipole deformation would resemble \locgeom. 

A direct study of this using equations of motion seemed more difficult this
time because the backreactions involve, among other things, brane
worldvolume terms.  To solve this problem we devised a set of duality
transformations that used aspects of U-dualities, gauge-gravity dualities
and certain strong-coupling dynamics. Using these, the resulting analysis
of the actual configuration turned out to be richer than expected.
Studying various limiting procedures gave us an indication that: 

\vskip.1in

$\bullet$ There are multiple ways to perform dipole deformations here,

\vskip.1in

\noindent resulting in different warped geometries like eqns (3.14) and 
(3.24). These metrics respectively differ from \locgeom\ precisely by the limits 
\sadno\ and \demand. Once such limits are applied, the 
 metrics take the conjectured form \locgeom, giving
us the final: 

\vskip.1in

$\bullet$ Dipole-deformed metrics given by eqns (3.18) and (3.25).  

\vskip.1in

\noindent We see that any of these metrics could be taken as the starting
point in the duality cycle of \gtone, in the regime where we would be
interested in considering the flavors together. The non-K\"ahlerity in type
IIA for each of these cases could be easily determined from the 

\vskip.1in

$\bullet$ Dipole-deforming $B_{NS}$ fields given by eqns (3.16) and (3.26)

\vskip.1in

\noindent respectively along with the RR fields (which we left for the
reader to derive). The above choices of background anti-symmetric fields
differ from the choices that we took in \gtone\  and \realm, because we are
in the regime where the deformations are done by the dipoles. We could
easily go away from this regime and study the theory with non-commutative
deformations, for example. All these analyses are easy to perform now,
because we have described a standard way to derive the metric configurations.
Of course, one definite advantage of dipole deformations, as emphasised
earlier in \dgg, is to observe that 

\vskip.1in

$\bullet$ The volumes of the two-cycle shrink for both kinds of dipole deformations.

\vskip.1in

\noindent The type II story is now more or less complete, although the
heterotic side is far from clear. The bundle structure and global metric
forming a possible {\it dual} to the wrapped NS5 branes in the heterotic
theory have already been evaluated in \gttwo. In
\dgg, we studied the local geometry {\it before} geometric transition and
found that the metric is a particular ${\bf C}^\ast$ fibration over Kodaira
surfaces. The heterotic NS5 branes wrap two-cycles of this geometry giving
rise to non-trivial torsion classes (see sec. (3.2) of \dgg). In this paper
we 

\vskip.1in

$\bullet$ Construct vector bundles on ${\bf C}^\ast$ fibration over Kodaira surfaces,

\vskip.1in

\noindent thus confirming that such a local solution may indeed be a
candidate manifold for gauge/gravity dualities, although a direct
calculation still needs to be done\foot{It may be helpful to use
string-string dualities to study such issues. A recent paper dealing with
Type IIA/Heterotic theory in the presence of torsion is \micuram. Although
our heterotic geometry is not U-dual to any type IIB background, it may
still be dual to some type IIA geometry. It will be interesting to exploit
this angle.}.  The last link of the story is to see whether the heterotic
manifold has a global completion in which the base tori are deformed to one
or more non-singular ${\bf P}^1$s. This and other issues will be addressed
elsewhere.

\centerline{\bf Acknowledgements}

\noindent We would like to thank Vincent Bouchard, Ron Donagi, 
Marc Grisaru, Anke Knauf, Thomas Nevins, Eric Sharpe, and Radu Tatar
for helpful conversations. KD would like to thank the Banff International
Research Station ({\bf BIRS}) where part of the work was done.  
The research of   
KD and RG are supported in part by NSERC grants. 
The research
of SK is supported in part by NSF grants DMS-02-44412 and DMS-05-55678.

\noindent {\bf Preprint Number:} ILL-(TH)-06-6

\listrefs

\bye


\noindent Begin by considering a D1-brane.  The metric is given by 
\eqn\Dstring{
ds_{D1}^2  = H^{-{1 \over 2}} ds_{01}^2 + H^{1 \over 2} ds_\perp^2}
in the string frame. The non-zero fields are the axion $\phi$ and the Ramond-Ramond B-field $B_{01}^{(2)}$. From \GauntlettCV we have that for a Dp-brane the dilaton is given by
\eqn\dil{e^{2\phi} = H^{- {(p - 3) \over 2}}}
so here we have $e^{2 \phi} = H$. Then we can move into the Einstein frame:
\eqn\einstein{\eqalign{d\tilde s_{D1}^2 & =  e^{- {\phi \over 2}} \left (H^{-{1 \over 2}} ds_{01}^2 + H^{1 \over 2} ds_\perp^2\right ) \cr & = H^{- {3 \over 4}} ds_{01}^2 + H^{1 \over 4} ds_\perp^2.}}
In the Einstein frame the form of the metric does not change under an S-duality, so that \einstein  also gives the metric of an F-string. A general  S-duality transformation is given by
\eqn\sduality{\eqalign{
{\cal M} &\rightarrow \Lambda {\cal M} \Lambda^T\cr
\left ( \matrix{B^{(1)}\cr B^{(2)}}\right )  &\rightarrow \left ( \Lambda^T \right )^{-1}\left ( \matrix{B^{(1)}\cr B^{(2)}}\right ) \cr
\tau &\rightarrow \frac{a \tau + b}{c \tau + d} 
}} 
where $ \Lambda = \left ( \matrix{a & b \cr c & d }\right ) \in SL(2, {\bf R})$ and $ \tau = \chi + \imath e^{-\phi}$ is the axion--dilaton.  Thus to go from a D1-string set-up with only $\phi$ and $B_{01}^{(2)}$ non-zero to an F-string set-up with $B_{01}^{(2)}$ zero and $B_{01}^{(1)}$ defined instead, $\Lambda$ would have to be of the form $$ \Lambda = \left ( \matrix { 0  & - 1 \cr 1 & 0 }\right ).$$ Then the R-R and NS-NS sectors would be simply interchanged (up to a sign).  More generally though, both $B_{01}^{(1)}$ and $B_{01}^{(2)}$ will be generated, and the S-duality will give an (F,D1) bound state. The S-duality can be written in terms of $m$ and $n$ to give the bound state of an (m,n) string, where $m$ is the number of units of NS-NS charge and $n$ is the number of units of R-R charge. Such an S-duality will produce both an axion and a dilaton.

\noindent This general solution is given by Lu and Roy \LuUC (with the metric written in the Einstein frame):
\eqn\schwarz{\eqalign{ds_E^2 &= H^{-3\over 4} ds_{01}^2 + H^{1 \over 4} ds_{23456789}^2,\cr
e^{\phi_B} & = e^{\phi_0} \tilde H H^{-\frac{1}{2}}, \cr
\chi_B & = \frac{mn(H -1) + \chi_0 \Delta e^{\phi_0}}{n^2H + (m - \chi_0 n)^2 e^{2 \phi_0}}.}}
With B-fields
\eqn\BSBfields{\eqalign{B_{01}^{(1)} & = - \Delta^{- \frac{1}{2}} e^{\phi_0} ( m - \chi_0 n) H^{-1} ;\cr
B_{01}^{(2)} & = \Delta^{- \frac{1}{2}} H^{-1} \left [ \chi_0 ( m - \chi_0 n) e^{\phi_0} - ne^{- \phi_0}\right ],}}
where 
\eqn\Htilde{\tilde H = \Delta^{- \frac{1}{2}} \left ( (m - \chi_0 n)^2 e^{\phi_0} + n^2 H e^{-\phi_0} \right ) }
and
\eqn\Deltamn{\Delta = ( m - \chi_0 n)^2 e^{\phi_0} + n^2 e^{-\phi_0}.} $\chi_0$ and $\phi_0$ are the asymptotic values of the axion and dilaton (as $r \rightarrow \infty$) respectively. We delocalise the harmonic function along all the directions along which we will be performing T-dualities. Beginning with 
\eqn\Hinitial{H = 1 + \frac{Q_1}{r^6}} where $r^2 = x_2^2 +  x_3^2 + ... x_9^2$, we find that the delocalised harmonic function is given by 
\eqn\Hdeloc{H = 1 - Q_7 ln r,}
where $r^2 = x_4^2 + x_5^2$, $Q_1 = 2^5 \pi^2 a^6 \Delta^{\frac{1}{2}}$ and $Q_7 = \Delta^{\frac{1}{2}}/2\pi$. $a = \alpha^{\frac{1}{2}}$ with $\alpha$ the string constant.

\noindent The T-dualities are carried out most easily in the string frame, as per the prescription given in \BergshoeffAS. The metric \schwarz can be rewritten in the string frame as
\eqn\schwarzstring{ds_{\rm{str}}^2 = e^{\phi_0 \over 2} \tilde H^{1 \over 2} H^{-1} ds_{01}^2 + e^{\phi_0 \over 2} \tilde H^{ 1 \over 2} ds_{\perp}^2.}
Because there are no cross-terms in the metric and no components of $B^{NS-NS}$ in the directions along which we are performing the T-dualities, the only change to the metric is that $g_{xx} \rightarrow g_{xx}^{-1}$ where x is the direction along which the T-duality transformation is being performed. Thus we can give the metrics of the (F/D2) and (F/D3) configurations immediately:
\eqn\F{\eqalign{
ds_{F/D2}^2& ~ = ~e^{\phi_0\over2} \tilde H^{1 \over 2} H^{-1} ds_{01}^2 + e^{-\frac{ \phi_0}  {2}} \tilde H^{- \frac{1}{2}} dx_6^2 + e^{\phi_0 \over 2} \tilde H^{1 \over 2} ds_\perp^2;\cr
ds_{F/D3}^2& ~ = ~e^{\phi_0\over2} \tilde H^{1 \over 2} H^{-1} ds_{01}^2 + e^{-\frac{ \phi_0}  {2}} \tilde H^{- \frac{1}{2}} ds_{6,7}^2 + e^{\phi_0 \over 2} \tilde H^{1 \over 2} ds_\perp^2.}}
The fields are given by
\eqn\Dtwofields{\eqalign{e^{\phi_A}& ~=  ~  e^{3 {\phi _0} \over 4} \tilde H^{3 \over 4} H^{- 1 \over 2},\cr
A_6 & ~=~  \frac{mn(H -1) + \chi_0 \Delta e^{\phi_0}}{n^2H + (m - \chi_0 n)^2 e^{2 \phi_0}}\cr
B_{01}^{(1)} &~ =~ - \Delta^{- \frac{1}{2}} e^{\phi_0} ( m - \chi_0 n) H^{-1}\cr
C_{601}&~=~\frac{2}{3}\Delta^{- \frac{1}{2}} H^{-1} \left [ \chi_0 ( m - \chi_0 n) e^{\phi_0} - ne^{- \phi_0}\right ]}}
for (F/D2) and
\eqn\Dthreefields{\eqalign{e^{\phi_B} & ~=~e^{\frac{\phi_0}{2}} \tilde H^{\frac{1}{2}} H^{- \frac{1}{2},}\cr
B_{01}^{(1)} &~ =~ - \Delta^{- \frac{1}{2}} e^{\phi_0} ( m - \chi_0 n) H^{-1}\cr
B_{67}^{(2)}&~=~ \frac{mn(H -1) + \chi_0 \Delta e^{\phi_0}}{n^2H + (m - \chi_0 n)^2 e^{2 \phi_0}}
}} with $D_{7601}$ non-zero for (F/D3).
In the string frame the metric transformation under an S-duality is given by $ g \rightarrow e^{- \phi} g$ while $ \phi \rightarrow - \phi$. Thus we can now write down the (D1/D3) metric (still in the string frame) and dilaton:
\eqn\DoneDthree{\eqalign{ds_{D1/D3}^2&~=~H^{- \frac{1}{2}} ds_{01}^2 + e^{- \phi_0} H^{\frac{1}{2}} \tilde H^{-1} ds_{67}^2 + H^{\frac{1}{2}}ds_\perp^2 \cr e^{\phi_B}&~=~e^{- \frac{\phi_0}{2}} \tilde H^{- \frac{1}{2}} H^{\frac{1}{2}}.
}}The B-fields are interchanged under the S-duality and the four-form is left unchanged, such that the non-zero field components are now $$B_{67}^{(2)}, B_{01}^{(1)}, D_{0167}.$$

\newsec{D5/D7: the bound state}
\noindent After the 4 T-dualities outlined above we obtain the metric of a D5/D7 bound state:
\eqn\DfiveDseven{ds_{D5/D7}^2 = H^{- \frac{1}{2}} ds_{012389}^2 + H^{\frac{1}{2}} \tilde H^{-1} e^{- \phi_0} ds_{67}^2 + H^{\frac{1}{2}}ds_{45}^2}
with dilaton
\eqn\bsdilaton{e^{\phi_B} = \left (H \tilde H e^{\phi_0} \right )^{- \frac{1}{2}},}
and non-zero axion, $B_{67}^{(1)}$ and $B_{45}^{(2)}$. (This field has to come from $F_{012389}$)

\newsec{D5/D7}
\subsec{A conifold ansatz}
\noindent We now want  to place the D5/D7 bound state in a conifold background. As in \DasguptaYD  we choose to place the conifold along (4,5,6,7,8,9). With its metric given by
\eqn\conifold{ds_{\rm conifold}^2 = f_1 ds_{45}^2 + f_2 ds_{89}^2 + f_3(dx_6 + A dx_4 + B dx_8)^2 + f_4dx_7^2,}
and make the following ansatz for the D5/D7 bound state in a conifold background:
\eqn\conansatz{\eqalign{ds^2 = &H^{-\frac{1}{2}}
 ds_{0123}^2 + H^{\frac{1}{2}}f_1 ds_{45}^2 + H^{- \frac{1}{2}} f_2 ds_{89}^2  + H^{\frac{1}{2}}\tilde H^{-1} e^{- \phi_0} f_3 ( dx_6 + A dx_4 + Bdx_8)^2 \cr &+ H^{\frac{1}{2}}\tilde H^{-1} e^{\phi_0} f_4 dx_7^2.}}
\noindent The fields remain $\chi$, $B_{67}^{(1)}$ and $B_{45}^{(2)}$ with the dilaton given by $e^{\phi_B} = (H \tilde H e^{\phi_0} )^{- \frac{1}{2}}$. The D7-brane is charged by the axion, the D5-brane by the Ramond-Ramond B-field, and the NS-NS B-field is left-over from the original D1/D3 bound state. 

\subsec{Flux-twisting and deformations}
To obtain the desired fluxes we perform a T-duality transformation along $x^6$, rotate $x^5$ and $x^6$. This will give the desired cross-terms in the metric which upon T-dualising back to IIB will map into B-field components. In what follows the labelling of the B-fields will be as shown below:

$$\matrix {\rm{ IIB} & \rightarrow & \rm{IIA} & \rightarrow &\rm {IIA} & \rightarrow & \rm {IIB}\cr \cr
\rm {B} & \rm{T_6} & \rm{b} & \rm{rotation} & \rm{\tilde{b}} & \rm{T_6} &\rm{ \tilde {B}}}$$

\noindent In detail, the set-up after this T-duality is given by

\eqn\prerotmetric{\eqalign{ds_{IIA}^2 = &H^{- \frac{1}{2}} ds_{0123}^2 + H^{\frac{1}{2}} f_1 ds_{45}^2 + H^{- \frac{1}{2}} f_2 ds_{89}^2 + H^{\frac{1}{2}} \tilde H^{-1} e^{- \phi_0} f_4 dx_7^2\cr& + H^{- \frac{1}{2}} \tilde H e^{\phi_0} f_3^{-1} (dx_6 + B_{67}^{(1)} dx_7)^2,}}

\noindent and 
\eqn\prerotfields{\eqalign{e^{\phi_A} &~=~H^{-\frac{3}{4}} f_3^{- \frac{1}{2}}, \cr
A_6 &~=~\chi ,\cr
A_7 &~=~\chi B_{67}^{(1)} ,\cr
b_{64}^{(1)} &~=~A,\cr
b_{68}^{(1)} &~=~B,\cr
b_{74}^{(1)}&~=~2A B_{67}^{(1)} ,\cr
b_{78}^{(1)}&~=~ 2 B B_{67}^{(1)} ,\cr
C_{745}&~=~B_{67}^{(1)} B_{45}^{(2)}, \cr
C_{645}&~=~ \frac{2}{3} B_{45}^{(2)}.}}
\noindent We perform the following rotation:
\eqn\rotation{\eqalign{x^6 &\rightarrow  \cos \theta x^6 \cr x^5 &\rightarrow  \sin \theta x^6 + \sec \theta x^5,}}
giving us
\eqn\postrotmetric{\eqalign{ds_{IIA}^2 = &H^{- \frac{1}{2}} ds_{0123}^2 + H^{\frac{1}{2}} f_1 dx_{4}^2 + H^{- \frac{1}{2}} f_2 ds_{89}^2 + H^{\frac{1}{2}} \tilde H^{-1} e^{- \phi_0} f_4 dx_7^2\cr& + H^{- \frac{1}{2}} \tilde H e^{\phi_0} f_3^{-1} (\cos \theta dx_6 + B_{67}^{(1)} dx_7)^2 + H^{\frac{1}{2}} f_1 (\sec \theta dx_5 n+ \sin \theta dx_6)^2,}}
and 
\eqn\postrotfields{\eqalign{e^{\phi_A} &~=~H^{-\frac{3}{4}} f_3^{- \frac{1}{2}}, \cr
A_6 &~=~\chi \cos \theta  ,\cr
A_7 &~=~\chi B_{67}^{(1)} ,\cr
b_{64}^{(1)} &~=~A \cos \theta ,\cr
b_{68}^{(1)} &~=~B \cos \theta ,\cr
b_{74}^{(1)}&~=~2A B_{67}^{(1)} ,\cr
b_{78}^{(1)}&~=~ 2 B B_{67}^{(1)} ,\cr
C_{745}&~=~B_{67}^{(1)} B_{45}^{(2)} \sec \theta, \cr
C_{746} &~=~B_{67}^{(1)} B_{45}^{(2)} \sin \theta, \cr
C_{645}&~=~ \frac{2}{3} B_{45}^{(2)}.}}
upon T-dualising back we obtain the final metric of a D5/D7 bound state on a conifold with fluxes:
\eqn\finalmetric{\eqalign{
ds^2 = &H^{- \frac{1}{2}} ds_{0123}^2 + H^{\frac{1}{2}} f_1\left ( dx_4^2 + \sec^2 \theta dx_5^2 \right ) + H^{- \frac{1}{2}} f_2 ds_{89}^2\cr
 &+\left  (H^{\frac{1}{2}} \tilde H^{-1} e^{- \phi_0}f_4 + H^{- \frac{1}{2}} \tilde H e^{\phi_0} f_3 \left (B_{67}^{(1)}\right )^2\right ) dx_7^2\cr 
 &+ \beta^{-1} \left (dx_6 + A \cos \theta dx_4 + B \cos \theta dx_8\right )^2 \cr & - \beta^{-1} \left (H^{\frac{1}{2}} f_1 \tan \theta dx_5 + H^{- \frac{1}{2}} \tilde H e^{\phi_0} f_3^{-1} B_{67}^{(1)} \cos \theta dx_7 \right )^2 ,}}
where $$ \beta = H^{\frac {1}{2}} f_1 \sin^2 \theta + H^{- \frac{1}{2}} \tilde H e^{\phi_0} f_3^{-1} \cos^2 \theta $$ and the field content is
\eqn\finalfields{\eqalign{
e^{\phi_B} &~=~\beta^{-\frac{1}{2}} H^{- \frac{3}{4}} f_3^{- \frac{1}{2}} \cr
\tilde \chi &~=~\chi \cos \theta \cr
\tilde B_{45}^{(1)} &~=~ 2 \beta^{-1} A \sin \theta H^{\frac{1}{2}} f_1 \cr
\tilde B_{85}^{(1)} &~=~ 2 \beta^{-1} B \sin \theta H^{\frac{1}{2}} f_1 \cr
\tilde B_{74}^{(1)}&~=~2AB_{67}^{(1)} \left (1 - \beta^{-1} \cos^2 \theta H^{- \frac{1}{2}} \tilde H e^{\phi_0} f_3^{-1}\right ) \cr
\tilde B_{78}^{(1)}&~=~2BB_{67}^{(1)} \left (1 - \beta^{-1} \cos^2 \theta H^{- \frac{1}{2}} \tilde H e^{\phi_0} f_3^{-1}\right ) \cr
\tilde B_{65}^{(1)} &~=~ \beta^{-1} \tan \theta H^{\frac{1}{2}} f_1\cr
\tilde B_{67}^{(1)} &~=~ \beta^{-1} H^{- \frac{1}{2}} \tilde H e^{\phi_0} f_3^{-1} \cos \theta B_{67}^{(1)}\cr
\tilde B_{65}^{(2)}&~=~\beta^{-1} \chi \sin \theta H^{\frac{1}{2}} f_1\cr
\tilde B_{67}^{(2)}&~=~ \chi B_{67}^{(1)} \left (\beta^{-1} \cos^2 \theta H^{- \frac{1}{2}} \tilde H e^{\phi_0} f_3^{-1} -1  \right) \cr
\tilde B_{45}^{(2)} &~=~ B_{45}^{(2)} + 2 \cos \theta \sin \theta \beta6{-1} \chi H^{\frac{1}{2}} f_1\cr
\tilde B_{47}^{(2)} &~=~B_{67}^{(1)} \left ( 2 \chi ( 1 + \beta^{-1} A \cos^3 \theta H^{ - \frac{1}{2}} \tilde H e^{\phi_0} f_3^{-1} - \frac {3}{2} B_{45}^{(2)} \sin \theta \right )\cr
\tilde B_{78}^{(2)} &~=~ 2 \chi \cos \theta B_{67}^{(1)} B \left ( 1 - \beta^{-1} \cos ^2 \theta H^{- \frac{1}{2}}\tilde H e^{\phi_0} f_3^{-1}\right ) \cr
\tilde B_{58}^{(2)} &~=~2 \cos \theta \sin \theta \beta^{-1} \chi h^{\frac{1}{2}} f_1\cr
D_{6578}&~=~ \frac{3}{2} \beta^{-1} H^{\frac{1}{2}}f_1 \sin \theta B B_{67}^{(1)} \chi\cr
D_{6745}&~=~\frac{3}{8} B_{67}^{(1)} B_{45}^{(2)} \sec \theta + \frac {3}{2} \beta^{-1} A \chi \sin \theta H^{\frac{1}{2}} f_1 B_{67}^{(1)}\cr& - \frac{3}{4} H^{- \frac{1}{2}} \tilde H e^{\phi_0} f_36{-1} \beta^{-1} B_{45}^{(2)} B_{67}^{(1)} \cos \theta - \frac{9}{8} \beta^{-1} B_{67}^{(1)} B_{45}^{(2)} \sin \theta \tan \theta H^{\frac{1}{2}} f_1
}}
where $B_{67}^{(1)}$ and $\chi$ on the right hand side are the original fields before twisting.  

\listrefs

\bye

Our scenario is more involved than the one considered in \difuli\ as we have non-trivial background topology,
branes and fluxes. A detailed discussion of this is given in sec 4.1. Secondly,
from our
computations, we can make a strong statement concerning the masses of the KK modes in the
presence of the NS field. We
explicitly show that the KK modes are heavier in the presence of the NS fluxes
which for confining theories was first observed in the
second paper of \nifuli. Being explicit, our solution does not have the potential problems
detailed in \land.
Our solution represents a concrete example of a
dipole deformed field theory, even though for all IR effects (that we are mostly concerned with) the
dipole deformations are not visible. Elaborating on the above analysis will be the subject of sec. 4.2.

Once we are in the realm of non-complex, non-K\"ahler manifolds,
we should also entertain possibilities of having {\it generalised}
complex structures {\it a l\`a} Hitchin-Gualtieri \hitchin,
\gualtieri. Our type II constructions have all the necessary
ingredients to realise these new configurations in string theory.
Some earlier attempts to discuss the {\it algebraic} aspects of
these manifolds have been presented in \minasian, \minalind. But a
full supergravity analysis of these new constructions still awaits
a thorough treatment. We will not attempt this here, and a more
detailed analysis on this will be presented soon in \gwyn.

Parallel to these developments are similar considerations for the
heterotic theory. We have already constructed examples that might
indicate possible gauge-gravity dualities also in heterotic theory
\realm, \gttwo. In \gttwo\ we gave an example of a manifold
without branes but only with fluxes (or {\it torsion} here) that
might form a gravity dual of the theory on heterotic NS5 branes.
The metric was a complex non-K\"ahler and non-compact manifold
that had some resemblance with the metric of \mn. In sec. 3.3 we
will discuss some new non-K\"ahler manifolds, some of which give
rise to the metric {\it before} geometric transition, i.e metrics
on which we can have wrapped five-branes. We will be able to
provide detailed discussion on the topological properties of these
manifolds, including a determination of the Betti numbers and the
cohomology classes.

Having such explicit background solutions with and without branes
still do not provide a convincing proof that they are related by
some gauge-gravity duality. We need more detailed analysis. One
possible way to in-principle confirm this is via a topological
theory argument, much like the one presented for type IIB \gv. As
we know, there is no ``standard'' topological theory or
topological twist in the heterotic theory. There is only a {\it
half-twist} \wittentop\ and therefore one would need to ask in the
half-twisted theories whether we can have dualities like \gv.
Existence of such a scenario would confirm possibilities of
gauge-gravity dualities in the heterotic theory. A recent attempt
to study correlation functions in (0,2) theories has been
addressed in \reczt, \wittwo. But to say something concrete, more
work is needed.

Finally one might also want to study {\it twisted} generalized
complex structures for a system like ours. An earlier work is
\kapuli, done for simple cases. For a background with non-trivial
topology a twisted version of a generalised complex structure may
be a hard problem to trace directly from a sigma-model point of
view but on the other hand knowing the explicit local geometry
might shed some light here. These aspects are still under
investigation.

\subsec{Formal outline of the paper}

In this paper we have tried to discuss many new aspects of
gauge-gravity dualities. Some of these aspects correspond to our
earlier studied models of geometric transitions. In sec. 2 we give
a detailed derivation of the local supersymmetry-preserving metric
with branes and fluxes, from equations of motion. The emphasis of
sec. 2.1 is to elucidate the non-trivial nature of the $U(1)$
fibration of the local metric. We show how warp factors and fluxes
conspire to give the right fibration. This analysis is done
without considering all the back-reactions of branes etc. In sec.
2.2 we study a fully supersymmetric configuration with D5s, D7s
and fluxes and their back-reactions. We give the possibility of
the existence of bound states of D5s on a single D7 brane that
could potentially occur at a point in the moduli space of our
configuration.

Section 3 is mostly a study of the corresponding heterotic
picture. The heterotic story that we present here (that has also
appeared in some of our earlier works \realm, \gttwo) is not in
any way {\it dual} to the type IIB background. Although the
original derivation was motivated by some duality arguments (see
\realm), the final configuration is deformed away from the
original result to a new metric that satisfies equations of motion
and is supersymmetric. The deformation is non-adiabatic and so
cannot be realised as a perturbation. As discussed earlier in
\realm, \gttwo\ there are two different heterotic backgrounds
conveniently classified as {\it before} and {\it after} geometric
transition. In \gttwo\ we analysed the background after geometric
transition. In sec. 3.1 we study the background before geometric
transition. This is a background given in terms of non-trivial NS5
branes wrapped on a two-cycle of the geometry. Our analysis show
that the resulting manifold is a new non-compact, non-K\"ahler
manifold that could even be complex. In this section we manage to
provide the local metric of the manifold, and in sec 3.2 we study
the torsion classes associated with this manifold. The manifold(s)
that we find are new, and in sec. 3.3 we study a family of such
manifolds including their mathematical structures and Betti numbers.

The analysis that we present in sec. 2 took all the branes and
fluxes into account. This analysis, however, could be thought of
as though we are {\it away} from the orientifold point. At the
orientifold point we have to carefully take the projections, and
this allows only some special $B_{NS}, B_{RR}$ fields. The choice
of these fields tells us that we have a {\it dipole} deformation
in the field theory. In sec 4 we elucidate this in great detail.
Due to non-trivial topology, branes and fluxes, the analysis turns
out to be particularly involved, and so we do this in two steps.
Step one is sec 4.1 where we study the system without
incorporating branes but keeping only non-trivial fluxes and
topology. Step two is sec. 4.2 wherein we put all the branes in
the geometry and study the possible deformations. With some effort
we manage to calculate the precise local metric with dipole
deformation\foot{Up to some possible subtleties that we will mention as we go
along.}. 
We also find something very interesting: the
decoupling of KK modes on the wrapped D5 branes. The two-cycle on
which we have wrapped D5 branes {\it shrinks} in volume due to the
background dipole deformation. We show this by evaluating the
volume both before and after deformation, and then calculating the
difference.

\noindent Finally in sec. 5 we give a short discussion and point
out possible future directions.

\newsec{New results on geometric transitions in type IIB theory}

This is a further continuation of our works \gtone, \realm\ and \gttwo, but now we would like to address issues
like solving equations of motion, possible dipole deformations and new non-K\"ahler manifolds.
Our earlier works were basically elaborating the story of geometric transitions by constructing precise
supergravity solutions that could be used to study the gauge/gravity dualities more consistently. Recall that
prior to our papers \gtone, \realm\ and \gttwo, there were {\it no}  supergravity descriptions for
geometric transitions. Most of the earlier descriptions were based on topological identifications that started
off with \gv\ (see also \ber)
and were soon incorporated in string theory by \vafai, \civ, \civd. Although many new developments were
reported using these identifications, a precise supergravity description was called for so that an explicit
quantitative analysis could be performed. This was not a concern for the other two parallel developments of
\ks\ and \mn\ that explored the same scenario from a sightly different angle because of the existence of
 supergravity backgrounds that formed the duals of cascading confining gauge theories.

\subsec{Precise supergravity analysis}

Our analysis of type IIB started with the dual of ${\cal N} = 1$ $SU(M)$ gauge theory that forms the far IR of a
gauge theory whose UV description involves $SU(N) \times SU(N+M)$ gauge fields with two 
different coupling constants\foot{There could be subtleties associated with the existence of Baryonic branches
that take us to different IR theories \minasianone, \seiberg. For our case we will ignore them here as we are only 
concerned with ${\cal N} = 1$ $SU(M)$ IR theories. Details on the other cases will be in the sequel to this paper.}. 
In the gravity description the far UV picture is captured by going to large radial distances i.e $r \to \infty$ in a
non-compact K\"ahler geometry. The IR picture, on the other hand, is captured when $r$ is small and this is also dual
to pure ${\cal N} =1$ gauge theory. Clearly the UV description can become complicated by various factors:

\noindent $\bullet$ Existence of flavors: Flavors can drastically
modify the UV picture for our case. In the model studied in
\gttwo, we showed that there are two different flavors that could
be considered in the story $-$ fundamental and bi-fundamental  $-$
which partake in the full UV description. The fundamental flavors
come from the seven-branes (not necessary all local) and the
bi-fundamental flavors come from the three-branes (necessarily
local). At low energies these flavors are either massive or reduce
in number by a renormalisation group flow and Seiberg dualities.
At high energies they can modify the story in interesting ways, so
we need to consider them carefully.

\noindent $\bullet$ Existence of KK modes: One of the clear
distinctions between the geometric transition picture and the
Klebanov-Strassler model is the existence of KK modes in the UV
for the former case. Recall that the UV of the geometric
transition is a {\it six} dimensional theory whereas the
Klebanov-Strassler model remains four-dimensional throughout. Once
the theory becomes six-dimensional i.e. the effect of the ${\bf
P}^1$ starts showing up, we have to consider the full theory on
the wrapped D5 branes. This would mean that from a
four-dimensional point of view we have to take into account the
full tower of KK states on the sphere. This becomes a formidable
problem.

Because of these issues, we see that the full $r \to 0$ to $r \to
\infty$ geometry is complicated. In the Klebanov-Strassler case
many of the above issues could be avoided, so a global geometry
can be easily considered. For our case we cannot ignore the KK
modes, so this will definitely make the UV behavior different from
the Klebanov-Strassler case. What about the flavors? Again,
clearly the bi-fundamental matter is the core of the story and
could not be avoided (even in the Klebanov-Strassler model), so
the question would be regarding the fundamental matter. As
discussed above, these are given by the local and non-local seven
branes. So can we ignore these seven-branes as we did for the
Klebanov-Strassler solution?

The model that we constructed in \gtone, \realm\ and \gttwo\
required us to take an F-theory solution which is a four-fold that
could be considered as a $T^2$ fibration over a K\"ahler base
${\cal B}$. The base ${\cal B}$ has at least one ${\bf P}^1$ that
is topologically non-trivial. On this two-cycle we can wrap $M$ D5
branes and this can easily give us ${\cal N} =1$ $SU(M)$ gauge
theory. However the F-theory torus also has to degenerate on the
base, and this will give us local and non-local seven-branes.

Having an underlying F-theory solution serves multi-fold purposes.
It can easily give us a type IIB background that preserves
supersymmetry in the presence of fluxes and branes. The base of
the fourfold can be made compact or non-compact; K\"ahler or
non-K\"ahler. In all cases we can have gauge theories preserving
minimal supersymmetry. The fourfold that we constructed in \gttwo\
had a K\"ahler base in the absence of branes and fluxes. In the
presence of fluxes and branes we know that the base could become
conformally K\"ahler or even non-K\"ahler. One might then expect
that the overall metric can be written as
\eqn\newmetnow{\eqalign{ds^2 = & ~F_0(\tilde r)~ ds^2_{0123} +
F_1(\tilde r) ~d\tilde r^2 + F_2(\tilde r)~ (d\tilde\psi + {\rm
cos}~\tilde\theta_1 d\tilde\phi_1 + {\rm cos}~\tilde\theta_2
d\tilde\phi_2)^2 + \cr & ~~~+ \Big[F_3(\tilde r)~d\tilde\theta_1^2
+ F_4(\tilde r)~{\rm sin}^2~\tilde\theta_1 d\tilde\phi_1^2\Big] +
\Big[F_5(\tilde r) ~d\tilde\theta_2^2 + F_6(\tilde r)~ {\rm
sin}^2~\tilde\theta_2 d\tilde\phi_2^2\Big],}} where the
$F_i(\tilde r)$ are the warp factors and we have labelled the
coordinates of the fourfold base as the radial coordinate $\tilde
r$, the two spherical coordinates ($\tilde\theta_i, \tilde\phi_i$)
and the $U(1)$ fibration as $\tilde\psi$.

There are a few important details regarding the above solution
that we should mention at this stage. First of all observe that we
have used {\it global} coordinates to write it. That would mean
that this solution is valid at $\tilde r \to \infty$ also. Whether
or not this could be the case still remains to be seen, because
our analysis from F-theory was done without considering localised
three-branes. Thus \newmetnow\ will have to be changed to reflect
the local behavior only.

Secondly, the way we constructed the metric tells us that the
background explicitly preserves supersymmetry. Thus this solution
is {\it different} from the one proposed in \pandoz\ as we
discussed in great detail in \gtone, \realm\ and \gttwo. The only
region where the two metrics \newmetnow\ and the one in \pandoz\
look similar in form is locally. The local behavior of both
metrics gives us \eqn\metformj{ds^2 = dr^2 + (dz + \Delta_1^0 {\rm
cot}~\langle\theta_1\rangle~ dx  + \Delta_2^0 {\rm
cot}~\langle\theta_2\rangle~dy)^2 + (d\theta_1^2 + dx^2) +
(d\theta_2^2 + dy^2)} Here ($r, z, x, \theta_1, y, \theta_2$) are
the local coordinates measured from a chosen point ${\bf P}_0$ in
our six-dimensional space \newmetnow, where \eqn\choipoi{{\bf
P}_0~ = ~r_0, \langle\psi\rangle, \langle\phi_1\rangle,
\langle\theta_1\rangle, \langle\phi_2\rangle,
\langle\theta_2\rangle} \noindent with $\Delta^0_i$ in the above
local metric defined as \eqn\deltanow{\Delta_1^0 ~ = ~
\sqrt{F_2(r_0)\o {F_4(r_0)}}, ~~~~~~~~~ \Delta_2^0 ~ = ~
\sqrt{F_2(r_0)\o {F_6(r_0)}}.} Imagine now that we choose our
point ${\bf P}_0$ not in the space \newmetnow, but at a point in
the space with the metric of \pandoz. How does the behavior of
$\Delta^0_i$ change? One can easily show by solving the background
equations of motion that the behavior of $\Delta^0_i$ is now:
\eqn\behofdel{\Delta^0_1 ~ = ~ \sqrt{\gamma_0'\over \gamma_0}~r_0,
~~~~~~ \Delta_2^0 ~ = ~ \sqrt{\gamma_0'\over \gamma_0 + 4a^2}~r_0}
with $a^2$ being the resolution factor. Thus up to re-definitions
of $\Delta^0_i$ the local behaviors are exactly identical!

There are still a few loose ends that we need to clarify before moving ahead. All have to do with our
metric \newmetnow\ and its local version \metformj.

\noindent $\bullet$ The metric \newmetnow\ could in general be
K\"ahler or non-K\"ahler. However our earlier derivations from
F-theory in \gttwo\ have only considered a K\"ahler base. Is it
possible to construct a non-K\"ahler base from our simple F-theory
derivation of \gttwo?

\noindent $\bullet$ We have not determined the warp factors $F_i$ in our metric \newmetnow. Of course
demanding spacetime supersymmetry will put some condition on these warp factors. Is it possible to predict
the susy constraints on the metric?

\noindent $\bullet$ Observe that our local metric has a $z-$
fibration that is indeed constant. This is because we haven't
taken the effects of the underlying seven-branes into account. In
our earlier papers \gtone, \realm\ and \gttwo\ we commented that
these constant fibrations will become non-constant. Can we predict
this from the background equations of motion?

All the above questions require detailed analysis. So let us start
with the first one. We now want to construct a fourfold whose base
is a generic non-K\"ahler space. Later we will make this space
also non-compact. We begin by considering a cubic hypersurface
$Y\subset{\bf P}^4$ satisfying the equation \eqn\fgequa{ x_1 f +
x_2 g  =0,} with $f$ and $g$ quadratic and general.  There are
conifolds at the four points \eqn\conifold{ x_1 = x_2 = f = g =
0.} They can be resolved by blowing up the surface $S\subset Y$
defined by $x_1=x_2=0$ to get a new threefold $X$.  Blowing up a
divisor doesn't change $Y$ at its smooth points, but repairs the
singularities at the conifolds.

Concretely, we introduce a variable $u=x_2/x_1$ to perform the
blowup and get \eqn\blowup{ f+ug=0,} which is smooth.  Over each
of the conifolds, \blowup\ is satisfied identically in $u$, so $u$
becomes a coordinate on the ${\bf P}^1$.  The other coordinate
patch on the ${\bf P}^1$s is given by $v=x_1/x_2$ and proceeding
similarly we complete the description of the blowup.

This $X$ is K\"ahler by standard facts in algebraic geometry. Or explicitly,
note that $X$ naturally embeds as a complex submanifold of ${\bf P}^4\times
{\bf P}^1$, which is K\"ahler, and its K\"ahler metric can be restricted to $X$.

Now $X$ contains four ${\bf P}^1$s. We can modify $X$ by flopping
only one of the ${\bf P}^1$s to obtain a new threefold $X'$ with
four ${\bf P}^1$s.  Now we no longer have an embedding into ${\bf
P}^4\times {\bf P}^1$ so the previous construction of a K\"ahler
class fails.  Indeed, it can be shown that $X'$ is not K\"ahler.
In the last paper, we had only one conifold, and flopping it
didn't destroy K\"ahlerity since in that case we could have
described the flop directly by using a different blowup.  If we
tried to do that here, we would end up having to flop all four
${\bf P}^1$s simultaneously.

Both of the resolutions $X,X'$ are essentially Fano.  More precisely, by the
adjunction formula, $c_1(X)=2H$, where $H$ is the hyperplane class of
$Y$ pulled back to $X$.  Similarly $c_1(X)=2H'$, where $H'$ is the
hyperplane class of $Y$ pulled back to $X'$.

So $c_1(X)\cdot C \ge 0$ for all curves C, and $c_1(X)\cdot C = 0$
only if $C$ is one of the four ${\bf P}^1$s coming from the resolved
conifolds.  So this is very similar to the example in our last paper.
The computation for $X'$ is identical.

This short computation  was intended to clarify the fact that we
may no longer be restricted to a K\"ahler base directly in type
IIB. A non-K\"ahler base could also lead to new gauge/gravity
dualities that have not been studied before. We will comment on
the possible topological string analysis for such a case in future
works.

For the time being observe that the metric \newmetnow, which forms
the base of the fourfold, cannot be regarded as the full global
metric because it doesn't show the existence of three- and seven-branes. On a
small patch in the neighborhood of ${\bf P}_0$ the metric is
\metformj. It seems that there may not exist a globally defined
coordinate for the system and the full global metric $-$ that
takes into account the D3s, D5s, D7s and the fluxes $-$ could only
be defined on patches. Nevertheless let us explore the constraints
on the warp factors $F_i(\tilde r)$ in \newmetnow\ and then we
shall restrict this on a given patch. The large $\tilde r$
behavior of $F_i$ for $i = 3,4,5,6$ can be expected to be
\eqn\smallrbe{\eqalign{&F_i(\tilde r) ~ = ~ {\tilde
r}^{k_i}~G_i(\tilde r), ~~~i = 3,4,\cr &F_j(\tilde r) ~ = ~
a^2~G_j(\tilde r), ~~~j = 5,6,}} where $a^2$ is the same
resolution parameter that we had in \behofdel. However we do not
require the large ${\tilde r}$ behavior, rather the small $r$
behavior. This can be easily arranged to be of the form
\eqn\smallev{\eqalign{F_i(\tilde r) &~ = ~ r_0^{k_i}~G_i(r_0) +
\Bigg(k_i~r_0^{k_i-1}~ G_i(r_0) + r_0^{k_i}~{\del G_i \over \del
\tilde r}\Big\vert_{\tilde r = r_0}\Bigg) r~ +\cr & + {r_0^{k_i}\o
2} \Bigg({k_i(k_i-2)\o r_0^2}~G_i(r_0) + {2 k_i\o r_0}~ {\del G_i
\over \del \tilde r} \Big\vert_{\tilde r = r_0} + {\del^2 G_i
\over \del \tilde r^2} \Big\vert_{\tilde r = r_0}\Bigg) r^2 ~ +
{\cal O}(r^3),}} where $i = 3,4$ in general. For $k_5, k_6$
defined as $$k_5 ~ = ~ k_6 ~ = ~ 2~{\rm log}_r~a$$ \noindent we
see that for $k_3 = k_4 \equiv \kappa$ the $\gamma$ defined in
\behofdel\ can be related to $G_i$ above only in the regime where
$G_3 \approx G_4 \equiv G$. In this regime the relation that
connects the space
\newmetnow\ with the one predicted by \pandoz\ is given by
\eqn\panda{\gamma'_0 ~r_0~ - ~ {\kappa ~\gamma_0} ~ - ~ r_0^{\kappa + 1}~G'(r_0)~ \to ~ 0,}
where an equality would correspond to exact identification. This is thus precisely the regime where we can trust
the metric of \pandoz.

In a generic situation when the equality in \panda\ is not
maintained, we have to worry about a couple of things. One of the
most important aspects is supersymmetry. As we discussed above,
both the global metric \newmetnow\ and the local version
\metformj\ preserve supersymmetry. However this conclusion was
extracted from our F-theory picture developed in \gttwo. The susy
model from F-theory {\it a priori} doesn't give any constraint on
the warp factors $F_i(r)$ because the F-theory solution is written
in terms of algebraic equations and not the metric. But we can use
the type IIB (2,2) sigma model on this background to derive
possible constraints. One of the simplest ways to start off is by
using the Poisson Sigma model \strobl\ and then include a
symmetric tensor. This has already been addressed in \gates,
\lindstrom, \minalind\ and the model can be written, in $(1,1)$
superspace, as
 \eqn\meto{S ~ =~ \int~ d^2 \xi d^2 \theta \big[{\bf \Psi}_{+\mu} {\bf
\Psi}_{-\nu}(G^{\mu\nu}+B^{\mu\nu}) + i {\bf
\Psi}_{(+\mu}D_{-)}{\bf \Phi}^\mu\big],}  where $G_{\mu\nu}$ is the
metric of \newmetnow\ and $B_{\mu\nu}$ is the NS B-field in this
background. It is easy to see that solving the equation of motion
of ${\bf \Psi}_{\pm \mu}$, we will get \eqn\psieqm{{\bf \Psi}_{\pm
\mu} ~ = ~ i D_{\pm}{\bf \Phi}^\nu (G_{\mu\nu}-B_{\mu\nu}),} which
when substituted back in \meto\ will give us the usual (2,2)
action of \gates. One might then ask about the susy variation of
${\bf \Phi}^\mu$ when we are on-shell for ${\bf \Psi}$. The susy
variation for ${\bf \Phi}$ is \eqn\susybac{\delta {\bf \Phi}^\mu ~
= ~ \epsilon^\pm~D_\pm {\bf \Phi}^\nu~J^{(\pm)\mu}_\nu - i
\epsilon^\pm {\bf \Psi}_{\pm
\rho}(G^{\nu\rho}-B^{\nu\rho})~\II^\mu_\nu, } where
$J^{(\pm)\mu}_\nu$ are two complex structures and $\II$ is the
identity matrix. Combining \susybac\ with \psieqm\ implies that
our background would preserve (2,2) supersymmetry if we chose
${\bf \Phi}^\mu$ in such a way that it satisfies
\eqn\phisat{\delta {\bf \Phi}^\mu ~ = ~ \epsilon^{\pm}D_\pm{\bf
\Phi}^\nu(J^\mu_\nu \pm \II^\mu_\nu).} The first ($\theta =0$)
components of ${\bf \Phi^{\mu}}$ have the usual interpretation as
complex coordinates $x^\mu$ for our space
\newmetnow\ while the components linear in $\theta$, $D_{\pm}
\bf{\Phi}\mid_{\theta =0}$, are the world-sheet fermions
$\lambda_\pm$. Thus, at $\theta =0$ eq.  \phisat\ gives
 \eqn\cmz{\delta x^\mu~=~ \epsilon^\pm\lambda_\pm^\nu (J \pm
\II)^\mu_\nu,} We could then use these components to construct {\it
primitive} fluxes in our space but will not do so here. It is also
important to notice that we have made no mention of the {\it
choice} of the complex structures. As we know there are two
allowed complex structures for our case \gttwo, \gates. For the
time being it is easy to see that there are three complex
one-forms given as: \eqn\oneforms{\eqalign{&\epsilon_0 ~=~
-\sqrt{F_1}~d\tilde r ~+~ i\sqrt{F_2}~ (d\tilde\psi + {\rm
cos}~\tilde\theta_1 d\tilde\phi_1 + {\rm cos}~\tilde\theta_2
d\tilde\phi_2), \cr & \epsilon_1 ~ = ~ \sqrt{F_3}~d\tilde\theta_1
~+~ i\sqrt{F_4}~ {\rm sin}~\tilde\theta_1~d\tilde\phi_1, ~~
\epsilon_2 ~ = ~ \sqrt{F_5}~d\tilde\theta_2 ~+~ i\sqrt{F_6}~ {\rm
sin}~\tilde\theta_2~d\tilde\phi_2.}} These one-forms are
particularly useful for constructing higher p-forms in IIB theory.
What we require for our case is to allow only primitive (2,1)
forms. This is possible if \eqn\primi{F_3~F_4 ~ - ~ F_5~F_6 ~ = ~
0,} which is motivated from the somewhat similar correspondence
for the background constructed in \pandoz. Clearly the metric of
\pandoz\ does not satisfy \primi\ and therefore breaks
supersymmetry \cvetic. So our minimal constraint should be \primi\
on the warp factors. One easy way to impose this on \smallrbe\
will be to consider \eqn\ggkk{(k_3, k_4) ~=~(4~{\rm log}_r~a -
k_4, k_4), ~~~~~~~~G_3~G_4~=~G_5~G_6,} where $a$ is the resolution
parameter for the resolved conifold as before. We will also have
to impose the condition \eqn\condg{a ~ \to ~ 0, ~~~ (G_5, G_6) ~
\to ~ \infty} such that ($F_5, F_6$) remain finite. In this limit
the metric of \pandoz\ becomes supersymmetric which is in fact the
metric of a fractional brane, and is therefore directly related to
the Klebanov-Strassler metric \ks. It is also easy to see that the
local metric \metformj\ satisfies the primitivity constraint
\primi\ and therefore preserves supersymmetry.

The way we have presented our local metric \metformj\ shows only a
constant $dz$ fibration. What we need for our analysis is a metric
with a non-constant $U(1)$ fibration so that it could be related
to our earlier metric of \gtone, \realm\ and \gttwo. So the
question is, under what condition does it allow non-trivial
fibration? To answer this, let us first assume that we can have a
generic local metric of the form \eqn\genlocmet{ds^2 ~ = ~ dr^2 +
\Big(dz + f_1(\theta_1)~dx + f_2(\theta_2)~dy\Big)^2 + \vert
dz_1\vert^2 + \vert dz_2\vert^2,} where $dz_1, dz_2$ form the two
tori (see discussions in \gtone, \gttwo).

We see the metric of \genlocmet\ is different from \metformj\ because of non-constant $f_1, f_2$. One immediate
question that might arise is why we are getting non-constant $f_i$ when by doing a local reduction from
\newmetnow\ we get \metformj. A very brief discussion of this was presented in \gttwo, and here we would like
to elaborate on it.

\noindent $\bullet$ First, observe that the metric \newmetnow\ is
not the global description for our case. A full global description
will require us to have $D5s, D3s$ and seven-branes. When we
ignore the $D3$ branes and keep the seven-branes far away\foot{To
be precise, this means that the wrapped five-brane metric does get
back-reacted, but the axion-dilaton are still negligible in a
local neighborhood near the five-branes.} the metric resembles
\newmetnow.

\noindent $\bullet$ Secondly, if we ignore the $D3$ branes and
also the seven-branes, but keep only the wrapped $D5$ branes then
the metric we get is the one predicted by \pandoz. This metric
breaks supersymmetry as we discussed above (see also \cvetic,
\gtone, \gttwo).

\noindent $\bullet$ Thirdly, for both cases the local behaviors
are almost identical except they differ in the details of the
$U(1)$ fibration. So a natural question would be to ask what
happens when we keep the seven-branes in the local vicinity of the
wrapped $D5$ branes.

\noindent $\bullet$ Finally, for the metric \newmetnow\ we should also determine the warp factors from the
equations of motion. In fact the equations of motion should be used to get the full global solution that
incorporates the $D3$ branes also.

Thus alternatively, in the absence of a full global solution, we
could impose equation of motion constraints to determine the
$f_i(\theta_i)$ in \genlocmet. This way we will know how much
control we have on the behavior of the axion-dilaton, at least
locally. Then a global solution could presumably be constructed by
connecting all the local patches.

It turns out, for our case, an analytical solution can be worked
out for the $f_i$ factors using a simple set of duality maps.
The map that we are interested in takes us to M-theory wherein the
analysis becomes tractable. It is not too difficult to see that
the $f_i$ of \genlocmet\ is mapped to a configuration of two
points $A$ and $B$ on a cylinder of length $l$ in M-theory such
that these two points form a codimension 4-surface in a Calabi-Yau
space. In  {\bf figure 1} below: 
\vskip.17in

\centerline{\epsfbox{m5cylin.eps}}\nobreak

\vskip.18in
\noindent we have denoted the surfaces as two points $A, B$ on an M-theory cylinder of length $l$
with the compact angular direction
related, as usual, to IIA coupling. On the other hand the
$f_i$ also map to four-form $G$-fluxes given as
\eqn\gflux{G ~ = ~ df_1 \oplus df_2.}
These co-dimensional surfaces should {\it not} be interpreted as any kind of dynamical branes in M-theory. Right now
they simply behave as localised sources of G-fluxes without having any world-volume dynamics. We will show that the
only possible way they could become dynamical is if the IIB warp factors $F_i$
in \newmetnow\ are taken into account. So the question would be to see how the warp factors change the story.

To proceed let us first assume that the warp factors $F_i(\tilde r)$ can be separated as products of two functions
in the following way
\eqn\prodoft{F_i(\tilde r)~ = ~ F_0^{-1}(\tilde r)~{\cal F}_i(\tilde r), ~~~~i ~\ne~ 0}
A physical motivation for this conjecture comes from the F-theory origin of our metric \newmetnow\ (see also
discussion in \dotd\ and \dotu)
and can be easily argued from the warped metric ansatz of \rBB\ using the analysis of
\sav.

One question would now be to ask what kind of $\tilde r$-behavior
we expect from $F_0$ and ${\cal F}_i$? We should look at the
metric \newmetnow\ for inspiration. In IIB \newmetnow\ can be
written as \eqn\metads{ds^2 ~ = ~ F_0~ds^2_{0123} +
F_0^{-1}~ds^2_{\cal M},} where ${\cal M}$ can be extracted from
\newmetnow. We see that in this form the metric fits into the
D-brane ansatz, so the behavior of $F_0$ can be predicted as
\eqn\fzero{F_0~=~ c_0 + \sum_i~{c_i \o \tilde r^{n_i}},} where the
$n_i$ are some integers with $c_0$ being basically
constant\foot{The conjecture \fzero\ is generic, but not generic
enough for higher-dimensional branes. As we know, sometime
delocalisation can change the behavior of the harmonic functions.
So if we remove the restriction of positivity on $n_i$ in \fzero\
then we might capture all possible solutions. Henceforth $n_i
\equiv \pm \vert n_i\vert$ unless mentioned otherwise.}. We also
know that the sum over $i$ terminates at some point so as to have
a physical model and generically $i$ cannot exceed some small
number.

Once we fix a possible behavior for $F_0$, we can try to work out
the possible $\tilde r$ dependence for ${\cal F}_i$. Better still,
we can try to determine the small $r$ behavior of ${\cal F}_i$.
The relation between $\tilde r$ and $r$ is already given in
\gttwo\ (see eq. 2.13) there) and therefore we will simply use
this to write the possible $r$ behavior of the warp factors. The
small $r$ expansion is given by \eqn\wurp{\eqalign{{\cal
F}_i(\tilde r)& ~ = ~ {\cal F}_i(r_0) ~+~ {r\o \sqrt{{\cal
F}_1(r_0)}}~ {\del {\cal F}_i \o \del \tilde r}\Big\vert_{\tilde r
= r_0} + {r^2\o 2{{\cal F}_1(r_0)}}~{\del^2 {\cal F}_i \o \del
\tilde r^2}\Big\vert_{\tilde r = r_0} + {\cal O}(r^3)\cr &~ =  ~
{\cal F}_i(r_0) ~+~ \alpha_i r + \beta_i r^2 + {\cal O}(r^3),}}
where we have kept terms up to second order in $r$ and $\alpha_i,
\beta_i$ can be easily identified.

At this point note that the local metric \metformj\ is in fact an
approximation where we ignore the $r$ dependence of the warp
factors ${\cal F}$ and keep only the constant terms ${\cal
F}_i(r_0)$. Putting in the $r$ dependence will give us
\eqn\metmet{{ds^2_{\cal M}~ = ~ {\cal A}~dr^2 + {\cal B}~(dz +
f_1~ dx + f_2~dy)^2~ + ({\cal C}~d\theta_1^2 + {\cal D}~ dx^2) +
({\cal E}~ d\theta_2^2 + {\cal F}~ dy^2),}} with the various
coefficients now defined as (we keep only up to $r^2$ terms)
\eqn\coffdeff{\eqalign{& {\cal A} = 1 + {\alpha_1\o {\cal
F}_1(r_0)} r +  {\beta_1\o {\cal F}_1(r_0)} r^2; \cr & {\cal B} =
1 + {\alpha_2\o {\cal F}_2(r_0)} r +  {\beta_2\o {\cal F}_2(r_0)}
r^2; \cr & {\cal C} = 1 + {\alpha_3\o {\cal F}_3(r_0)} r +
{\beta_3\o {\cal F}_3(r_0)} r^2; \cr & {\cal D} = \Bigg(1 + {\rm
cot}~\langle \theta_1\rangle~\sum_n~b_n \theta_1^n\Bigg) \Bigg(1 +
{\alpha_4\o {\cal F}_4(r_0)} r +  {\beta_4\o {\cal F}_4(r_0)}
r^2\Bigg); \cr & {\cal E} = 1 + {\alpha_5\o {\cal F}_5(r_0)} r +
{\beta_5\o {\cal F}_5(r_0)} r^2; \cr & {\cal F} = \Bigg(1 + {\rm
cot}~\langle \theta_2\rangle~\sum_n~c_n \theta_2^n\Bigg) \Bigg(1 +
{\alpha_6\o {\cal F}_6(r_0)} r +  {\beta_6\o {\cal F}_6(r_0)}
r^2\Bigg),}} where $c_n, b_n$ are small non-zero constants and
$\alpha_i, \beta_i$ are defined as before. It is easy to see that
for ($r, \theta_1, \theta_2$) $\to 0$ these warp factors are
essentially constants, as they should be. This is consistent with
our local metric ansatz. We also see that the $U(1)$ fibration is
no longer required to be constant and is given in terms of
$f_1(\theta_1), f_2(\theta_2)$. In fact we can write the form of
the $f_i$ also. They are given by \eqn\fivalue{f_1 ~ = ~
\sqrt{{\cal F}_2(r_0) \o {\cal F}_4(r_0)}\Bigg({\rm cot}~\langle
\theta_1\rangle + \sum_n~a_n \theta_1^n\Bigg), ~~f_2 ~ = ~
\sqrt{{\cal F}_2(r_0) \o {\cal F}_6(r_0)}\Bigg({\rm cot}~ \langle
\theta_2\rangle + \sum_n~d_n \theta_2^n\Bigg),} where again $a_n,
d_n$ are small constants. The above form of the $f_i$ is perfectly
consistent with \deltanow. All we now need is to determine the
coefficients $a_n, b_n, c_n$ and $d_n$ from the background
equations of motion and get a closed form for the series.

This, as it stands, is a formidable task and we shall see how far we can pursue this to get the kind of answer that
we want for our case. We start by considering some limits, and will show how the final answer would justify these
simple assumptions. From the warp factors \coffdeff\ we can easily construct new warp factors by introducing
algebraic relations between them. For example
\eqn\cerel{{\cal C} {\cal E} ~ = ~ 1 + C_0 r + C_1 r^2 + C_2 r^3 + {\cal O}(r^4),}
where we have kept the $r^3$ term in the series assuming that $C_2$ coefficient is well defined. The various
$C_i$ can be easily extracted from \coffdeff\ and are given by
\eqn\defcof{\eqalign{& C_0 ~ = ~ {\alpha_3\o {\cal F}_3(r_0)} + {\alpha_5\o {\cal F}_5(r_0)}, ~~~~C_2 ~ = ~
{\alpha_3 \beta_5 + \beta_3 \alpha_5\o {\cal F}_3(r_0){\cal F}_5(r_0)},\cr
& C_1 ~ = ~ {\beta_5\o {\cal F}_5(r_0)} + {\alpha_3\alpha_5\o {\cal F}_3(r_0){\cal F}_5(r_0)} +
{\beta_3\o {\cal F}_3(r_0)}.}}
The above is simply a redefinition: we haven't said anything yet. All we have to see are the constraints on the warp factors coming
from the equations of motion. There are numerous papers that study such systems (see for
example \tseytii\ and citations therein; and \pisin\ for more recent advances) so we will not go through them in
detail. The interested readers may want to see these references.

To simplify our ensuing analysis of the equations of motion we
will ignore the fluxes for the time being. This will not change
the expected behavior in any significant way because we will have
more than one way to verify the correctness of the analysis. The
background equations of motion thus put the following constraints
on the various coefficients: \eqn\conscoef{\alpha_i~ >~ \beta_i,
~~i = 3,..,6, ~~~~~~~(b_n, c_n)\big\vert_{n\ge 1}~ \to ~ 0} with
$b_0$ and $c_0$ arbitrary (but could be small); and both $\beta_1,
\beta_2$ are not required to be smaller than $\alpha_1, \alpha_2$. In
fact the equations of motion demand the following simple relations
between $\alpha_i$ and $\beta_i$: \eqn\relbetalbe{\eqalign{&
{\alpha_1\o {\cal F}_1(r_0)} - {\alpha_3\o {\cal F}_3(r_0)}-
{\alpha_5\o {\cal F}_5(r_0)} ~ = ~ 0, \cr & {\beta_1\o {\cal
F}_1(r_0)} - {\beta_5\o {\cal F}_5(r_0)} - {\beta_3\o {\cal
F}_3(r_0)}~ = ~
 {\alpha_3\alpha_5\o {\cal F}_3(r_0){\cal F}_5(r_0)}.}}
These relations should get modified as higher-order terms in $r$
are incorporated. However since we have imposed \conscoef\ the
terms that are higher order in $\beta_i$ will also get
subsequently reduced. Therefore the above equations will not get
corrected too much.

There are also a few more relations that do not directly appear from the equations of motion, but could be
justified nevertheless. They are of the form:
\eqn\morerel{\eqalign{
& {\alpha_3\o {\cal F}_3(r_0)} - {\alpha_4\o {\cal F}_4(r_0)} ~ \approx ~ 0, ~~~~ {\beta_3\o {\cal F}_3(r_0)}
- {\beta_4\o {\cal F}_4(r_0)} ~\approx ~ 0, \cr
& {\alpha_5\o {\cal F}_5(r_0)} - {\alpha_6\o {\cal F}_6(r_0)} ~ \approx ~ 0, ~~~~ {\beta_5\o {\cal F}_5(r_0)}
- {\beta_6\o {\cal F}_6(r_0)} ~\approx ~ 0.}}
These equations are only approximate and their exact forms are not known as finding them requires solving higher order
equations of motion. Furthermore these equations are valid only for the particular set-up of a resolved
conifold or a conifold.

The relations of $\alpha_2$ and $\beta_2$ with other coefficients
are a little tricky to work out from equations of motion as their
closed forms are difficult to derive. We have been able to work
out the relations only when $\alpha_i, \beta_i$ and ${\cal
F}_i(r_0)$ are very small. In that case there are perturbative
expansions that one could use to determine the results. After the
dust settles, the final results are somewhat similar to
\relbetalbe\ but a little more complicated:
\eqn\relbetnow{\eqalign{& {\alpha_2\o {\cal F}_2(r_0)} +
{\alpha_4\o {\cal F}_4(r_0)}+ {\alpha_6\o {\cal F}_6(r_0)} ~ = ~
0, \cr & {\beta_2\o {\cal F}_2(r_0)} + {\beta_6\o {\cal F}_6(r_0)}
+ {\beta_4\o {\cal F}_4(r_0)}~ = ~ {\alpha^2_4\o {\cal
F}^2_4(r_0)} + {\alpha^2_6\o {\cal F}^2_6(r_0)}+
{\alpha_4\alpha_6\o {\cal F}_4(r_0){\cal F}_6(r_0)},}} which could
be related to \relbetalbe\ using \morerel. However observe the
relative sign differences between \relbetalbe\ and \relbetnow.
This will be crucial later.

What we now require is to evaluate the complex structures of the
two tori in the metric \metmet. One can easily see that the
complex structures are of the form $\tau = i\tau_2$ with vanishing
real part. The imaginary part is given by
\eqn\cstar{\eqalign{\tau_2 ~ = ~& 1 + {1\o 2} \Big({\alpha_4\o
{\cal F}_4(r_0)} - {\alpha_3\o {\cal F}_3(r_0)}\Big)r ~ + \cr &
~~~~~~~~~~~+ {1\o 2} \Big({\alpha^2_3 - \beta_3{\cal F}_3(r_0) \o
{\cal F}^2_3(r_0)} - {\alpha_3\alpha_4\o {\cal F}_3(r_0){\cal
F}_4(r_0)} + {\beta_4\o {\cal F}_4(r_0)}\Big)r^2 + {\cal
O}(r^3).}} Applying now the constraints that we derived in
\morerel\ the complex structure will take the final form as
\eqn\tautwo{\tau_2 ~ = ~ 1 + {\cal O}(r^3)} and therefore gives us
square tori at least up to the order $r^2$. We believe this will
continue to hold to arbitrary orders in $r$, but we haven't
checked this as yet.

The readers may have already noticed that the above conclusion is perfectly consistent with our local metric
ansatz that we gave in \gtone, \realm\ and \gttwo. In fact our present analysis should be thought of as a
consistent derivation of this fact from first principles. What we now require is to evaluate the fibration
structure in \metmet\ to get our final form of the metric.

To do this we first apply the conditions \relbetalbe\ and \relbetnow\ in \metmet\ assuming of course the
approximate constraints \morerel. The identifications are a little tedious to entangle, but one can see the
following structure evolving:
\eqn\newstrt{\eqalign{&{\cal A} - {\cal C}\cdot {\cal E} ~ = ~ {\cal O}(r^3),\cr
& {\cal B}
- {{\cal D}^{-1} \cdot {\cal F}^{-1} \o
\big(1 + b_0~{\rm cot}~\langle \theta_1\rangle\big)\big(1 + c_0~{\rm cot}~\langle \theta_2\rangle\big)}
~ = ~ {\cal O}(r^3),}}
which are actually evaluated under two very specific conditions: (a) the coefficients $b_n, c_n$ for $n \ge 1$ are
neglibly small, and (b) the higher order ${\cal O}(r^p)$ terms for $p \ge 3$ approach zero quickly.

What conditions can we impose on the constants $b_0, c_0$? With the weak form of the constraints \morerel, we can only
say that $b_0, c_0$ are very small. If \morerel\ is an exact equality then it is easy to show that
\eqn\bncn{b_n ~ = ~ 0, ~~~~~ c_n ~ = ~ 0, ~~~~~n \ge 0.}
Under this condition we find some surprising simplification, with \newstrt\ reducing to the following condition on
${\cal B}$:
\eqn\calbco{{\cal B} - {\cal C}^{-1} \cdot {\cal E}^{-1} ~ = ~ {\cal O}(r^3),}
which will allow us to have some important simplification in the equations of motion for $f_i$, although we
should remember that \bncn\ is a strong condition and may not exactly hold for our background. However since
the weak condition \morerel\ implies that ($b_n, c_n$) are essentially very small, we cannot be too far off from our
results.

Our next task is to figure out the relation between the warp
factors ${\cal C}$ and ${\cal E}$. Observe that all other warp
factors in the metric \metmet\ are given in terms of either ${\cal
C}$ or ${\cal E}$ or both. The relation between ${\cal C}$ and
${\cal E}$ can be easily worked out if we demand supersymmetry. We
have already seen this earlier in \primi\ and in \ggkk. For our
present case the susy constraint on the metric \metmet\ gives us
the following relation between ${\cal C}$ and ${\cal E}$:
\eqn\cerel{{\cal C}~ = ~  {\cal E}\sqrt{1+\tau_{2(2)}^2 \o
1+\tau_{2(1)}^2},} where $\tau_{2(i)}$ for $i = 1,2$ are the
complex structures of the two tori, evaluated earlier in \cstar\
(for one of the tori). On the other hand, for a more generic
conifold of the form \eqn\genconi{(XY)^l ~ = ~ (ZW)^m} and their
resolved cases, the above equations \morerel\ (and \cerel) will
pick up a relative factor of ${m\o l}$ in all the relations as
\eqn\exrelll{{\alpha_3\o {\cal F}_3(r_0)} - {m\o
l}\Big({\alpha_5\o {\cal F}_5(r_0)}\Big)~ \ge ~ 0, ~~~~~ {\rm for}
~~~  l ~\ge~m} with similar coefficients for others. Observe that
when $l \ne m$ then there is no equality between the coefficients.
In our present analysis we will restrict ourselves to the simplest
case of $l = m$ with \genconi\ resolved by blowing up a ${\bf
P}^1$.

Under the weak form of the constraint \morerel\ the two complex
structures are indeed equal and they both form square tori with
$\tau_{(1)} = \tau_{(2)} = i$. Therefore \cerel\ will simplify
drastically giving rise to {\it one} warp factor $-$ say ${\cal
C}(r)$ $-$ in terms of which which the metric \metmet\ could be
written. This implies that the final metric for our case that
satisfies all the equations of motion can be written as
\eqn\metmetnow{{ds^2_{\cal M}~ = ~ {\cal C}(r)^2~dr^2 + {\cal
C}(r)^{-2}~\Big(dz + f_1(\theta_1)~ dx + f_2(\theta_2)~dy\Big)^2~
+ {\cal C}(r)~\vert dz_1\vert^2 + {\cal C}(r)~\vert dz_2\vert^2,}}
with ${\cal C}(r)$ defined as before and $dz_i$ the two tori with
complex coordinates \eqn\ccord{dz_1 ~ = ~ d\theta_1 + i dx, ~~~~~~
dz_2 ~ = ~ d\theta_2 + idy} We would like to remind the reader
that the above metric not only satisfies all the equations of
motion, but also satisfies the supersymmetry constraints. This
should be contrasted with the metric of \pandoz\ which satisfies
the equations of motion but not the susy constraints. We also see
that the metric is exactly of the form predicted in \gtone,
\realm\ and \gttwo.

\noindent To complete the picture we have to answer the following questions now:

\noindent $\bullet$ How do we show that the background is K\"ahler?

\noindent $\bullet$ What are the values of the coefficients $f_i(\theta_i)$ appearing above in \metmetnow?

\noindent $\bullet$ Can we determine the warp factors $F_0$ and ${\cal F}_i$ in the original metric \newmetnow?

\noindent $\bullet$ We haven't introduced the effects of the
seven-branes. How are the seven-branes affecting the story here?

\noindent $\bullet$ Finally, how is the M-theory picture that we
gave earlier modified once we know the warp factors correctly?

We will start by making some comments on the K\"ahlerity issue of
the metric. Recall that in some of our earlier works \beckerD,
\bbdg, \bbdgs, \bbdp, \bd\ we have constructed non-K\"ahler
manifolds that have a somewhat similar fibration structure as above.
However the details of the fibration differ, and also there were
$U(1) \times U(1)$ fibrations instead of the single $U(1)$
fibration presented here. The examples therein were compact and
mostly in the heterotic theory, whereas here the manifolds are in
type IIB and are non-compact. In addition to that the heterotic
examples have a topology of a non-trivial ${\bf T}^2$ fibration
over ${\bf K3}$ bases. Here we will have a non-trivial ${\bf S}^1$
fibration over a ${\bf T}^2 \times {\bf T}^2$ base.

A further analysis on the issue of K\"ahlerity can only come after
we have evaluated all the unknown coefficients in the metric
\metmetnow. A formal analysis of the equations connecting the
fibration coefficients $f_i$ with the warp factors will give us a
simple result where $a_1 = d_1 = 1$ and $a_n = d_n = 0$ for $n \ne
1$ in \fivalue. But we can do a little better than that. As we
pointed out earlier in \gflux, the $f_i$ map to four-form G-fluxes
in M-theory. We will presume that these fluxes can be globally
defined because then they would consistently couple with the
points {\bf A} and {\bf B} discussed in fig. 1 above, to form
sources. This would in turn mean that the $f_i$ now satisfy the
standard source equations {\it globally}. The resulting analysis
turns out to be straightforward, but long and tedious. Interested
readers may want to see related details in \rustse, \tduality,
\tscvetic, \pisin\ etc. The final results are two decoupled
equations connecting the various coefficients with the warp factor
${\cal C}$ as: \eqn\fcrela{\eqalign{&{\del f_1 \o \del \theta_1} -
{\del {\cal C} \o \del r} + f_1~{\rm cot}~\theta_1 ~ = ~ 0; \cr &
{\del f_2 \o \del \theta_2} - {\del {\cal C} \o \del r} + f_2~{\rm
cot}~\theta_2 ~ = ~ 0.}} To solve the above equations we will
assume that the radial coordinate is small. This is justifiable
because we are in the local regime where $r$ is indeed small.
Secondly we will take $\alpha_3 >> \beta_3$. To justify this we
need to go back to \conscoef\ where we argued the weaker form
$\alpha_3 > \beta_3$. However since ${\cal O}(r^2)$ terms are
small, this could be justified and hence ${\cal C}$ will become
simpler: \eqn\calcnow{{\cal C} ~ = ~ 1 + \Bigg({1 \o {\cal
F}_3(r_0) \sqrt{{\cal F}_1(r_0)}} {\del {\cal F}_3 \o \del
r}\Big\vert_{r = r_0}\Bigg)~r ~ \equiv ~ 1 + Q~r,} where we have
written the partial derivative w.r.t. $r$ instead of $\tilde r$ to
avoid clutter, and $Q$ is defined accordingly. Thus plugging
\calcnow\ into the two equations \fcrela\ we get our two fibration
coefficients $f_i$ for $\theta_i \ne 0$
as \eqn\finow{f_1(\theta_1)~ = ~ Q~{\rm
cot}~\theta_1, ~~~~~~ f_2(\theta_2)~ = ~ Q~{\rm cot}~\theta_2,}
which is {\it exactly} what we had predicted earlier in \gtone,
\realm\ and \gttwo\foot{At least up to the coefficient $Q$ which,
since it is a constant, can be absorbed in the definitions of $dx,
dy$.}! The final metric therefore takes the following form:
\eqn\metnownow{\eqalign{ds^2_{\cal M}~ = ~ &{\cal C}(r)^2~dr^2 +
{\cal C}(r)^{-2}~\Big(dz + Q~{\rm cot}~\theta_1~ dx + Q~{\rm
cot}~\theta_2~dy\Big)^2~ + \cr & ~~~~~~~~~~~~~~~~~~ + {\cal
C}(r)~(d\theta_1^2 + dx^2) + {\cal C}(r)~(d\theta_2^2 + dy^2),}}
which is the same as our predicted local metric in the limit where
$r \to 0$ and ${\cal C} \to 1$. Thus we have justified all the
choices made in understanding the duality cycle for geometric
transition.

Now before we go back to the issue of K\"ahlerity of our metric,
let us ask how the wrapped five-branes and the seven-branes show
up in our metric \metnownow. We have already seen how to put
five-branes in our setup (see sec. 3 of \gtone). In our present
formulation, the existence of five-branes would be signalled by
the warp factor $F_0(r)$ in \prodoft. In fact singularities of
$F_0$ will be related to the presence of localised five-brane
charges. For the local region, where we are located near $\tilde r
= r_0$ and have access only to the small neighborhood governed by
the coordinates ($r, x, y, \theta_i, z$) we will not detect the
singularity and $F_0$ will be essentially constant.

Clearly then the global behavior with wrapped D5 branes has to
incorporate these issues. The seven-branes on the other hand, not
only introduce the singularities (associated with the position of
the seven branes) but also change the {\it topology} of the
underlying space by converting the tori to spheres \gsvy. For
example if the F-theory seven-branes are kept at a point on the
($x, \theta_1$) torus, then a large number of such seven-branes
will compactify that direction to an approximate spherical
topology. So our original metric \newmetnow\ with ${\cal F}_i$
given as \wurp\ is unlikely to capture the global picture.

\subsec{Analysis of the complete background: Branes and Fluxes}

One might also wonder about the case when we introduce back the D3
branes along with the D5s and the seven-branes. To solve for the
metric in the presence of all these branes is quite formidable,
and at present no known solutions exist. Some attempts to address
this issue with D3s and D7s have been discussed earlier in \afm\
 based on a F-theory picture advocated in \dmftheory. Thus
we cannot go too far in $r$ in the original metric
\newmetnow\ without hitting an $F_i$ singularity, and we cannot go too far in the angular direction without
encountering a possible topology change. Thus choosing a local patch like \metnownow\ from \newmetnow\ seems
like the only known solution for the system.

Let us analyse this a bit more. From \gttwo\ we know that the F-theory torus can degenerate over a
2d surface given by $dz_1 = dx + id\theta_1$. In the presence of all the branes, the generic metric along the
$z_1$ direction is given by
\eqn\sevmet{ds^2_1 ~ = ~ \vert f(r,z_1, \bar z_1)~dz_1\vert^2 ~ = ~ {\cal C}(r) \big\vert e^{\phi\o 2} dz_1\big\vert^2
~~~{\longrightarrow}~~~ {\cal C}(r) \vert dz_1\vert^2,}
when the dilaton $\phi \to 0$. Thus moving the seven branes far away in the $z_1$ space will imply that the warp
factors are given simply by ${\cal C}(r)$ and have no angular dependence. This is again consistent with our earlier
choice in \gtone, \realm\ and \gttwo.

Now that we know the precise local metric and also the effects of
moving the seven branes, we should re-analyse our configuration.
In  {\bf figure 2} below:

\vskip.17in

\centerline{\epsfbox{path.eps}}\nobreak

\vskip.18in \noindent the local patch in six-dimensional space is
denoted by a cubical space. We denote the ($x, \theta_1$)
direction as the x-axis, the ($y, \theta_2$) direction as the
y-axis, and the ($z, r$) direction as the z-axis. This way the
full six-dimensional space can be represented. In this space the
seven-branes are two dimensional surfaces that partition the patch
into two regions. Clearly the seven-branes are stretched along the
($y, \theta_2, z, r$) directions and are points in the ($x,
\theta_1$) direction. The D5 branes wrap the ($y, \theta_2$)
direction and appear as 1D lines inside the cube. It is easy to
see that the D5 branes are parallel to the seven-branes and can be
moved {\it away} from them  along the ($x, \theta_1$) direction.
The D3 branes (which wouldn't be present if we wanted to study
only the IR of gauge theory, but would be there in the full UV
story) appear as points inside the cubical space.

In the figure above it is easy to see that there could be strings
stretched between the seven branes and the five-branes and also
between the D3 branes. These strings that stretch between the
seven-branes and the D3s and D5s give rise to fundamental
multiplets in ${\cal N} = 1$ gauge theory. Similarly the strings
stretched between the D5s and D3s give rise to bi-fundamental
multiplets. The fundamental multiplets can be made very massive by
moving the seven-branes away, as can be easily seen by embedding
this patch in the full global geometry. In  {\bf figure 3} below:
\vskip.17in

\centerline{\epsfbox{sphere.eps}}\nobreak

\vskip.18in \noindent our local patch is shown inside the full
geometry. We have also kept the patch near the origin to emphasis
the IR behavior of our configuration. The seven-branes and the D3
branes can be moved out of the patch to construct pure ${\cal N}
=1$ gauge theory. The origin of the space will have the topology
of a resolved conifold.

What does all this say about the supersymmetry of our model? As
long as our local metric is of the form \metnownow\ with ${\cal C}
\to 1$ we would preserve supersymmetry with both five-branes and
seven-branes. From an F-theory point of view, the fourfold could
be constructed with a $T^2$ fiber degenerating along the ($x,
\theta_1$) direction in the local geometry, as shown in \gttwo.
This way we would know the full global topology but not the global
metric. Only the local metric is known so far. The metric
\newmetnow, although written in terms of global coordinates,
cannot give us the global metric because the metric has no
information about the three-branes and seven-branes. Once we get
the local patch \metnownow\ we can insert the other branes and
combine all the cubical patches (see figure above) to get the full
global picture.

Furthermore, from the figure above, we see that there is also an
interesting regime where we can have light fundamental multiplets
in the IR. This will be the case when the five-branes are near the
seven-branes. In fact the five-branes could possibly dissolve in
the seven-branes as first Chern class of gauge bundles. For such a
thing to happen the topology of the wrapped seven-branes is
important. In the local geometry the seven-branes wrap the torus
($y, \theta_2$) along with the ($r, z$) direction. The local
geometry along the ($r, z$) direction at a constant value of
($x,y$) is given by \eqn\rzmet{ds^2_{rz}~ = ~ (1+Qr)^2~dr^2~ +
~{dz^2 \o (1+Qr)^2} ~ = ~ dR^2 ~+~ {dz^2 \o 1+2QR},} where $R$ and
$r$ are related in the standard way: $R = r + {Qr^2\o 2}$. Both
$R$ and $r$ are local variables, and so the $z$ circle would
decrease as we move away from the origin. This behavior is only
local of course and in the absence of a global metric it is
difficult to predict the behavior of the ($r,z$) metric. In case,
however, the $z$ behavior continues to persist globally, then
topologically the ($r,z$) metric will become the metric of a
squashed sphere, and then the D5 brane charges $Q_5$ will be given
by \eqn\qfive{Q_5 ~ = ~ \int_{{\bf P}^1} {\rm tr}~F ~ \equiv~2\pi
c_1(F),} which is the first Chern class of the vector bundles on
the seven-branes. The trace is over the adjoint representation of
a subgroup of the full global group (which could be as big as $E_8
\times E_8$).

The possibility of the existence of a bound state in our system
may give us a possible hint of the existence of an {\it obviously}
supersymmetric configuration that has a close resemblance to our
present setup. Imagine we start with a F-theory configuration with
only seven-branes and no five-branes. We could then isolate {\it
one} seven-brane out of the full bunch and put fluxes on it {\it a
l\`a} \qfive\ to create the five-branes. This is a supersymmetric
configuration \witp\ as the strings between five-branes and the
seven-brane become tachyonic resulting in a negative energy that
effectively reduces the total energy of the system to its bound
state energy. The other strings that connect the five-branes with
the rest of the seven-branes are naturally massive (for details on
this see the second reference of \witp). This would also mean,
going back to the solution of \pandoz\ and taking a metric like
\pandoz\ with additional seven-branes, that we can {\it soak} the
five-branes on an isolated $D7$ brane to form a bound state
preserving supersymmetry. Since most of the seven-branes are far
away from the bound system, the axion-dilaton in the local
neighborhood of the $D5$ branes is negligible. Thus locally the
metric would resemble the solution of \pandoz\ but the global
picture would be different, as one might expect. In this way we
can resolve all the issues in our model which would be difficult
to study in the models presented in the literature.

So far we have been ignoring the fluxes. It is time now to take
them into account. Of course choosing fluxes that satisfy
equations of motion will not suffice. We need fluxes that also
preserve supersymmetry. A generic choice of fluxes {\it at} and
{\it away} from the orientifold point is given earlier in \gttwo.
Here, for simplicity, we shall consider only a simple choice. For
$B_{NS}$ and $B_{RR}$ we choose \eqn\bnsbrr{\eqalign{&B_{NS} ~ = ~
{\cal B}_{y\theta_1}(\theta_2), ~~~~~~ B_{RR} ~ = ~ {\tilde{\cal
B}}_{xz}(r),\cr & H_{NS} ~ = ~ {\cal H}_{y\theta_1\theta_2} ~ = ~
\del_{[\theta_2}{\cal B}_{y\theta_1]}, ~~~ H_{RR} ~ = ~
{\tilde{\cal H}}_{xzr} ~ = ~ \del_{[r}\tilde{\cal B}_{xz]},}}
where the complete anti-symmetrisation implies the situation when
we are away from the orientifold point. It is easy to see that the
supersymmetry condition on fluxes, \eqn\susyflux{{\tilde{\cal
H}}_{xzr} ~ = ~ \ast_6~{\cal H}_{y\theta_1\theta_2,}} can be
preserved with $\ast_6$ being the Hodge dual in our
six-dimensional local metric. Both the NS and the RR fluxes
survive the orientifold projection: $\Omega\cdot (-1)^{F_L}\cdot
{\cal I}_{x\theta_1}$ but could be defined away from the
orientifold point also. Finally the seven-form field strength on
the wrapped D5 branes is given by \eqn\hseven{{\cal H}_7 ~ = ~
\del_{[\theta_1}{\cal C}_{0123y\theta_2]} ~ = ~ \ast_{10}~
\tilde{\cal H}_{xzr}} and therefore gives the five-form sources
${\cal C}_{0123y\theta_2}$ on the wrapped D5 branes with
$\ast_{10}$ being the Hodge dual in the full ten-dimensional
space.

Before moving ahead we would like to make the following
observation: from the orientation of the $B_{NS}$ field \bnsbrr,
we see that one component of the $B_{NS}$ field is along the
direction of the five-branes, whereas the other component is
orthogonal to it. With the existence of the field strength ${\cal
H}_{y\theta_1\theta_2}$ we are guaranteed that the $B_{NS}$ field
cannot be gauged away, and therefore will give rise to {\it
dipole} deformations of ${\cal N} = 1$ gauge theory \difuli! This
has recently been addressed as $\beta$-deformation of gauge
theories in \nifuli. One of the immediate advantages of this is to
decouple KK modes. We will discuss this more later in the paper.

Coming back to our earlier discussion of the cylindrical
configuration in M-theory, we can now quantify the flux \gflux.
These G-fluxes are in principle {\it different} from the F-theory
G-fluxes with components \eqn\fgflux{G ~ = ~
g_1(\theta_2)~d\theta_1 \wedge d\theta_2 \wedge dy \wedge dx^a ~
+~ g_2(r) ~ dx \wedge dz \wedge dr \wedge dx^3,} where $x^a$
denotes the eleventh direction and $g_1, g_2$ are sufficiently
different from $f_1, f_2$, but their values are not yet
determined. Existence of two different G-fluxes for the same type
IIB picture implies that there could exist two different {\it
dual} configurations. Both pictures may cover varying amounts of
information for the type IIB set-up. These two dual configurations
should {\it not} be confused with the {\it gravity} dual already
present in type IIB! Existence of so many dual configurations also
means that we might be able to interpolate between them.

But there is more to that. There are {\it three} different dual
configurations that are related to the ${\cal N} = 1$ gauge
theory: M-theory with an MQCD-like brane configuration \dotu,
M-theory compactification on a $G_2$ manifold \bobby, and an
F-theory compactification on a fourfold \dotd. The configuration
with G-fluxes \fgflux\ is defined exclusively on an elliptically
fibered fourfold \gttwo, \gtone, whereas the other one with
G-fluxes \gflux\ can interpolate between the two descriptions
\dotu\ and \bobby. In the limit where the description is
completely in terms of branes {\it a l\`a} \dotu, one can show
that the weakly coupled type IIA description is when the
co-dimension four surfaces in CY, $A$ and $B$, are connected by a
D-brane or an anti D-brane (see fig. 1 above). We can go from one
picture to another by simply crossing $A$ and $B$ i.e. $l~ \to~-l$.
In general $A$ and $B$ can move along the $x^a$ circle as well in
M-theory. In the limit when the co-dimension four surfaces behave
as NS5 branes, we know that for weakly coupled type IIA, $l = 0$
is a susy preserving fixed point where $A$ and $B$ are on top of
each other \dotu.

The third dual background of M-theory on a $G_2$ structure manifold is by now well known
from the supergravity solution presented in \gtone, \realm, \gttwo\ that gives the original conjecture of \vafai\
a firmer footing.


\newsec{The heterotic background: Torsion and Non-K\"ahler manifolds}

In \realm\ we showed how we can construct possible gauge/gravity
dual models in the heterotic and type I $SO(32)$ theories. The
construction relied on identifying a possible orientifold corner
of type IIB theory and then U-dualising the picture to go to the
heterotic side. Following duality chains we were able to give a
{\it local} description of the gravity dual of wrapped heterotic
NS5 branes in \realm. In \gttwo\ we realised that there is a
possible way to construct the full global picture of the gravity
dual. In fact we were able to construct the explicit metric using
some special identifications, and found that under some
simplifying assumptions on the warp factors the global metric
looks very similar to the Maldacena-Nunez type metric \mn\ in the
IR and to the \grana\ type metric in the UV. For a more generic
choice of the warp factors the metric may not resemble any known
configuration, so would give rise to a new class of supergravity
solutions with a well defined UV and IR behavior.

In \gttwo, as discussed above, we found the UV description completely. This is of course the metric {\it after}
geometric transition (in the language of \gtone). Here we would like to address the issue of global
completion for the metric {\it before} geometric transition. But before that let us clarify a few subtle points.

\noindent $\bullet$ The heterotic background that we proposed in
\realm\ and \gttwo\ is {\it not} U-dual to the type IIB background
that we discussed above. The orientifolding effect used in type
IIB is very different from the one used to go to the U-dual
heterotic background. The existence of {\it two} different
orientifolding possibilities for the same type IIB background
reflects the differences between the existence of isometries and
the existence of invariances. What we saw in \realm\ was that the
isometry directions do not always imply metric invariances. In
fact even in the local limit the metric was not invariant under
orientifolding of two isometry directions. This is where our
background differs from the one studied earlier in \sav, \beckerD,
\bbdg, \bbdgs. In those works, invariance and isometry went hand
in hand, and therefore orientifolding was easy. Here an {\it
obvious} orientifolding {\it does not} lead us to the heterotic
background. The standard orientifolding gives us a supersymmetric
background with seven branes that we were able to use effectively
to study ${\cal N} = 1$ gauge theories. This orientifold operation
may be related to the orientifold of T-dual brane configuration
proposed in \urangapark. The orientifolding that we used to go the
the heterotic side in \realm, \gttwo\ keeps only part of the
original metric invariant.

\noindent $\bullet$ Our next venture was to identify the invariant
part of the metric that should also solve the background equations
of motion, along with the susy conditions. Analysing the
superpotential showed that the invariant part is in fact the {\it
trivial} $U(1)$ fibration over the ${\bf T}^2 \times {\bf T}^2$
base inside our local geometry \metnownow. Recall that the
original type IIB theory has the topology of a non-trivial $U(1)$
fibration over a ${\bf T}^2 \times {\bf T}^2$ base locally. The
type IIB tori ($x, \theta_1$) and ($y, \theta_2$) start off as
square tori, but then are boosted to provide the correct mirror
backgrounds. On the other hand, the two base tori in the heterotic
side are succinctly constructed as ($x,y$) and ($\theta_1,
\theta_2$) with possible non-trivial complex structures between
them (see the discussion in \realm). Minimising the type IIB
superpotential also hints that the heterotic tori could be more
general than square tori to preserve supersymmetry.

\noindent $\bullet$ The semi-toroidal geometry ${\bf T}^2 \times
{\bf T}^2 \times {\bf S}^1 \times {\bf \IR}^+$ with square tori
unsurprisingly turns out not to be the most generic solution. This
is of course consistent with earlier studies on flux
compactifications \sav, \beckerD, \kachruone. A particular choice
of fluxes may lead to tori with non-trivial complex structures. In
fact there is already a complex structure inherited from the
parent metric \metnownow \eqn\cmps{\tau_1 ~ = ~ {1\o 2}\Bigg({\rm
sin}~2\langle\theta_1\rangle ~{\rm cot}~\langle\theta_2\rangle +
i~{2~{\rm sin}^2\langle\theta_1\rangle \o
\sqrt{\langle\alpha\rangle}}\Bigg), ~~~~\tau_2 ~\approx~ i,} which
should be allowed for specific choices of the background $B_{NS}$
fields ($b_{x\theta_1}, b_{y\theta_2}$) in the notation of \realm.
The constant $\langle\alpha\rangle$ is defined in \gtone. On the
other hand, after geometric transition the parent metric allows
the following complex structure to be inherited by the two tori:
\eqn\cmpsnow{\tau_1 ~\approx~ i, ~~~~~\tau_2 ~=~ {a}_1~{\rm
cot}~\langle\theta_1\rangle~ {\rm cot}~\langle\theta_2\rangle +
i\sqrt{{a}_2 - {a}_3~{\rm cot}^2~\langle\theta_1\rangle~{\rm
cot}^2~\langle\theta_2\rangle},} where the $a_i$ are some
constants that could be determined from the duality cycle of
\gtone\ and \realm. Thus we see that the semi-toroidal geometry
has some inherent complex structures already for the two ${\bf
T}^2$.

\noindent $\bullet$ The inherited complex structure before geometric transition dualises in the heterotic theory
to a more complicated fibration of the first tori. In fact this non-trivial fibration is the key reason why the
heterotic metric fails to retain K\"ahlerity \realm. The local analysis of the system was presented in \realm.
So here we will first explore a convenient representation of the local metric that explicitly shows the wrapped
brane configuration (or its possible generalisation) and later see how far we can go to elucidate the full global
picture in a somewhat similar
vein as we explored the global metric after geometric transition in \gttwo.

\noindent Our starting point is the U-dual metric of \realm\ that
is an explicit solution of the heterotic equations of motion. The
proposed local metric for our space is worked out in \realm\ and
will be given as \eqn\metinHH{\eqalign{ds^2 ~ = ~& d_1~ (dy -
b_{yj}~d\zeta^j)^2 + d_2~ (dx - b_{xi}~d\zeta^i)^2  + d_3~ dr^2~ +
\cr ~&~~~ - 2 d_4~(dx - b_{xi}~d\zeta^i)(dy - b_{yi}~d\zeta^j) +
d_5~{dz^2} + d_6 \vert d\chi_2\vert^2,}} \noindent which is
similar to the type I metric that we had in \realm, as it should
be. The various coefficients appearing in \metinHH\ are defined as
follows: \eqn\metdefcoco{\eqalign{&
d_1~=~\sqrt{\langle\alpha\rangle}\big(1 + {\rm
cot}^2~\langle\theta_1\rangle\big),\cr
&d_2~=~\sqrt{\langle\alpha\rangle}\big(1 + {\rm
cot}^2~\langle\theta_2\rangle\big),\cr & d_3 ~ = ~ {\gamma'(r_0)
\sqrt{H(r_0)} \o \sqrt{\langle\alpha\rangle}}, ~~~~~~ d_5 ~ = ~
{1\o \sqrt{\langle\alpha\rangle}}, \cr & d_4 ~ = ~
\sqrt{\langle\alpha\rangle}~{\rm cot}~\langle\theta_1\rangle~{\rm
cot}~\langle\theta_2\rangle.}} The metric has the usual fibration
along the $dx$ and the $dy$ directions and the base is given by
the ($\theta_1, \theta_2, r, z$) coordinates. The background $B$
field (whose components may depend on $r, \theta_1$ and $\theta_2$
only) can be written in terms of the U-dual IIB B-fields with legs
along ($x, \theta_1$) and ($y, \theta_2$) directions. These
B-fields $-$ that will serve as the torsion $-$ and the coupling
constant are given by \eqn\bandcoup{\eqalign{& H^{\rm het}~ \equiv
~ H~=~ {\cal H}^b_{xz\theta_1}~ dy ~\wedge ~dz ~\wedge ~d\theta_2
- {\cal H}^b_{yz\theta_2}~ dx ~\wedge ~dz ~\wedge ~d\theta_1 ~ +
\cr & ~~~~~~~~~~~~~ + ~{\cal H}^b_{xzr}~ dy ~\wedge ~ dz ~\wedge~
dr~ -~ {\cal H}^b_{yzr}~dx ~\wedge ~dz ~\wedge ~dr; \cr & ~~~~
g^{\rm het} ~ =~ {1\o \sqrt{\langle\alpha\rangle}}.}} We make the
following observations:

\noindent $\bullet$ Along with the solution
to the DUY equation, \metinHH\ and \bandcoup\ will specify the complete background.

\noindent $\bullet$ The fibrations in \metinHH\ are due to $b_{x\theta_1}, b_{y\theta_2}$ which do
survive the orientifold action for this case\foot{Recall that the orbifold actions that led to \metinHH\ and
\bandcoup\ are along the ($x, y$) directions. The orbifold actions in sec. 2.1 are along ($x, \theta_1$) that will
not allow these components of B-fields.}.

\noindent $\bullet$ The B-fields that we took in \gttwo\ to study the global geometry after geometric transition
in the heterotic theory are along ($b_{x\theta_2}, b_{y\theta_1}$).

\noindent $\bullet$ As expected the local metric has only constant coefficients. This is consistent with the
fact that the dilaton \bandcoup\ is a constant (see \rstrom, \hull, \sav, \beckerD\ for details).

{}From all the above discussion we see that we can keep all the components of the B-fields: ($b_{x\theta_1},
b_{x\theta_2}, b_{y\theta_1}, b_{y\theta_2}$) and study the deformation of the metric. It is a straightforward
exercise to work out the deformation $s_i$, and the final metric with deformations is given by:
\eqn\korothdef{\eqalign{G & = \pmatrix{G_{xx} & G_{xy} &
G_{xz} & G_{x\theta_1} & G_{x\theta_2} \cr \noalign{\vskip -0.20
cm}  \cr G_{xy} & G_{yy} & G_{yz} & G_{y\theta_1} & G_{y\theta_2}
\cr \noalign{\vskip -0.20 cm}  \cr G_{xz} & G_{yz} & G_{zz} &
G_{z\theta_1} & G_{z\theta_2} \cr \noalign{\vskip -0.20 cm}  \cr
G_{x\theta_1} & G_{y\theta_1} & G_{z\theta_1} &
G_{\theta_1\theta_1}& G_{\theta_1\theta_2} \cr \noalign{\vskip
-0.20 cm}  \cr G_{x\theta_2} & G_{y\theta_2} & G_{z\theta_2} &
G_{\theta_1\theta_2} & G_{\theta_2\theta_2}} \cr \noalign{\vskip
-0.25 cm}  \cr & = \pmatrix{d_2 & -{d_4}
& 0 & s_1 - d_2~b_{x\theta_1} & s_2 + {d_4}~b_{y\theta_2}
\cr \noalign{\vskip -0.20 cm}  \cr
 -{d_4} & d_1 & 0 & s_3 + {d_4}~b_{x\theta_1}
 & s_4 - d_1~b_{y\theta_2} \cr
\noalign{\vskip -0.20 cm}  \cr 0 & 0  & d_5 & 0
& 0 \cr \noalign{\vskip -0.20 cm}  \cr s_1 - d_2~b_{x\theta_1}
&  s_3 + {d_4}~b_{x\theta_1} & 0 & d_6 + {\cal A}_1
 & -{d_4}~b_{x\theta_1} b_{y\theta_2} - {d^{-1}_4} s_i s_j \cr
\noalign{\vskip -0.20 cm}  \cr
s_2 + {d_4}~b_{y\theta_2} & s_4 - d_1~b_{y\theta_2} & 0
 & -{d_4}~b_{x\theta_1} b_{y\theta_2} - {d^{-1}_4} s_i s_j   &
d_6 + {\cal A}_2}}} where we have already defined the $d_i$ in
\metdefcoco, and the various deformations in the metric are now
given as \eqn\fotki{\eqalign{& s_1~=~ {d_4}~b_{y\theta_1}, ~~~~~
s_2~=~ -{d_2}~b_{x\theta_2}, ~~~~~ s_3~=~ -{d_1}~b_{y\theta_1},
~~~~~ s_4~=~ {d_4}~b_{x\theta_2},\cr & {\cal A}_1~=~
{d_1}~b^2_{y\theta_1} + {d_2}~b^2_{x\theta_1} - 2
d_4~b_{y\theta_1}~b_{x\theta_1}, ~~ {\cal A}_2~=~
{d_1}~b^2_{y\theta_1} + {d_2}~b^2_{x\theta_2} - 2
d_4~b_{y\theta_2}~b_{x\theta_2},}} along with the following
definition: $s_i s_j~=~ s_1 s_4 - s_2 s_4 - s_1 s_3$. The above
construction will be the most generic metric that we can study
with constant background dilaton. It is also easy to see that if
the $\chi_2$ torus had a non-trivial complex structure i.e ${\rm
Re}~\tau_2 \ne 0$, then the only changes we would need to
incorporate are \eqn\changes{\eqalign{&G_{\theta_1 \theta_2}~\to~
G_{\theta_1 \theta_2} + d_6~{\rm Re}~\tau_2; \cr & G_{\theta_2
\theta_2}~ \to ~ G_{\theta_2 \theta_2} - d_6 \big(1 -
\vert\tau_2\vert^2 \big).}} It is also clear that the B-field in
\bandcoup\ will now have to change to incorporate other
components. This is easy to work out, so we shall not do it here.
Instead we want to concentrate on various interesting cases that
can be studied from the above metric \korothdef\ by going to
different allowed limits.

\subsec{Heterotic NS5-branes wrapped on a two-cycle of a torsional
manifold}

This is the situation where we switch on all the components of
type IIB $B_{NS}$-fields. In terms of our orientifold construction
this would be the case when we are away from the orientifold
point. The $d_i$ coefficients of the metric \korothdef\ are
constants at least in the local limit and therefore could be fixed
at some values by coordinate redefinition and scalings of $B$
fields. In fact what we require is to have \eqn\didef{d_1 ~ = ~
d_4 \Bigg({b_{x\theta_1} \o  b_{y\theta_1}}\Bigg), ~~~~~~~ d_2 ~ =
~ d_4 \Bigg({b_{y\theta_1} \o  b_{x\theta_1}}\Bigg),} leaving us a
metric that is independent of coordinate reversal $\zeta_i ~ \to
~-\zeta_i$ where $\zeta_i$ are the local coordinates, as well as
invariant under type IIB {\it orbifold} action ($x, y$) ~$\to$~
($-x, -y$). However this parity transformation makes sense only if
the matrix \eqn\kommet{\Theta ~ = ~
\pmatrix{b_{x\theta_1}&b_{x\theta_2}\cr
b_{y\theta_1}&b_{y\theta_2}}} has a vanishing determinant  ${\rm
det}~\Theta ~ = ~ 0$.

Our aim in allowing a parity invariant metric is to convert our
background heterotic metric \korothdef\ to the following form:
\eqn\korothkr{ds^2~=~ \vert dz_1\vert^2 + \vert dz_2\vert^2 +
d\tilde z^2 + dr^2,} where $dz_1 = d\tilde x + \tau_3 ~d\tilde y$
and $dz_2 = d\tilde\theta_1 + \tau_4 ~d\tilde\theta_2$. The
tilde-coordinates are defined as \eqn\cordee{d\tilde x ~ = ~
\sqrt{d_2}~dx, ~~~~d\tilde y ~ = \sqrt{d_2}~dy,
~~~~d\tilde\theta_i ~ = ~ \sqrt{d_6}~d\theta_i,} where $d_i$ are
defined in \metdefcoco. The complex structures of the two tori are
now \eqn\fotcom{\tau_3 ~ = ~{1\o 2\sqrt{d_2}} \Big(-d_4 +
i\sqrt{4d_1 d_2 - d_4^2}\Big),~~~~~~ \tau_4 ~ = ~ \tau_2} with all
the coefficients defined earlier in \fotki. Observe that for the
$dz_2$ torus, we get back the original complex structure.

The metric \korothkr\ is not quite the metric that we might expect
for the wrapped five branes. This is because the coefficients of
the metric are strictly constant. To see what could possibly
change when we make the coefficients non-constant, we have to
follow a series of steps that would take us to type IIB and back
{\it not} as an orientifold operation but through sigma-model
identification. This should be reminiscent of what we did for the
heterotic case in \gttwo\ (see sec. 3.1 and 3.2 therein). First,
however, let us change our definition of $dz$. We want to define
\eqn\zdef{d\tilde z ~ = ~ dz + a~{\rm
cot}~\langle\theta_1\rangle~d\tilde x  + b~{\rm
cot}~\langle\theta_2\rangle~d\tilde y,} which, being a total
derivative, doesn't change anything in the original metric
\korothkr. The metric \korothkr\ can be rewritten as
\eqn\kurkut{ds^2~=~ dr^2 + \big(dz + a~{\rm
cot}~\langle\theta_1\rangle~d\tilde x
 + b~{\rm cot}~\langle\theta_2\rangle~d\tilde y\big)^2 + \vert dz_1\vert^2 +
\vert dz_2\vert^2,}
which looks very close to the local metric \metformj\ except (a) we now have non-trivial complex structures
on the two tori, and (b) our tori are ($x, y$) and ($\theta_1, \theta_2$) whereas in \metformj\ the
tori are ($x, \theta_1$) and ($y, \theta_2$). Of course both in \kurkut\ and \metformj\ the definitions
of base tori are not crucial, so a formal identification of the metrics will help us to rewrite \kurkut\ in such a
way as to reflect the wrapped brane metric (at least locally).

In sec. 2.1 we saw how we could go from \metformj\ to \metnownow\
by solving the equations of motion with a warped metric ansatz.
Now the question is, can we follow the same argument for \kurkut\
also? This is where sigma-model identification of heterotic and
type IIB helps. More specifically, the steps that we would like to
follow are:

\noindent $\bullet$ Define the heterotic sigma model on this background with metric \korothkr\
along with torsion and gauge bundle.

\noindent $\bullet$ Use the sigma model identification by redefining vector bundles with torsional connection
{\it only} in the left-moving sector. The right moving sector remains unchanged.

\noindent $\bullet$ Absence of anomalies and Chern-Simons corrections tells us that the resulting sigma model
should be viewed as the sigma model in type IIB theory on the same metric \korothkr.

\noindent $\bullet$ The heterotic vector bundle will now have one-to-one correspondence with {\it torsional}
curvature in type IIB theory.

\noindent $\bullet$ The type IIB metric \korothkr\ can then be manipulated, as before, to get the final metric of
the form \metnownow.

\noindent $\bullet$ The metric \metnownow\ could then be transferred back to the heterotic side by performing
the reverse transformations from type IIB to the heterotic  side.

The above set of steps generically give us a non-K\"ahler manifold
in type IIB with torsion, instead of a conformally K\"ahler. The
background preserves minimal supersymmetry. That this is not a
contradiction can be easily shown: because of the existence of an
underlying type IIB U-duality symmetry a supersymmetric
non-conformally K\"ahler manifold can exist in the presence of
torsion and zero RR three-forms (see \gttwo\ for details).

To implement the above set of steps, we need the sigma model of heterotic theory on the background \korothkr.
We will follow the notations of \hull, where the left and right moving fermions are called $S^\rho$ and
$\Psi^A$ respectively. The worldsheet interactions of these fermions with the background gauge fields are given
by the following terms of the lagrangian:
\eqn\kotmot{S_{\rm interaction}~=~ {i\o 8\pi \alpha'}
\int \Big(S^\rho \omega^{ab} \sigma_{ab}^{\rho\sigma} S^\sigma -i F_{ij(AB)}
\sigma^{ij}_{\rho\sigma} S^\rho S^\sigma \Psi^A \Psi^B + \Psi^A A_i^{(AB)} \bar\del X^i \Psi^B \Big),}
where by definition $F_{ij(AB)} = F^a_{ij} M^a_{AB}$ and the index $a$ labels the adjoint of the gauge group;
and along with \kotmot\ there is the standard kinetic term
\eqn\kinu{S_{\rm kinetic} ~ = ~ {1\o 8\pi \alpha'} \int \Big(\del X \bar\del X \cdot ({\bf g + B}) +
i S \cdot \del S + i \Psi \cdot \bar\del \Psi\Big),}
where indices are contracted accordingly. The total action is of course the sum of the two actions \kinu\ and
\kotmot. The supersymmetry of the sigma model at this stage is just ($0,1$) and {\it not} ($0,2$) as one might
have naively expected.
If we now employ the following identification:
\eqn\lcdas{A_i^{AB} ~ = ~ \pmatrix{\omega_{i}^{ab}&{\bf 0} \cr \noalign{\vskip -0.20 cm} \cr
{\bf 0}&{\bf 0}}, ~~~~~~~~
\Psi^A ~ = ~ \pmatrix{S^{\dot q} \cr \Psi^{9} \cr ...\cr ...\cr \Psi^{32}},}
then it is easy to show that the interaction term \kotmot\ changes to
\eqn\motu{{\tilde S}_{\rm interaction} ~ = ~ {i\o 8\pi \alpha'}
\int \Bigg(S^\rho \omega^{ab} \sigma_{ab}^{\rho\sigma} S^\sigma +
S^{\dot \rho} \omega^{ab} \sigma_{ab}^{\dot \rho \dot\sigma} S^{\dot\sigma} - {i\o 2} {\cal R}_{ijkl}
\sigma^{ij}_{\dot\rho \dot\sigma}\sigma^{kl}_{\kappa \gamma} S^{\dot\rho} S^{\dot\sigma} S^{\kappa} S^{\gamma} \Bigg).}
On the other hand the kinetic term \kinu\ remains more or less unchanged. The only change therein is
\eqn\ktmt{\Psi \cdot \bar\del \Psi ~ \to ~ S^{\dot\sigma} ~\bar\del S^{\dot\sigma} + \sum_{A = 9}^{32}~
\Psi^A \bar\del \Psi^A,}
where $\Psi^9, .... \Psi^{32}$ are completely decoupled because of our choice of $A_i$ in \lcdas. Now defining
$\tilde S \equiv {\tilde S}_{\rm kinetic} + {\tilde S}_{\rm interactions}$ and
$S_{\rm het} \equiv {S}_{\rm kinetic} + {S}_{\rm interactions}$ we see that
\eqn\ktmtsen{S_{\rm het} ~ \to ~ {\tilde S} ~ = ~ S_{\rm IIB} ~ + ~ {i\o 8\pi \alpha'} \sum_{A = 9}^{32}~
\Psi^A \bar\del \Psi^A,}
where $S_{\rm IIB}$ is the type IIB worldsheet lagrangian. Thus up to the decoupled $\Psi^A$ fermions, the
heterotic background maps to the type IIB background.

This is a very interesting situation now. The heterotic metric
\kurkut\ is now the metric in type IIB also. Thus all the
manipulations that we performed in type IIB (in sec. 2.1) should
apply to this metric also. In particular the final metric that we
will get in type IIB will {\it not} be K\"ahler. In fact there is
no reason for the metric to be even conformally K\"ahler.
Furthermore we expect the metric of the two-cycle on which we have
wrapped NS5 branes to be of the form \eqn\metnsfive{ds_{\rm
2-cycle}^2 ~ = ~ \Big(\vert\tau_3\vert^2 + a_o \Big)~d\tilde y^2 ~
+ ~ \Big(\vert\tau_4\vert^2 + b_o \Big)~d\tilde \theta_2^2,} where
$\tau_{3, 4}$ have been defined earlier in \fotcom\ and $a_o, b_o$
are the possible corrections (to be determined below).

In the limit $\tau_{3, 4} = i$ the metric \korothkr\ (or its equivalent \kurkut) can be made K\"ahler. We are now
in the realm of our earlier calculations of sec 2.1. We then expect that the background equations of motion will
yield the following metric in type IIB:
\eqn\formet{\eqalign{ds^2 ~ = ~ &{\cal D}(r)^2~dr^2 + {\cal D}(r)^{-2}~\Big(dz + Q_o~{\rm cot}~\theta_1~ dx
+ Q_o~{\rm cot}~\theta_2~dy\Big)^2~ + \cr
& ~~~~~~~~~~~~~~~~~~~~~ + {\cal D}(r)~(dy^2 + dx^2 + d\theta_2^2 + d\theta_1^2),}}
where ${\cal D}(r) ~ = ~ 1 + Q_o~ r$ is a linear function of $r$ and we have written in terms of un-tilded coordinates
to avoid clutter.

In the absence of an RR background, a K\"ahler (or a conformally K\"ahler) geometry generically breaks supersymmetry.
Therefore only a non-K\"ahler deformation of the above background along with $H_{NS}$ fluxes will preserve supersymmetry.
We have already discussed the possibility of the existence of such a background from F-theory (in sec. 2.1), and here
our ansatz for such a metric will be to deform the K\"ahler background \formet\ by switching on a non-trivial
complex structure $\tau_{3,4}$ such that the final type IIB metric is
\eqn\gulu{\eqalign{ds^2 ~ = ~ &{\cal D}(r)^2~dr^2 + {\cal D}(r)^{-2}~\Big(dz + Q_o~{\rm cot}~\theta_1~ dx
+ Q_o~{\rm cot}~\theta_2~dy\Big)^2 ~+ \cr
& ~~~~~~~~~~~~~~~~~~~~~~~~~~ + ~{\cal D}_1(r)~\vert dx + \tau_3~dy\vert^2 +
{\cal D}_2(r)~\vert d\theta_1 + \tau_4~d\theta_2 \vert^2,}}
where the functional form for ${\cal D}_i(r)$ could in general be non-linear, and the metric of the two cycle
on which we have wrapped NS5 will have the following form for $r \to 0$ (i.e in our local patch):
\eqn\fibrme{ds_{\rm 2-cycle}^2 ~ = ~ \Big(Q_o^2~{\rm cot}^2~\theta_2  + \vert\tau_3\vert^2 \Big)~dy^2 ~ +
~ \vert\tau_4\vert^2 ~d\theta_2^2.}
Therefore, once we know the local type IIB geometry, we can again use our sigma-model identification to go back to the
heterotic side! Then the final local metric in the heterotic theory for the wrapped NS5 branes is given by
the non-K\"ahler metric \gulu\ with a torsion. After geometric transition we can infer the {\it full} global metric,
which is derived in \gttwo.

Another possibility is when we allow some of components of the
$B_{NS}$ fields to vanish. For example we could consider
\eqn\bvane{b_{x\theta_2} \ne 0,~~~b_{y\theta_1} \ne 0, ~~~
b_{x\theta_1} ~ = ~b_{y\theta_2} = 0.} In this case the previous
simplifying relations between the $d_i$ cannot be imposed,
resulting in extra cross-    terms in the metric. We can simplify
the ensuing analysis a little by first observing that the B-fields
are defined on compact spaces, and therefore are periodic. We can
then write them in terms of angular coordinates $\theta$ and
$\tilde\theta$ in the following way: \eqn\ththepr{\theta ~ = ~
{\rm arctan}~\Bigg[{b_{y\theta_1}\sqrt{d_1}\o \sqrt{d_6}}\Bigg],
~~~ \tilde\theta ~ = ~ {\rm
arctan}~\Bigg[{b_{x\theta_2}\sqrt{d_2}\o \vert\tau_2\vert
\sqrt{d_6}}\Bigg],} where $\tau_2$ represents the non-trivial
complex structure discussed earlier in \cmps. In fact before
geometric transition, the inherited complex structure is $\tau_2 =
i$ \cmps. Here we would like to switch on a non-zero real part of
$\tau_2$ so as to simplify some of the following analysis. Of
course to switch on such a complex structure we have to change the
torsion a bit. Assuming that it is indeed possible to do this, we
find that the following choice of $\tau_2$ simplifies the metric
quite a bit: \eqn\simtau{{{\rm Re}~\tau_2 \o \vert\tau_2\vert}~ =
~ {1\o 2} {d_4\o \sqrt{d_1d_2}}~ {\rm tan}~\theta~{\rm
tan}~\tilde\theta.} The geometry of the two two-tori is very
interesting now. In our previous case discussed above we see that
the ($\theta_1, \theta_2$) and the ($x, y$) tori are decoupled
\gulu. Now the ($\theta_1, \theta_2$) torus is non-trivially
fibered over the other ($x, y$) torus\foot{Readers should observe
that this is the simplest solution for this case. For our previous
example, there could also be non-trivial fibration of the
($\theta_1, \theta_2$) torus over ($x, y$) base, but the metric
presented in \gulu\ is the simplest case.}. One can work out the
fibration precisely, and it turns out to have the following form:
\eqn\fimen{ds^2_{\rm fib} ~ = ~ d_6~{\rm
sec}^2~\theta\Big[d\theta_1 + {\rm sin}~2\theta (a~dx -
b~dy)\Big]^2 + d_6~\vert\tau_2\vert^2~{\rm
sec}^2~\tilde\theta\Big[d\theta_2 - {\rm sin}~2\tilde\theta
({\tilde a}~dx - {\tilde b}~dy)\Big]^2,} which in fact would be
more complicated if ${\rm Re}~\tau_2 = 0$. For ${\rm Re}~\tau_2$
satisfying \simtau, there are no $d\theta_1 d\theta_2$ cross
terms. The coefficients appearing in the metric are defined in
terms of $d_i$ as (we have taken $d_4$ here as one-half of the
original choice): \eqn\abab{\eqalign{&a ~=~ {d_4 \o 4\sqrt{d_1
d_6}}; ~~~~~~~~~~~~~ b ~=~ {\sqrt{d_1} \o 2\sqrt{d_6}}; \cr &
{\tilde b} ~=~ {d_4 \o 4 ~\vert\tau_2\vert \sqrt{d_2 d_6}};
~~~~~~~ {\tilde a} ~=~ {\sqrt{d_2} \o 2
~\vert\tau_2\vert\sqrt{d_6}}.}} The base torus ($x, y$) is now no
longer as simple as \fotcom. In fact the original complex
structures of \fotcom\ don't give the full metric. The ($x, y$)
torus metric is \eqn\xytorus{ds^2_{\rm xy} ~ = ~ \vert d\tilde x +
\tau_3 ~d\tilde y\vert^2 - \sigma_o ~\vert d\tilde x +
\tau_5~d\tilde y\vert^2,} where $a, {\tilde a}$ are defined in
\abab\ and the tilde-coordinates ($\tilde x, \tilde y$) are scaled
by $d_2$ from the original coordinates ($x, y$) given in \cordee.
The shift $\sigma_o$ in \xytorus\ is \eqn\sigmao{\sigma_o~=~ 4d_6
d_2^{-1}\big[a^2~{\rm sin}^2~\theta + \vert\tau_2\vert^2~{\tilde
a}^2~{\rm sin}^2~ \tilde\theta\big],} with $\tau_5 \equiv {\rm
Re}~\tau_5 + i~{\rm Im}~\tau_5$ defined in terms of the above
variables as \eqn\taufive{\eqalign{&{\rm Re}~\tau_5 ~ = ~
-{ab~{\rm sin}^2~\theta + \vert\tau_2\vert^2~{\tilde a}{\tilde
b}~{\rm sin}^2~\tilde\theta\o a^2~{\rm sin}^2~\theta +
\vert\tau_2\vert^2~{\tilde a}^2~{\rm sin}^2~\tilde\theta}; \cr &
{\rm Im}~\tau_5 ~ = ~ {(a{\tilde b} - {\tilde a}b){\rm
sin}~2\theta ~{\rm sin}~ 2\tilde\theta\o  d_6\big[a^2~{\rm
sin}^2~\theta + \vert\tau_2\vert^2~{\tilde a}^2~{\rm
sin}^2~\tilde\theta\big]},}} which together with \fimen\ captures
the full ${\bf T}^2 \otimes {\bf T}^2$ structure of the base where
$\otimes$ denotes the non-trivial fibration of ($\theta_1,
\theta_2$) torus on the ($x, y$) torus. The metric for our case
now is not \korothkr\ but a more complicated one given as
\eqn\korothnow{ds^2 ~ = ~ ds^2_{\rm fib} ~ + ~ ds^2_{\rm xy} ~ +
d\tilde z^2 + dr^2,} where $ds^2_{\rm fib}$ and $ds^2_{\rm xy}$
are given above in \fimen\ and \xytorus\ respectively. The
solution \korothnow, however, is still {\it not} the full metric
with wrapped $NS5$ branes and torsion. This is of course similar
to the case encountered earlier where \korothkr\ was not the full
metric whereas \gulu\ was. Here too our ansatz will be that the
final heterotic metric can be written as
\eqn\gulukola{\eqalign{ds^2 ~ = ~ &{\cal H}(r)^2~dr^2 + {\cal
H}(r)^{-2}~\Big(dz + Q_o~{\rm cot}~\theta_1~ dx + Q_o~{\rm
cot}~\theta_2~dy\Big)^2 ~+ \cr & ~~~~~~~~~~~~ + ~{\cal
H}_1(r)~\Big(\vert dx ~+~ \tau_3~dy\vert^2 -\sigma_o ~ \vert dx +
\tau_5~dy\vert^2\Big) ~+ ~{\cal H}_2(r)~ds^2_{\rm fib},}} where as
before ${\cal H}(r)$ is a linear function in $r$, and ${\cal
H}_i(r)$ could generically be non-linear. Similarly the metric of
the two-cycle on which we have wrapped NS5 branes will be more
complicated than the one derived before in \fibrme, and is given
by: \eqn\metto{\eqalign{ds^2_{\rm 2-cycle} ~ = ~& \Big[{\cal
H}^{-2}~ Q_o^2~{\rm cot}^2~\theta_2 + {\cal H}_1
~\vert\tau_3\vert^2 + ({\cal H}_2 - {\cal H}_1)({\tilde d}_1~{\rm
sin}^2~\theta + {\tilde d}_2~{\rm sin}^2~\tilde\theta)\Big]~dy^2 ~
+ \cr & ~~~~~~~~~~~~~ + {\cal H}_2 ~d_6~\vert\tau_2\vert^2~{\rm
sec}^2~\tilde\theta~d\theta_2^2,}} where ${\tilde d}_1\equiv
d_1^2, {\tilde d}_2 \equiv {d_4^2 \o 4 d_2}$. For our local patch
where $r \to 0$ we expect \metto\ to reduce to
\eqn\mette{ds^2_{\rm 2-cycle} ~ = {}^{\rm lim}_{\epsilon \to 0}~
 \Big[Q_o^2~{\rm cot}^2~\theta_2  + \vert\tau_3\vert^2 +
\epsilon (d_1'~{\rm sin}^2~\theta + d_2'~{\rm
sin}^2~\tilde\theta)\Big]~dy^2 + ~d_6~\vert\tau_2\vert^2~{\rm
sec}^2~\tilde\theta~d\theta_2^2,} where we have assumed that
$\epsilon = {\cal H}_2 - {\cal H}_1$ is a very small quantity.
Thus \mette\ is almost similar to the metric of the two-cycle
discussed earlier for \fibrme, which means that the two-cycle
doesn't change too much locally even if we consider different
choices of the $B$ fields. Once we know the two-cycle, a geometric
transition will take us to the dual gravity theory whose {\it
global} metric was derived earlier in \gttwo.

The local heterotic geometry before geometric transition therefore has the following topology: the four-dimensional
base ${\cal B}$
is a non-trivial $T^2$ fibration over another $T^2$. The first $T^2$ is parametrised by the coordinates
($\theta_1, \theta_2$) and the other $T^2$ is parametrised by ($x, y$) with a non-trivial complex structure.
In addition to that there is an overall $U(1)$ fibration of $dz$ over the base ${\cal B}$. Of course both the
tori and the $U(1)$ directions are warped differently along the radial direction $r$ and the local topology
is of the form
\eqn\loctop{ \big({\bf T}^2~ \otimes ~{\bf T}^2 \big)~\otimes ~ {\bf S}^1 ~\times ~ {\bf \IR}^+}
where $\otimes$ denotes a non-trivial fibration, and the $NS5$ branes wrap two cycles in the manifold
${\bf T}^2~ \otimes ~{\bf T}^2$. Geometrically the non-trivial fibration makes the metric non-K\"ahler, and
the torsion is caused by the sources of the $NS5$ branes. We will discuss more details of this manifold 
and a family of them in sec. 3.3.

\noindent Alternatively one can see that there is a non-trivial
${\bf T}^3$ torus over the base \xytorus. The metric of the ${\bf
T}^3$ is \eqn\tthreet{\eqalign{ds^2_{T^3}~ = &~ {\cal
H}(r)^{-2}\big(dz + \alpha_1\cdot dx + \sigma_1\cdot dy\big)^2 ~ +
\cr & ~~~~~~~ + {\cal H}_3(r)~ \big(d\theta_1 + \alpha_2\cdot dx +
\sigma_2\cdot dy\big)^2 + {\cal H}_4(r)~ \big(d\theta_2 +
\alpha_3\cdot dx + \sigma_3\cdot dy\big)^2,}} where ${\cal H}(r)$
is a linear function of $r$, and the other two warp factors are
defined as \eqn\warhde{{\cal H}_3 ~ = ~ d_6~{\cal H}_2~{\rm
sec}^2~\theta, ~~~~~~ {\cal H}_4 ~ = ~
d_6~\vert\tau_2\vert^2~{\cal H}_2~{\rm sec}^2~\tilde\theta,} while
the warp factor ${\cal H}_2$ has already appeared in \gulukola.
The other coefficients $\alpha_i, \sigma_i$ can be easily
extracted from \gulukola\ and \fimen. They are given by:
\eqn\toula{\pmatrix{\alpha_1&\alpha_2&\alpha_3 \cr \noalign{\vskip
-0.20 cm} \cr \sigma_1&\sigma_2&\sigma_3} ~ = ~ \pmatrix{{\rm
cot}~\theta_1& ~a~{\rm sin}~2\theta & - {\tilde a}~{\rm
sin}~2\tilde\theta \cr \noalign{\vskip -0.20 cm} \cr {\rm
cot}~\theta_2& -b~{\rm sin}~2\theta & ~{\tilde b}~{\rm
sin}~2\tilde\theta},} where ($a, b, {\tilde a}, {\tilde b}$) are
defined in \abab. {}From the metric \tthreet\ we see that the
${\bf T}^3$ fibration is only approximate. The $dz$ fibration also
depends on the other ($\theta_1, \theta_2$) torus, and therefore a
global geometry is more likely to be an extension of \loctop\
instead of \tthreet. One should also note that this ${\bf T}^3$
fibration is {\it not} the ${\bf T}^3$ fibration on which we can
perform mirror transformation.

\subsec{Analysis of the torsion classes}

We would now like to classify the heterotic non--K\"ahler metric \gulukola. Such
non-K\"ahler backgrounds are conveniently classified in terms of their intrinsic torsion
or their so--called torsion classes. These torsion classes correspond to the decomposition
of the intrinsic torsion into $SU(3)$ representations, because four--dimensional supersymmetry
 requires the internal manifold to have an $SU(3)$ structure \gauntlett, \lust. See \gstructure\ for
  a rather mathematical discussion of manifolds with G--structure.
These manifolds are characterized by a globally defined $SU(3)$
invariant spinor that is constant w.r.t. a torsional connection.
This reduces the structure group of the six-dimensional manifold
from $SO(6)$ to $SU(3)$ and the intrinsic torsion decomposes under
$SU(3)$ into five classes, see e.g. \lust, \salamon:
 ${\cal W}_1\oplus{\cal W}_2\oplus{\cal W}_3\oplus{\cal W}_4\oplus{\cal W}_5$.

The failure of the torsional connection to be the Levi--Civita connection is measured in the
failure of fundamental 2--form and holomorphic 3--form to be closed. Defining a set of real
vielbeins $\{e_i\}$ one can define an almost complex structure via a set of complex vielbeins $\{E_i\}$ as
\eqn\almcompl{E_1=e_1+i~e_2~,~~~~ E_2=e_3+i~e_4~,~~~~ E_3=e_5+i~e_6, }
which give rise to a (1,1)--form w.r.t. this almost complex structure
\eqn\defJ{J=e_1\wedge e_2+e_3\wedge e_4+e_5\wedge e_6\,.}
Similarly, one defines a holomorphic 3--form
\eqn\defOmega{\Omega=\Omega_+ + i~\Omega_-
  =(e_1+i~e_2)\wedge(e_3+i~e_4)\wedge(e_5+i~e_6)\,,}
where $\Omega_+$ and $\Omega_-$ are the real and imaginary part of $\Omega$, respectively.
$J$ and $\Omega$ fulfill the compatibility relations
\eqn\compatible{\eqalign{& J\wedge \Omega_+=J\wedge \Omega_-=0;\cr
  & \Omega_+ \wedge \Omega_-={2\o 3}J\wedge J\wedge J\,.}}
The torsion classes ${\cal W}_i$ are then determined
by the following equations
\eqn\deftorsform{\eqalign{& d\Omega_\pm \wedge J=\Omega_\pm \wedge dJ = {\cal W}_1^\pm
    ~ J\wedge J\wedge J,\cr
  & d\Omega_\pm^{(2,2)} = {\cal W}_1^\pm~J\wedge J+{\cal W}_2^\pm\wedge J, \cr
  & dJ^{(2,1)}~=~(J\wedge{\cal W}_4)^{(2,1)}+{\cal W}_3\,,}}
so ${\cal W}_1$ is given by two real numbers, ${\cal W}_1^+$ and ${\cal W}_1^-$.
${\cal W}_2$ is a (1,1) form and ${\cal W}_3$ is a (2,1) form.
With the definition of the contraction
\eqn\contraction{\rightharpoonup~:~\bigwedge{}^k ~T^*\otimes \bigwedge{}^n ~T^*
  \longrightarrow \bigwedge{}^{n-k} ~T^*,}
and the convention $e_1\wedge e_2~\rightharpoonup~e_1\wedge e_2\wedge e_3\wedge e_4
  ~=~e_3\wedge e_4$ one defines
\eqn\defwfour{\eqalign{& {\cal W}_4={1 \o 2}~ J \rightharpoonup dJ, \cr
   &  {\cal W}_5={1\o 2}~ \Omega_+ \rightharpoonup d\Omega_+\,.}}


We now want to study the intrinsic torsion of the heterotic metric \gulukola. This metric can be brought into the form
\eqn\hettori{\eqalign{ds^2 ~ = ~ &{\cal H}(r)^2~dr^2 + {\cal H}(r)^{-2}~\Big(dz + Q_o~{\rm cot}~\theta_1~ dx
+ Q_o~{\rm cot}~\theta_2~dy\Big)^2 \cr
&~~+ ~{\cal H}_1(r)~(1-\sigma_0)~~\vert dx ~+~ \tau_6~dy\vert^2 \cr
&~~+~ {\cal H}_2(r)~d_6~\Big( {\rm sec}^2~\theta\Big[d\theta_1 + {\rm sin}~2\theta (a~dx - b~dy)\Big]^2 \cr
&~~~~~~~~ +  \vert\tau_2\vert^2~{\rm sec}^2~\tilde{\theta}\Big[d\theta_2 - {\rm sin}~2\tilde{\theta} (\tilde{a}~dx -
\tilde{b}~dy)\Big]^2\Big),}}
where the new complex structure $\tau_6$ of the $(x,y)$--torus is a function of $\tau_3,~\tau_5$ and $\sigma_0$ and determined by
\eqn\bigtau{\rm{Re}~\tau_6 ~ = ~ {\rm{Re}~\tau_3-\sigma_0\rm{Re}~\tau_5 \over 1-\sigma_0}~,~~~~~\vert\tau_6\vert^2 ~=~
 {\vert\tau_3\vert^2-\sigma_0\vert\tau_5\vert^2 \over 1-\sigma_0}~.}
The $(\theta_1,\theta_2)$--torus is non--trivially fibered over the $(x,y)$--torus, as mentioned before.

In the following we will assume that the original B--fields $b_{x\theta_2},~b_{y\theta_1}$
(or equivalently $\theta,~\tilde{\theta}$) are functions of $r$ only. This translates into an $r$--dependence of $\tilde{a},~\tilde{b},~
 \sigma_0,~\tau_2,~\tau_5$ and $\tau_6$, whereas the coefficients $d_i,~a,~b,~Q_0$ and $\tau_3$ remain constant.
 Since $\tilde{a},~\tilde{b},~\sigma_0$ and $\tau_5$ are actually given through $\vert\tau_2\vert$ and $\theta,~\tilde{\theta}$,
 we have only five independent functions left: ${\cal H}_1(r),~{\cal H}_2(r), ~\theta,~\tilde{\theta}$ and $\vert\tau_2\vert$
 (but recall that $\tau_2$ is related to $\theta$ and $\tilde{\theta}$ via \simtau). ${\cal H}$ is linear in $r$
 and therefore completely determined by two real constants.

Due to its toroidal structure, there is a very natural choice for the complex structure on this metric. We define real vielbeins
\eqn\realvielb{\eqalign{e_1 ~=~ & {\cal H}(r)~dr~,~~~~~~~~~~~~~~~~~
e_2 ~=~ {\cal H}(r)^{-1}~\big(dz + Q_o~{\rm cot}~\theta_1~ dx  + Q_o~{\rm cot}~\theta_2~dy\big),\cr
e_3 ~=~ & \sqrt{{\cal H}_1(r)~(1-\sigma_0)}~~(dx ~+~ {\rm Re}~\tau_6~dy)~,~~~~
e_4 ~=~ \sqrt{{\cal H}_1(r)~(1-\sigma_0)}~~{\rm Im}~\tau_6~dy,\cr
e_5 ~=~ & \sqrt{{\cal H}_2(r)~d_6}~~{\rm sec}~\theta~\Big[d\theta_1 + {\rm sin}~2\theta ~(a~dx - b~dy)\Big],\cr
e_6 ~=~ & \sqrt{{\cal H}_2(r)~d_6}~~\vert\tau_2\vert~{\rm sec}~\tilde{\theta}~\Big[d\theta_2 - {\rm sin}~2\tilde{\theta}~ (\tilde{a}~dx - \tilde{b}~dy)\Big]}}
and the usual almost complex structure is induced by choosing complex vielbeins \almcompl.

Let us first study ${\cal W}_4$ and ${\cal W}_5$.  As it was shown
in \lust\  supersymmetry for a heterotic solution with torsion
requires ${\cal W}_1={\cal W}_2=0$ together with
\eqn\susylust{{\cal W}_5 ~=~ -2~{\cal W}_4.} In terms of real
vielbeins $e_i$ one finds with the definition \defwfour,
\eqn\wfour{\eqalign{{\cal W}_4 ~=~ &{1\o 2{\cal H}}~\Big({\rm
tan}\theta~\tilde{\theta}' + {\rm
tan}\tilde{\theta}~\tilde{\theta}'+{{\cal H}_2' \o {\cal
H}_2}+{{\cal H}_1' \o {\cal H}_1}+{(\vert\tau_2\vert)' \o
\vert\tau_2\vert}+{(\sigma_0)' \o \sigma_0-1}+ {({\rm Im}~\tau_6)'
\o {\rm Im}~\tau_6}\cr &-~ {b ~Q_0~{\rm csc}^2\theta_1~\sin
2\theta+\tilde{a}~Q_0~{\rm csc}^2\theta_2~\sin 2\tilde{\theta} \o
{\cal H}_1(\sigma_0-1)~{\rm Im}~\tau_6}\Big)~~e_1;\cr {\cal W}_5
~=~ & -{1\o 2{\cal H}}~\Big({\rm tan}\theta~\theta' + {\rm
tan}\tilde{\theta}~\tilde{\theta}'+{{\cal H}_2' \o {\cal
H}_2}+{{\cal H}_1' \o {\cal H}_1}+{(\vert\tau_2\vert)' \o
\vert\tau_2\vert}+{(\sigma_0)' \o \sigma_0-1}+ {({\rm Im}~\tau_6)'
\o {\rm Im}~\tau_6}-{{\cal H}'\o {\cal H}}\Big)~~e_1 \cr &
+~{({\rm Re}~\tau_6)' \o 2{\cal H}~{\rm Im}~\tau_6}~~e_2.}} Here,
the prime indicates a derivative w.r.t. $r$. We will not attempt to
solve the supersymmetry conditions fully. We would find five
differential equations for the five independent functions
mentioned above when imposing \susylust\ and ${\cal W}_1={\cal
W}_2=0$. But they are highly non-trivial, even if we assume linear
$r$--dependence for ${\cal H}$ and ${\cal H}_{1,2}$. However, it
is immediately obvious that for \susylust\ to be fulfilled the
part of ${\cal W}_5$ that is proportional to $e_2$ has to vanish.
We therefore find the first non--trivial constraint
\eqn\consttau{{\rm Re}~\tau_6 ~=~ {\rm const}.} This means that
the ($x, y$) torus is not quite a square torus and would have been
square if ${\rm Re}~\tau_6$ were identically zero. On the other
hand \consttau\ translates into \eqn\condone{\eqalign{{\rm const}
~=~ &-{1\o 2}\Bigg[4 d_1
d_2-d_4^2+{\sin^22\theta~\sin^22\tilde{\theta}(-d_2+4d_6~P(r))(\tilde{a}b-a\tilde{b})^2
\o d_2d_6~P(r)}\cr & ~~~-~ 16d_2~P(r)~\bigg(d_1-{d_4^2\o
4d_2}-{d_4~\tilde{P}(r)\o 8\sqrt{d_2}~P(r)}\bigg)\Bigg]^{1/2}~
\big(d_2^2-4d_2d_6~P(r)\big)^{-1},}} where we have introduced the
abbreviations
\eqn\defp{P(r)~=~a^2\sin^2\theta+(\tilde{a})^2\vert\tau_2\vert^2\sin^2\tilde{\theta}~,~~~~
\tilde{P}(r)~=~ab~\sin^2\theta+\tilde{a}\tilde{b}~\vert\tau_2\vert^2~\sin^2\tilde{\theta}.}
To fulfill this condition, the $r$--dependences of $\theta$,
$\tilde{\theta}$, $\vert\tau_2\vert$, $\tilde{a}$ and $\tilde{b}$
need to be carefully balanced.

The remaining torsion classes are evaluated using only the constraint \consttau. For the real and imaginary part of ${\cal W}_1$ one obtains
\eqn\wone{\eqalign{{\cal W}_1^+ ~=~ & \bigg(Q_0~{\rm csc}^2\theta_1~\cos\theta~{\rm Re}\tau_6 +
2d_6{\cal H}_2~\Big[a\cos2\theta~{\rm sec}\theta~{\rm Im}\tau_6~\tilde{\theta}' \cr
&~~~~~+ \vert\tau_2\vert \Big(\tilde{b}~\sin\tilde{\theta}
- \tilde{a}'~{\rm Re}\tau_6~\sin2\tilde{\theta}
+ \cos2\tilde{\theta}~{\rm sec}\tilde{\theta}~\big(\tilde{b}-\tilde{a}~{\rm Re}~\tau_6\big)~\tilde{\theta}'\Big)\Big]\bigg)\times\cr
&\times~\Big(6{\cal H}~{\rm Im}~\tau_6~\sqrt{d_6~{\cal H}_1{\cal H}_2~(1-\sigma_0)}\Big)^{-1};\cr
{\cal W}_1^- ~=~ & \Big(\vert\tau_2
\vert~\cos\theta~\big[Q_0~{\rm csc}^2\theta_1~{\rm Im}\tau_6+2d_6{\cal H}_2(b-a~{\rm Re}
\tau_6)\cos2\theta~{\rm sec}^2\theta~\theta'\big]\cr
& ~~~~-d_6 {\cal H}\vert\tau_2\vert^2{\rm Im}\tau_6~{\rm sec}\tilde{\theta}~\big[\tilde{a}'
~\sin2\tilde{\theta}+2\tilde{a}~\cos2\tilde{\theta}~\tilde{\theta}'\big]-Q_0~{\rm csc}^2\theta_2~\cos\tilde{\theta}\Big) \times\cr
& \times ~\Big(6{\cal H}\vert\tau_2\vert~{\rm Im}~\tau_6~\sqrt{d_6~{\cal H}_1{\cal H}_2 ~ (1-\sigma_0)}\Big)^{-1}.}}

It turns out that ${\cal W}_2$ has only two nonzero components; $E_2\wedge\overline{E}_3$ and $E_3\wedge\overline{E}_2$, which are furthermore
  identical for the ${\cal W}_2^+$--part and have opposite sign for the ${\cal W}_2^-$ part. Their precise values are determined to be
\eqn\wtwo{\eqalign{{\cal W}_2^+ ~=~ & \Big[{\rm Im}\tau_6
~\Big(-\vert\tau_2\vert'+ \vert\tau_2\vert\big(\tan\theta~\theta'-
\tan\tilde{\theta} ~ \tilde{\theta}'\big)\Big) +
\vert\tau_2\vert~({\rm Im}\tau_6)'\Big]~
{i~(E_2\wedge\overline{E}_3+E_3\wedge\overline{E}_2) \o 4{\cal
H}\vert\tau_2\vert~{\rm Im}\tau_6},\cr {\cal W}_2^- ~=~ &
\Big[{\rm Im}\tau_6 ~\Big(-\vert\tau_2\vert'+
\vert\tau_2\vert\big(\tan\theta~\theta'- \tan\tilde{\theta} ~
\tilde{\theta}'\big)\Big) - \vert\tau_2\vert~({\rm
Im}\tau_6)'\Big]~
{(E_2\wedge\overline{E}_3-E_3\wedge\overline{E}_2) \o 4{\cal H}
\vert\tau_2\vert~{\rm Im}\tau_6}.}} Note that these two-forms do
not just differ by an overall factor: there is a sign difference
in the $({\rm Im}\tau_6)'$ term as well. The last torsion class
${\cal W}_3$ is given by the (2,1)--form
\eqn\wthree{\eqalign{{\cal W}_3 ~=~ &
{X_1(\overline{\tau}_6)+i~X_2 (\overline{\tau}_6) \o
X_3}~E_1\wedge E_2\wedge \overline{E}_3 ~+~ {X_1(\tau_6)-i~X_2
(\tau_6) \o X_3}~E_1\wedge E_3\wedge \overline{E}_2\cr
 +~ & {X_1(\overline{\tau}_6)-i~X_2 (\overline{\tau}_6) \o X_3}~E_2\wedge E_3\wedge \overline{E}_1,}}
where $\overline{\tau}_6={\rm Re}\tau_6-i~{\rm Im}\tau_6$ and
we have introduced \eqn\defxonetwo{\eqalign{X_1(\tau_6) ~=~ &
Q_0~{\rm csc}^2\theta_2~\cos\tilde {\theta} + 2d_6{\cal H}_2
\vert\tau_2\vert~\cos2\theta~{\rm sec}\theta~\theta'~
(a~\tau_6-b),\cr X_2(\tau_6) ~=~ & Q_0~{\rm
csc}^2\theta_1~\vert\tau_2\vert\cos\theta~\tau_6 - d_6{\cal
H}_2\vert\tau_2\vert^2{\rm \sec}
\tilde{\theta}~\big[\sin2\tilde{\theta}~(\tilde
{a}'\tau_6-\tilde{b}') +
2\cos2\tilde{\theta}~(\tilde{a}\tilde{\theta}'~\tau_6
-\tilde{b})\big],\cr X_3 ~=~ & 8{\cal H}\vert\tau_2\vert
~\sqrt{d_6{\cal H}_1{\cal H}_2(1-\sigma_0)}~{\rm Im}~\tau_6.}}
These non--vanishing torsion classes show that our heterotic
background \gulukola\ will in general be non--K\"ahler. The
supersymmetry conditions ${\cal W}_1={\cal W}_2=0$ and ${\cal W}_5
~=~ -2~{\cal W}_4$ result in five differential equations that
 are given through
\eqn\susycond{\eqalign{{\cal W}_1^+~=~ & {\cal W}_1^- ~=~ 0, \cr
{\cal W}_2^+\vert_{E_2\wedge\overline{E}_3} ~=~ & {\cal W}_2^-\vert_{E_2\wedge\overline{E}_3} ~=~ 0, \cr
{\cal W}_5\vert_{e_1} ~=~ & -2 {\cal W}_4\vert_{e_1}.}}
Solving these together with the constraint \condone\ should in principle allow for a determination of the
functions ${\cal H}_1(r),~{\cal H}_2(r), ~\theta,~\tilde{\theta}$ and $\vert\tau_2\vert$, but we were not able
 to find an analytic solution. In the following section, we will provide a detailed mathematical analysis of these non-K\"ahler manifolds.

\subsec{A family of non-K\"ahler manifolds}

In this section we will first quickly describe the global geometry 
corresponding to \gulukola\ (or \hettori\ in an expanded form), 
then we give the
analysis that leads to it, and finally we analyze the
geometry.

The non-K\"ahler complex threefold ${\bf X}$ is desribed in terms of a
non-K\"ahler complex surface ${\bf S}$ called a {\it primary Kodaira
surface\/} which was already known to mathematicians.  The surface
${\bf S}$ has a non-trivial holomorphic fibration over ${\bf T}^2$ with ${\bf T}^2$
fiber characterized as being twisted by translations in a topologically
non-trivial manner.  This will be made more precise below.
The desired complex threefold ${\bf X}$ is a
non-trivial holomorphic ${\bf C}^*$ fibration over ${\bf S}$.  It admits
a nowhere vanishing holomorphic 3~form $\Omega$.  
This geometry will be described
in more detail following the analysis that led to it.

We know that ${\bf X}$ is built from a holomorphic base ${\bf T}^2$ in two
steps: first construct a complex surface ${\bf S}$ by two ${\bf S}^1$
fibrations over ${\bf T}^2$, leading to a holomorphic ${\bf T}^2$ fibration
over ${\bf T}^2$.  Then we take an ${\bf S}^1$ fibration ${Y}$ over ${\bf S}$ and 
take ${X}={Y}\times{\bf R}$.
The holomorphic ${\bf T}^2$ fibrations over ${\bf T}^2$ are completely
classified.  A convenient reference for the results are in \BPV\
Section V.5.

Let ${\bf B}$ denote the base ${\bf T}^2$ with its holomorphic
structure as an elliptic curve ${\bf C}/\Lambda$, where $\Lambda$ is a
lattice ${\bf Z}\oplus \tau_B{\bf Z}$.  Let ${\bf E}$ denote the fiber
${\bf T}^2$ with its holomorphic structure as an elliptic curve ${\bf
C}/L$, where $L$ is a lattice ${\bf Z}\oplus \tau_E{\bf Z}$.

We need to study holomorphic automorphisms of ${\bf E}$ in order to
build non-trivial holomorphic bundles with fiber ${\bf E}$.  The
translation automorphisms of ${\bf E}$ can be identified with ${\bf
E}$ itself by associating to $e\in {\bf E}$ the translation
automorphism $t_e$ of ${\bf E}$ defined by $t_e(z)=z+e$.  Letting
${\rm Aut}({\bf E})$ denote the group of holomorphic automorphisms of 
${\bf E}$, and
${\bf E}\subset {\rm Aut}({\bf E})$ the translation subgroup just described, 
then it is
well-known that $({\rm Aut}({\bf E}))/{\bf E}$ is a finite group ${\bf Z}_n$.
Usually $n=2$ and these automorphisms are just the ${\bf Z}_2$
subgroup generated by the inversion $z\mapsto -z$ of ${\bf E}$, but $n$ can
be larger if
\eqn\esime{{\bf E}\simeq {\bf
C}/({\bf Z}\oplus i{\bf Z})~~~~~{\rm or}~~~~~ {\bf E}\simeq {\bf C}/({\bf Z}\oplus
\omega{\bf Z})}
with $\omega={\rm exp}(\pi i/3)$.  For the first elliptic curve in 
\esime, we have
$n=4$ with the ${\bf Z}_4$ generated by the automorphism $z\mapsto iz$.
For the second elliptic curve in \esime, we have
$n=6$ with the ${\bf Z}_6$ generated by the automorphism $z\mapsto \omega z$.
This
description makes ${\rm Aut}({\bf E})$ into a disconnected 1~dimensional
complex manifold, identified with $n$ disjoint copies of ${\bf E}$.

We now turn to the description of non-trivial holomorphic ${\bf E}$ fibrations
over ${\bf B}$.  
The surface ${\bf S}$ is constructed by choosing open sets $U_i\subset {\bf B}$
covering the base ${\bf B}$ and gluing the trivial products $U_i\times
{\bf E}$ and $U_j\times {\bf E}$ using non-trivial holomorphic automorphisms of
$(U_i\cap U_j)\times {\bf E}$. 

We introduce $w$ as the coordinate on ${\bf B}$.  When studying $U_i\times
{\bf E}$, we will use the coordinates $(w,z_i)$.  In other words, we continue
to use the coordinate $w$ on $U_i\subset {\bf B}$, and modify notation 
slightly by using $z_i$ instead of $z$ as the coordinate on ${\bf E}$.  The
introduction of this subscript allows us to describe a gluing 
$U_i\times {\bf E}$
with $U_j\times {\bf E}$ by identifying the common $(U_i\cap U_j)\times
{\bf E}$ subset via an identification which necessarily takes the form
\eqn\gluing{
\left(w,z_j\right) = \left(w,\left(\rho_{ij}(w)\right)(z_i)\right)}
Here, $\rho_{ij}(w)$ is a holomorphic automorphism of ${\bf E}$ depending
holomorphically on $w$.  In other words, the mapping
$\rho_{ij}:U_i\cap U_j\to {\rm Aut}({\bf E})$ is holomorphic.  

Let ${\cal A}({\bf E})$ denote the trivial holomorphic fiber bundle over
${\bf B}$ with fiber ${\rm Aut}({\bf E})$, i.e.\ ${\cal A}({\bf E})=
{\bf B}\times {\rm
Aut}({\bf E})$ as a complex manifold.  Then
$\rho_{ij}$ is a holomorphic section of the bundle ${\cal A}({\bf E})$
over the open set $U_i\cap U_j$.

\noindent We have the obvious compatibility condition:
\eqn\compati{\rho_{jk}(w)\circ\rho_{ij}(w)=\rho_{ik}(w),}
valid for $w\in U_i\cap U_j\cap U_k$.  In other words, $\{\rho_{ij}\}$
is a Cech cocycle representing a cohomology class
\eqn\cech{\rho\in H^1({\bf B},{\cal A}({\bf E})).}
The quotient map $\pi:{\rm Aut}(E)\to {\bf Z}_n$ induces a class
$\pi(\rho)\in H^1({\bf B},{\bf Z}_n)$.\foot{$\pi(\rho)$ is a natural shorthand for 
the Cech cohomology class represented by the cocycle whose
value over $U_i\cap U_j$ is $\pi(\rho_{ij}(w))\in{\bf Z}_n$.}

The first relevant result is that if $\pi(\rho)$ is non-trivial,
then ${\bf S}$ is K\"ahler (actually projective algebraic).  Such an
algebraic surface is called a {\it hyperelliptic surface\/}.
Details are in \BPV\ (pp147--8). Since the ${\bf S}$ that we need is
non-K\"ahler, we require that $\pi(\rho)\in H^1({\bf B},{\bf Z}_n)$ is
the trivial cohomology class.

Let ${\cal E}$ be the trivial holomorphic bundle over ${\bf B}$ with
fiber ${\bf E}$. The exact sequence
\eqn\eseqone{0\to {\cal E}\to {\cal A}({\bf E})\to {\bf Z}_n\to 0}
determines a corresponding exact sequence of cohomology groups
\eqn\seqco{H^1({\bf B},{\cal E})\to H^1({\bf B},{\cal A}({\bf E}))\to 
H^1({\bf B},{\bf Z}_n).}
Since the second map takes $\rho$ to $\pi(\rho)$, assumed to be
trivial, from this sequence we see that $\rho$ arises from a
cohomology class in $H^1({\bf B},{\cal E})$. In other words, we can and
will assume that the $\rho_{ij}$ take values in the translation
subgroup ${\bf E}\subset {\rm Aut}({\bf E})$. We let $\sigma_{ij}(w)\in {\bf E}$ by
the element of ${\bf E}$ corresponding to these translations, i.e.
\eqn\transrep{
\rho_{ij}(w)=t_{\sigma_{ij}(w)},}
where $t_e$ continues to denote the translation automorphism by an element 
$e\in {\bf E}$.  Then the compatibility condition \compati\ implies
the compatibility condition
\eqn\comptrans{
\sigma_{jk}(w)+\sigma_{ij}(w)=\sigma_{ik}(w),}
valid for $w\in U_i\cap U_j\cap U_k$.
The condition
\comptrans\ says that the $\sigma_{ij}$ define a 
cohomology class $\sigma\in H^1({\bf B},{\cal E})$.

From the description of ${\bf E}$ as the quotient of ${\bf C}$ by the lattice 
$L={\bf
Z}+\tau_E{\bf Z}$, we have an exact sequence
\eqn\latt{0\to L\to {\bf C}\to {\bf E}\to 0}
which determines the cohomology coboundary mapping
\eqn\latmap{H^1({\bf B},{\cal E})\to H^2({\bf B},L).}
The resulting class in $H^2({\bf B},L)$ will
 be denoted by $c(\rho)$. It is a generalization of the well-known
 Chern class of a $U(1)$ bundle; in fact, if one chooses an isomorphism
 $L\simeq {\bf Z}^2$ (e.g.\ the one determined by 1 and $\tau_E$),
 then $H^2({\bf B},L)\simeq H^2({\bf B},{\bf Z})^2$ and $c(\rho)$ is
 identified with the Chern classes of the two $S^1$ bundles used
 to construct $S$.

 In \BPV\ (page 146), it is shown that if $c(\rho)=0$, then $S$ is
 a complex torus, hence K\"ahler.  Thus we require that
 $c(\rho)\ne0$, or equivalently that the $T^2$ bundle is topologically
non-trivial.
 This is the situation of a {\it primary Kodaira surface\/}
 considered in \BPV\ (pp 146--7).  The invariants are
\eqn\coin{H^1({\bf S},{\bf Z})={\bf Z}^3,~~~ H^2({\bf S},{\bf Z})={\bf Z}^4\ {\rm or\ }
{\bf Z}^4\oplus {\bf Z}_m.} 
These invariants can also be computed by the Gysin sequence as has been used
in \GP.
Note that since $H^1({\bf S},{\bf R})$
is odd-dimensional, ${\bf S}$ is not K\"ahler.

Furthermore, ${\bf S}$ admits a nowhere vanishing
holomorphic 2-form.\foot{In \BPV\ this is described as
the triviality of the canonical bundle.}  This can be made explicit.
From \gluing\ and \transrep\ we have the coordinate transformation
\eqn\transition{
z_j=z_i+\sigma_{ij}(w).}
It follows immediately from \transition\ that
\eqn\twoform{
dz_j\wedge dw= dz_i\wedge dw.}
Thus the holomorphic 2-forms $dz_i\wedge dw$ in $U_i\times{\bf E}$ agree
on their overlaps and patch together to give a well-defined holomorphic
2-form $dz\wedge dw$ on ${\bf S}$, even though $dz$ is not a well-defined
1-form.

Next, we consider a non-trivial $S^1$ bundle
$Y$ over ${\bf S}$; then finally put $X=Y\times {\bf R}$.  Thus the fiber
$S^1$ is promoted to $S^1\times{\bf R}\simeq {\bf C}^*$.\foot{Note that
${\bf C}^*$ has a unique holomorphic structure up to isomorphism,
so we use the standard one.} We
study holomorphic ${\bf C}^*$ bundles ${\bf X}$ over ${\bf S}$.  We explicitly
assume that the ${\bf C}^*$ bundle structure is a principal bundle.\foot{Our
duality chain indicates a similar structure.}

We cover ${\bf S}$ by open sets $V_\alpha$ and build ${\bf X}$ by
gluing together the open sets $V_\alpha\times {\bf C}^*$.  Since ${\bf S}$
is non-K\"ahler, methods for constructing metrics on ${\bf X}$ from the
metric on ${\bf S}$ will not produce a K\"ahler metric.  Presumably
${\bf X}$ does not admit any exotic K\"ahler metrics, but we have not
definitively ruled out that possibility.

Let
$t_\alpha$ be the ${\bf C}^*$ coordinate in $V_\alpha\times{\bf C}^*$.
The principal bundle ansatz is that ${\bf C}^*$ multiplications are
used to perform the gluing, i.e.\ that the coordinates are related by
\eqn\xglue{
t_\beta=\kappa_{\alpha\beta}(w,z)t_\alpha,}
where $\kappa_{\alpha\beta}:V_\alpha\cap V_\beta\to {\bf C}^*$ is
holomorphic.  Then \xglue\ implies that
\eqn\threeform{
dw\wedge dz\wedge \frac{dt_\beta}{t_{\beta}}
=dw\wedge dz\wedge \frac{dt_\alpha}{t_{\alpha}}}
in $(V_\alpha\cap V_\beta)\times {\bf C}^*$, so these holomorphic 3-forms
patch to give a nowhere vanishing holomorphic 3-form $\Omega=
dw\wedge dz\wedge dt/t$.

\smallskip
It remains to check the existence of non-trivial holomorphic principal
${\bf C}^*$ fibrations over ${\bf S}$.  The transition functions
$\kappa_{\alpha\beta}$ satisfy
\eqn\kappatrans{
\kappa_{\beta\gamma}\kappa_{\alpha\beta}=\kappa_{\alpha\gamma}}
in $V_\alpha\cap V_\beta\cap V_\gamma$, so define a cohomology class
$\gamma\in H^1({\bf S},{\cal O}_{\bf S}^*)$.  Here ${\cal O}_{\bf S}^*$
denotes the sheaf of nowhere vanishing holomorphic functions on arbitrary open
subsets of ${\bf S}$, and has been introduced since $\kappa_{\alpha\beta}$ is
a holomorphic section of ${\cal O}_{\bf S}^*$ over $V_\alpha\cap V_\beta$.
So we only have to show that
$H^1({\bf S},{\cal O}_{\bf S}^*)$ is nontrivial.

For this purpose, we have the exponential sequence of sheaves on ${\bf S}$
\eqn\exp{
0\to {\bf Z}\to {\cal O}_{\bf S}\to {\cal O}_{\bf S}^*\to 0,}
where ${\bf Z}$ denotes the sheaf of locally constant integer-valued functions
and ${\cal O}_{\bf S}$ are the holomorphic functions.  The first non-trivial 
map in \exp\ is the inclusion, and the second map takes a holomorphic function
$f$ to ${\rm exp}(2\pi i f)$.

The cohomology sequence associated to \exp\ includes the segment
\eqn\exples{
\cdots\to H^1({\bf S},{\bf Z})\to H^1({\bf S},{\cal O}_{\bf S})
\to H^1({\bf S},{\cal O}_{\bf S}^*)
\to H^2({\bf S},{\bf Z})\to H^2({\bf S},{\cal O}_{\bf S})\to \cdots,}
so that $H^1({\bf S},{\cal O}_{\bf S}^*)$ contains
$H^1({\bf S},{\cal O}_{\bf S})/H^1({\bf S},{\bf Z})$ as a subgroup.

In \BPV, $H^1({\bf S},{\cal O}_{\bf S})\simeq H^{0,1}({\bf
S})$\foot{The Dolbeault isomorphism used here is valid for
non-K\"ahler manifolds.}  has been computed to be ${\bf C}^2$.  In our
situation, we can even explicitly exhibit two independent
$\bar\partial$-closed $(0,1)$ forms on ${\bf S}$: the global form
$d\bar{w}$ coming from the base ${\bf B}$, and the form $d\bar{z}$ on
the fiber, which is globally well-defined by \transition\ and the
holomorphicity of $\sigma_{ij}$.

Combining this calculation with \coin, we see that $H^1({\bf S},
{\cal O}_{\bf S}^*)$ contains a subgroup of the form ${\bf C}^2/{\bf Z}^3$,
which is certainly nontrivial as claimed.

We can also show that we have holomorphic ${\bf C}^*$ fibrations with 
topologically nontrivial structure.  The topology can be computed from 
the coboundary mapping $\delta:H^1({\bf S},{\cal O}_{\bf S}^*)
\to H^2({\bf S},{\bf Z})$ from \exples.  The image in 
$H^2({\bf S},{\bf Z})$ of the cohomology class
in $H^1({\bf S},{\cal O}_{\bf S}^*)$ representing our holomorphic fibration
is just the chern class of the original $S^1$ bundle.  So if we specify a
non-trivial class $c\in H^2({\bf S},{\bf Z})$ corresponding to a topologically
nontrivial $S^1$ bundle and want to know if the corresponding 
$S^1\times {\bf R}={\bf C}^*$ fibration admits a holomorphic principal
bundle structure, we ask if $c$ is in the image of $\delta$.  By
\exples, this is equivalent to asking if the image of $c$ in 
$H^2({\bf S},{\cal O}_S)$ vanishes.  But $H^2({\bf S},{\cal O}_S)$ is
a vector space, so if, for example, we are in the case with torsion,
$H^2({\bf S},{\bf Z})={\bf Z}^4\oplus {\bf Z}_m$ as in \coin, the torsion
classes $c\in {\bf Z}_m$ must map to 0 as complex vector spaces have no
torsion classes.  Hence these classes are in the image
of $\delta$ and the corresponding $S^1\times {\bf R}$ fibration
supports a holomorphic ${\bf C}^*$ structure.  We conclude that there are
certainly examples of complex threefolds ${\bf X}$ with the properties
dictated by the duality chain.

\smallskip
In summary, the fibration structured dictated by the duality chain led us
to construct specific manifolds ${\bf X}$ with (integrable) complex structures.
Not only do we show that these exist, but we see that these manifolds have
nowhere vanishing holomorphic 3-forms, something that we didn't require
in the construction, a good check.
We leave the study of the metric and gauge bundle to future work.

\medskip
We close this section with a brief description of the topology of ${\bf X}$.
Since ${\bf R}$ is contractible, it follows that ${\bf X}$ is homotopic to $Y$,
hence 
\eqn\homot{H^*({\bf X},{\bf Z})\simeq H^*(Y,{\bf Z}).}
The Gysin sequence for the $S^1$ bundle ${\bf Y}$ reads
\eqn\gysin{
\cdots\to H^i({\bf S},{\bf Z})\to H^{i+2}({\bf S},{\bf Z})\to 
H^{i+2}(Y,{\bf Z})\to H^{i+1}({\bf S},{\bf Z})\to \cdots}
The first map in \gysin\ is cup product with the first chern class $c_1$
of the $S^1$ bundle ${\bf Y}$.  The second map in \gysin\ is the pullback map
associated with the projection ${\bf Y}\to {\bf S}$.

Without knowing more than the nontriviality of $c_1$, there is not
enough information in \gysin\ to even determine the cohomology ranks.
So that we can say something more definite, let us suppose that the
$S^1$ fibration is general enough that the first map in $c_1$ has the
maximum possible rank, namely the minimum of the ranks of $H^i({\bf
S},{\bf Z})$ and $H^{i+2}({\bf S},{\bf Z})$.  From \coin, \homot,
\gysin, and Poincar\'e duality on ${\bf S}$, it is computed in this
situation that, ignoring possible torsion, 
\eqn\cohom{H^i({\bf X},{\bf Z})=\cases{{\bf Z}&i=0,\ 5\cr
{\bf Z}^3&i=1,2,3,4\cr
0& {\rm otherwise}}}

\newsec{Dipole deformations in the geometric transition setup}

There is one important issue that we have overlooked for some time
in the setup of geometric transition in type IIB theory. This is
the appearance of dipole deformations in these theories. Recall
that due to inherent orientifold action, the allowed choices of
the $B$ fields are given by \bnsbrr. The $B_{NS}$ field is thus
oriented along ${\cal B}_{y\theta_1}$ and therefore has one leg
along the $D5$ branes wrapped on the two-torus ($y, \theta_2$). As
we now know from \difuli\ this situation is  ripe for dipole
theories (see also \ramdam\ and the second reference of \robbins).
The ${\cal B}_{y\theta_1}$ gives rise to dipole deformations of
our geometric transition setup from the five-brane point of view.
These dipoles are therefore {\it not} visible in the IR of our
gauge theory, and their presence is fully manifested once we go to
the far UV of the theory where we expect six-dimensional gauge
theory. For our case, assuming that we choose a scale where we do
not integrate out the dipole degrees of freedom, we can then ask
two questions:

\noindent $\bullet$ Can we calculate the precise supergravity
metric for our case now? Recall that now our ingredients will be
seven-branes, wrapped five-branes and the $B$ fields generating
the dipole deformations {\it on} a non-trivial background that
locally looks like a K\"ahler resolved conifold.

\noindent $\bullet$ Can we see what happens to this background after we perform a geometric transition? Clearly
there would be a gravitational dual to this theory because (a) it is ${\cal N} = 1$ gauge theory, and
(b) it has dipole deformations. So far we have been pursuing both cases individually in \gtone, \realm, and
\gttwo\ for the gravity duals of confining gauge theories; and in \difuli, \robbins\ for the gravity duals
of dipole theories. We would like to know what happens now when we combine these\foot{Recall that
in \gttwo\ we computed the mirror type IIA picture, where the dipole deformations in type IIB theory results
in specific non-K\"ahler deformations in type IIA.}.

\vskip.17in

\centerline{\epsfbox{newboxes.eps}}\nobreak

\vskip.18in

\noindent In  {\bf figure 4} above we have represented the whole
set-up. The middle set of boxes (and the operations therein) have
already been dealt with in \gtone, \gttwo. The dotted line that
takes us to the heterotic theory is not {\it a priori} connected
to type IIB because we used a local orientifolding operation to go
to the heterotic side and then moved away from the orientifold
point to study the global story. The gauge theory (i.e the NS5
branes side) is basically what we addressed above. The gravity
dual was studied in \gttwo. Whether the two sides are connected by
a possible geometric transition is still unknown, and so we have
used {\bf A} to represent this.

The part of the figure that takes us from the wrapped $D5$ branes picture to an equivalent IIA brane configuration
is the starting point of the dipole deformation. In fact this procedure
addresses the first of the two questions mentioned above, namely, determining the supergravity
solution for our wrapped D5 brane configuration with dipole deformations on the world-volume of the
branes. This is already a complicated enterprise, because our dipole theory is now embedded in a much more
non-trivial background. Due to the complexity of the problem, we will address this in two steps:

\noindent $\bullet$ First we will
see how our type IIB metric ansatz \metmet\ with warp factors \coffdeff\ changes when we incorporate dipole
deformations on the background without
$D5$ (or $D7$)-branes. We will only be able to work this order-by-order in the $B$ field
${\cal B}_{y\theta_1} \equiv b$ by treating the dipole deformation as a perturbation.

\noindent $\bullet$ Secondly, we will insert back all the branes
in our framework and study the final dipole deformed metric. The
analysis will be done up to some orders in $r$ and $b$ as above,
so that this would capture the essential feature of the whole
story. The usual constraint of an order-by-order expansion is that
we cannot tell how higher order terms modify the result.
Nevertheless, we will be able to pursue both avenues in enough
detail so that the final picture can be presented clearly and
unambigously.

The second question of determining the gravity duals of these new
theories is represented in the figure above by the lines joining
the last two boxes with letters {\bf B} and {\bf C} as the
underlying operation (as yet unknown). Clearly, after we go to the
IIA brane configuration, and then do a dipole twist followed by a
T-duality we {\it do not} come back to the same configuration.
This is of course expected from our earlier studies (see \difuli)
as we go to  dipole theories. This is a very interesting scenario
because it opens up, for the first time, the possibility of
studying gravity duals of dipole theories in the setting of the
geometric transition! In fact the whole set of transformations of
\gtone\ could possibly be done to get the gravity duals of these
theories. We can ask many questions here:

\noindent $\bullet$ How do we see the non-locality of dipole theories in the gravity duals of these theories?

\noindent $\bullet$ In the type IIA mirror configuration, we map directly to a non-K\"ahler manifold
instead of another dipole theory. What happens to the dipoles of the original theory?

\noindent $\bullet$ What are the operations {\bf B} and {\bf C}?
Are they in any way connected to the geometric transitions?

\noindent $\bullet$ What happens in M-theory? How do we go to the gravity duals in IIA using M-theory? Is there
an equivalent operation to the flop here too?

In this paper we will only be able to address the supergravity solution for the wrapped D5 branes with
dipole deformations (this is not the gravity dual!) following the two-step procedure that we mentioned
above. The rest of the questions will be addressed in the sequel to this paper \gwyn.

Before we go ahead with the  dipole story, let us mention one more
thing regarding the dipoles studied earlier in \difuli. The dipole
theories that we studied before were associated with vanishing
beta functions, i.e. conformal theories\foot{Even though they have
a length scale, the dipole length!}. The gravity duals of these
theories were therefore determined from the {\it near-horizon}
geometries \difuli\ exactly in the same way as for other CFTs
\maldacena. On the other hand, the theories that we are studying
here are confining gauge theories and have non-zero beta
functions. The gravity duals of these theories follow a somewhat
different route as we saw earlier in \ks, \mn, \gtone, \gttwo,
\realm. Once we make a dipole deformation to these theories the
gravity duals follow yet other different routes mentioned as operations {\bf
B} and {\bf C} in the figure above. These details will be
relegated to the sequel to this paper.

\subsec{Supergravity solution without branes}

As we mentioned earlier, our basic point is to treat the dipole deformations perturbatively. In previous sections
we have managed to study the local metric from analysing the equations of motion, without carefully considering the
backreactions of fluxes and seven branes on the geometry. Here we would like to study the local metric when we
put in everything like the branes and fluxes along with a non-trivial background geometry.

{}From the very look of it, this is a pretty complicated problem.
Therefore we will attack it in a few simple steps: First we
analyse the background with back-reactions from B-fields. These
B-fields would eventually be responsible for the dipole
deformations on the wrapped D5 branes. Next we insert back the D7
branes in the geometry by including their back-reactions. And
finally we create bound states of D5 branes on a single D7 brane
by switching on first Chern classes. This way we will have
explicit supersymmetric solutions for the system. Of course this
is not the only way to get susy solutions here. As we discussed in
much detail earlier, F-theory allows us to study a supersymmetric
solution with separated D5s, D7s and primitive fluxes. Once we
separate the D7 branes we are not bound to remain at the
orientifold point, and there we can have all possible components
of the B-fields. The local supergravity solution of the system
with B-fields along the five-brane directions has been given
earlier in \metnownow\ without carefully considering the full
back-reactions. Here we want to see possible corrections to this
metric when we consider all the branes and fluxes.

We begin by first removing the seven-branes, but keeping the
fluxes. We shall assume that the metric ansatz for our case looks
similar to \metmet\ i.e \eqn\metmat{{ds^2_{\cal M}~ = ~ {\cal
A}~dr^2 + {\cal B}~(dz + f_1~ dx + f_2~dy)^2~ + ({\cal
C}~d\theta_1^2 + {\cal D}~ dx^2) + ({\cal E}~ d\theta_2^2 + {\cal
F}~ dy^2),}} with the coefficients having the same expansion as
\coffdeff\ although we might have to go beyond $r^2$ terms in
\coffdeff\ to see the full back-reaction. We will deal with these
details as we go on. As a starter we need new combinations of the
warp factors as \eqn\newcom{{\cal E}(1-{\cal F}) ~ = ~D_0 -D_1~r -
D_2~r^2 - D_3~r^2 + ...} where ${\cal E}, {\cal F}$ have the same
expansion as \coffdeff\ before. The coefficients $D_i$ in the
expansion above are defined as \eqn\ddefe{\eqalign{& D_0 ~ = ~ 0,
~~~D_1~=~ {\alpha_6\o {\cal F}_6}~r, ~~~ D_2 ~ = ~ {1\o {\cal
F}_6} \Bigg(\beta_6 + {\alpha_5 \alpha_6 \o {\cal
F}_5(r_0)}\Bigg),\cr & D_3~=~{1\o {\cal F}_6(r_0)}~\Bigg(\gamma_6
~+~{\alpha_5 \beta_6\o {\cal F}_5(r_0)} ~+~ {\beta_5 \alpha_6 \o
{\cal F}_5(r_0)}\Bigg),}} where $\gamma_6$ is an ${\cal O}(r^3)$
term in ${\cal F}$ \coffdeff. In addition to that, the
coefficients $\alpha_{5,6}$ and $\beta_{5,6}$ have the same
equivalence relations as in \morerel, although $-$ and this is
very crucial $-$ $\alpha_{3,4}$ and $\beta_{3,4}$ are no longer
related by \morerel\ because of the back-reactions of the
B-fields. The coefficients $\gamma_{5,6}$ are expected to satisfy
the following approximate relation \eqn\gfgs{{\gamma_5\o {\cal
F}_5(r_0)} ~-~ {\gamma_6\o {\cal F}_6(r_0)} ~\approx~ 0} where now
$\gamma_5$ is an ${\cal O}(r^3)$ term in ${\cal E}$ \coffdeff,
much like $\gamma_6$ above. In \gwyn\ we will discuss a toy
example where an exact equality in \gfgs\ (and also in \morerel)
is realised.

The reader may notice that we haven't said anything too new so far. Let us now switch on a $B_{NS}$ field
${\cal B}_{y\theta_1} \equiv b$ and define an expansion of the form
\eqn\bswde{ \kappa_o ~\equiv ~ 1 ~+~ a_1~b^2~{\cal E}(1-{\cal E}) ~+~ a_2~b^4~{\cal E}^2(1-{\cal E})^2 ~+~ .....}
with constant coefficients $a_{1,2}$ that we will be kept arbitrary in this paper.  
The expansion $\kappa_o$ will help us
see the corrections to the equations \relbetnow\ due to the B-field $b$. For this we need, instead of \cerel, a more
involved structure,
\eqn\invstr{\kappa_o {\cal D} {\cal E} ~ = ~ 1 ~+~ E_0 r ~+~ E_1 r^2 ~+~ E_2 r^3 ~+~ E_3 r^4 ~+~ {\cal O}(r^5),}
where we are assuming that the $E_3$ coefficient is well defined. The various $E_i$ can be written in terms of
($\alpha_4, \alpha_5, \alpha_6$), ($\beta_4, \beta_5, \beta_6$), ($\gamma_4, \gamma_5, \gamma_6$)
and $b$ as
\eqn\eidef{\eqalign{&E_0 ~ = ~ {\alpha_5 \o {\cal F}_5(r_0)} + {\alpha_4 \o {\cal F}_4(r_0)} -
{a_1 b^2 \alpha_6 \o {\cal F}_6(r_0)},\cr
& E_1 ~ = ~ {a_2 b^4 \alpha_6 - a_1 b^2 \beta_6 \o {\cal F}_6(r_0)} + {\beta_5 \o {\cal F}_5(r_0)} +
{\beta_4 \o {\cal F}_4(r_0)} -{2 a_1 b^2 \alpha_5 \alpha_6 \o {\cal F}_5(r_0) {\cal F}_6(r_0)} ~ + \cr
&~~~~~~~~ - {a_1 b^2 \alpha_6 \o {\cal F}_6(r_0)} + {\alpha_5 \alpha_4 \o {\cal F}_5(r_0) {\cal F}_4(r_0)}, \cr
& E_2~=~ {a_2 b^4 \alpha_5 \alpha_6 \o {\cal F}_5(r_0) {\cal F}_6(r_0)} +
{a_2 b^4 \alpha_4 \alpha_6 \o {\cal F}_4(r_0) {\cal F}_6(r_0)} -
{2 a_1 b^2 \alpha_5 \beta_6 \o {\cal F}_5(r_0) {\cal F}_6(r_0)} -
{a_1 b^2 \beta_6 \alpha_4 \o {\cal F}_6(r_0) {\cal F}_4(r_0)}~ - \cr
&~~~~~~~~ - {a_1 b^2 \alpha^2_5 \alpha_6 \o {\cal F}^2_5(r_0) {\cal F}_6(r_0)} -
{2 a_1 b^2 \alpha_5 \alpha_6 \alpha_4 \o {\cal F}_5(r_0) {\cal F}_6(r_0) {\cal F}_4(r_0)} -
{a_1 b^2  \gamma_6 \o {\cal F}_6(r_0)} -{a_1 b^2 \beta_5 \alpha_6 \o {\cal F}_5(r_0) {\cal F}_6(r_0)} ~ + \cr
&~~~~~~~~ - {a_1 b^2 \beta_4 \alpha_6 \o {\cal F}_4(r_0) {\cal F}_6(r_0)} + {\gamma_5\o {\cal F}_5(r_0)} +
{\gamma_4\o {\cal F}_4(r_0)} + {\alpha_4 \beta_5 + \alpha_4 \gamma_5 + \alpha_5 \beta_4 \o
{\cal F}_5(r_0) {\cal F}_4(r_0)}, \cr
& E_3~=~ {a_2 b^4 (\alpha_5 \alpha_6 + \alpha_6 \beta_5) - a_1 b^2 (2\alpha_5 \gamma_6 + 2\beta_5 \beta_6 +
\gamma_5 \gamma_6
+ \alpha_6 \gamma_5) \o {\cal F}_5(r_0) {\cal F}_6(r_0)} ~ + \cr
&~~~~~~~~ + {a_2 b^4 \beta_6 - a_1 b^2 \gamma_6 \o {\cal F}_6(r_0)} -
{a_1 b^2 (\gamma_6 \alpha_4 + \alpha_6 \gamma_4) \o
{\cal F}_6(r_0) {\cal F}_4(r_0)} - {a_1 b^2(\alpha_5^2 \beta_6 + 2 \beta_5 \alpha_6 \alpha_5) \o
{\cal F}^2_5(r_0) {\cal F}_6(r_0)} ~ + \cr
&~~~~~~~~ - {a_1b^2(2\alpha_5 \beta_6 \alpha_4 + 2 \beta_5 \alpha_6 \alpha_4 +
\alpha_6 \alpha_4 \gamma_5 + 2 \alpha_5 \alpha_6 \beta_4) -a_2 b^4 \alpha_6\alpha_5 \alpha_4 \o
{\cal F}_4(r_0) {\cal F}_5(r_0) {\cal F}_6(r_0)} ~ + \cr
&~~~~~~~~ -{a_1 b^2 \alpha^2_5 \alpha_6 \alpha_4 \o {\cal F}_4(r_0) {\cal F}^2_5(r_0) {\cal F}_6(r_0)}
+{a_2b^4 \alpha_6 \beta_4 - a_1 b^2 \beta_6 \beta_4 \o {\cal F}_4(r_0){\cal F}_6(r_0)} +
{\beta_4 \beta_5 + \gamma_4 \alpha_5 \o {\cal F}_4(r_0){\cal F}_6(r_0)},}}
where we see that even in the absence of a $B$ field we expect higher order corrections to
\defcof\ given here as
\eqn\dccorr{\eqalign{&E^0_0 ~=~ C_0, ~~~~ E^0_1 ~ = ~ C_1, ~~~~
E^0_3 ~ = ~ C_3 + {\gamma_3 \alpha_5 \o {\cal F}_3(r_0){\cal F}_6(r_0)}, \cr
& E^0_2 ~ = ~ C_2 + {\gamma_5\o {\cal F}_5(r_0)} +
{\gamma_3\o {\cal F}_3(r_0)} + { \alpha_3 \gamma_5 \o {\cal F}_5(r_0) {\cal F}_3(r_0)},}}
with $C_3 = {\beta_3 \beta_5 \o {\cal F}_3 {\cal F}_5}$, and $E^0_i \equiv {}^{\rm lim}_{b \to 0} E_i$; and
using \morerel.
We see that the higher order corrections typically
arise from $\gamma_i, i = 3, 5, 6,$ terms in the absence of $B$ fields.

Now to make connections with the B-fields and the coefficients of
the proposed metric \metmat\ we need to first define the
coefficients to higher orders in $r$ than what we took earlier in
\coffdeff. Let us first consider the coefficient ${\cal B}$ in
\metmat. This is given by \eqn\bnuy{{\cal B} = 1 + {\alpha_2\o
{\cal F}_2(r_0)} r +  {\beta_2\o {\cal F}_2(r_0)} r^2 +
{\gamma_2\o {\cal F}_2(r_0)} r^3 + {\delta_2\o {\cal F}_2(r_0)}
r^4 + {\cal O}(r^5).} As one would expect, the connection between
these coefficients and the $E_i$ defined above is much more
involved here than \relbetnow. This is of course expected, as the
B-fields will back-react on the geometry and distort it from the
simple form \metmetnow\ (where we didn't consider the
back-reactions carefully). As we will see, the distortion is order
by order in the parameter $b$, and so we will not be too far from
our original choice of metric \metmetnow. However our initial
expectation of $\alpha_2, \beta_2$ in \relbetnow\ changes to
\eqn\relbetnownow{\eqalign{& {\alpha_2\o {\cal F}_2(r_0)} +
{\alpha_5\o {\cal F}_5(r_0)}+ {\alpha_4\o {\cal F}_4(r_0)} ~ = ~
{a_1 b^2 \alpha_6 \o {\cal F}_6(r_0)}, \cr & {\beta_2\o {\cal
F}_2(r_0)} + {\beta_5\o {\cal F}_5(r_0)} + {\beta_4\o {\cal
F}_4(r_0)}~ = ~ {\alpha^2_5\o {\cal F}^2_5(r_0)} + {\alpha^2_4\o
{\cal F}^2_4(r_0)}+ {\alpha_5\alpha_3\o {\cal F}_5(r_0){\cal
F}_4(r_0)}~ + \cr &~~~~~~~~~~~~~~~~~~~~ - {a_2 b^4 \alpha_6 - a_1
b^2 (\beta_6+\alpha_6) \o {\cal F}_6(r_0)} + {a^2_1 b^4 \alpha^2_6
\o {\cal F}^2_6(r_0)} -{2 a_1 b^2 \alpha_4 \alpha_6 \o {\cal
F}_4(r_0) {\cal F}_6(r_0)},}} which in the limit $b \to 0$ starts
to look like \relbetnow\ once we incorporate the identifications
in \morerel.

The other two coefficients $\gamma_2$ and $\delta_2$ are too
complicated to be written in terms of other coefficients of the
metric \metmat. Therefore we will use the $E_i$'s in \eidef\ to
write them down. Unfortunately the analysis turns out to be very
involved, and only under some simplifying assumptions have we been
able to find the following relations between the coefficients:
\eqn\jimkaal{\eqalign{& {\gamma_2\o {\cal F}_2(r_0)} + E_2 + E_0^3
~= ~ 2 E_0 E_1, \cr & {\delta_2\o {\cal F}_2(r_0)} + E_3 + 3 E_0^2
E_1~ =~ E_0^4 + 2 E_0 E_2 + E_1^2,}} where one could write the
values of $E_i$ to get more direct relations. It will be a
formidable exercise to disentangle any information out of this.
Fortunately, this is not the complete story. We can find more
relations between the coefficients in \metmat\ to decipher the
structure of the metric in terms of at least one or more warp
factors (and the $b$ field). One such set of connections is an
immediate modification of \relbetalbe\ as
\eqn\relbetalbenow{\eqalign{& {\alpha_1\o {\cal F}_1(r_0)} -
{\alpha_4\o {\cal F}_4(r_0)}- {\alpha_5\o {\cal F}_5(r_0)} ~ = ~
0;\cr & {\beta_1\o {\cal F}_1(r_0)} - {\beta_5\o {\cal F}_5(r_0)}
- {\beta_4\o {\cal F}_4(r_0)}~ = ~
 {\alpha_4\alpha_5\o {\cal F}_4(r_0){\cal F}_5(r_0)},}}
where, when we apply \morerel\ we get back \relbetalbe\ from above. Of course we cannot apply \morerel\ to this
case when we have a non-trivial $b$ factor, and so \relbetalbenow\ gives rise to new connections between the
various coefficients.

So far we could get relations for all the expansions in \coffdeff\
except ($\alpha_3, \beta_3$, ...). Since the first line of
\morerel\ is not enough to get the relations, we have to see how
\morerel\ is corrected by $b$. Again the analysis is done order by
order in $b$, and we see the following relations emerging from our
calculations to correct our original evaluation of \morerel:
\eqn\albethga{\eqalign{& {\alpha_3\o {\cal F}_3(r_0)} -
{\alpha_4\o {\cal F}_4(r_0)} = F_0,\cr &{\beta_3\o {\cal
F}_3(r_0)} - {\beta_4\o {\cal F}_4(r_0)} = {F_0 \alpha_4\o {\cal
F}_4(r_0)} + F_0^2 -F_1,}} with $F_0, F_1$(as expected) dependent
on the $b$ field in the following way: \eqn\fzfo{F_0 = {a_1 b^2
\alpha_6\o {\cal F}_6(r_0)}, ~~~~~ F_1 = {a_2 b^4 \alpha_6\o {\cal
F}_6(r_0)} - a_1 b^2 \Bigg({\beta_6\o {\cal F}_6(r_0)} +
{\alpha_5\alpha_6\o {\cal F}_5(r_0){\cal F}_6(r_0)}\Bigg),} where
both $F_0, F_1$ vanish when $b \to 0$ resulting in getting
\morerel\ from \albethga. One can also see that up to possible
constants $a_{1,2}$ the corrections to \morerel\ are known in
powers of $b$. Although these corrections are evaluated using some
approximations, they will be helpful later to fix the precise
back-reactions due to $b$ on the geometry. We also see that the
corrections are dependent on $\alpha_6, \beta_5$, etc. This may
look a little counter-intuitive but will become clearer later when
we fix the geometry. And finally, the other coefficients in
\coffdeff\ are related as \eqn\cdrelcha{\eqalign{& {\gamma_3\o
{\cal F}_3(r_0)} - {\gamma_4\o {\cal F}_4(r_0)} = {F_0\beta_4\o
{\cal F}_4(r_0)} + {F_0^2 \alpha_4\o {\cal F}_4(r_0)} - {F_1
\alpha_4\o {\cal F}_4(r_0)} - 2 F_0 F_1 + F_0^3 + F_2 ,\cr
&{\delta_3\o {\cal F}_3(r_0)}-{\delta_4\o {\cal F}_4(r_0)} = {F_0
\gamma_4\o {\cal F}_4(r_0)} + {F^2_0\beta_4\o {\cal F}_4(r_0)}-
{F_1\beta_4\o {\cal F}_4(r_0)} - {(2F_0 F_1 - F_0^3 -
F_2)\alpha_4\o {\cal F}_4(r_0)} ~ + \cr &
~~~~~~~~~~~~~~~~~~~~~~~~~~~~ + 2F_0 F_2 - F_3 + F_1^2 - 3 F_0^2
F_1 + F_0^4,}} where we could go beyond these orders; but that
will not be necessary for our purposes. We also see that the
expansions are defined in terms of $F_0, F_1$ and two new terms
$F_2$ and $F_3$. They are defined as \eqn\newff{\eqalign{&F_2 ~ =
~ a_1 b^2\Bigg({\gamma_6\o {\cal F}_6(r_0)} + {\alpha_5\beta_6 +
\beta_5 \alpha_6 \o {\cal F}_5(r_0){\cal F}_6(r_0)}\Bigg), \cr &
F_3~=~{a_2 b^4 \beta_6\o {\cal F}_6(r_0)} + {a_2 b^4
\alpha_5\beta_6 - a_1b^2(\alpha_5 \gamma_6 + \beta_5 \beta_6 +
\gamma_5 \alpha_6) \o {\cal F}_5(r_0){\cal F}_6(r_0)} - {a_1 b^2
\gamma_6\o {\cal F}_6(r_0)}.}} The above set of relations should
be enough to get some relations between the warp factors in the
metric \metmat. The first thing to look for are the complex
structures of the two tori ($y, \theta_2$) and ($x, \theta_1$).
The complex structures are different from the ones that we
calculated earlier in \cstar. In fact we don't even expect them
 to be identical. They  are:
\eqn\cstnno{dz_1 = dx + i\tau_{(1)} d\theta_1,~~~~ dz_2 = dy +
i\tau_{(2)} d\theta_2,} where $\tau_{(i)}$ are real numbers. It
turns out that only $\tau_{(1)}$ is affected by the $b$ field, and
not $\tau_{(2)}$ (we will provide a reason later). Therefore
$\tau_{(1)}$ is more involved than the other, and is given here by
\eqn\tuone{\eqalign{\tau_{(1)} ~ = ~ & 1 + {F_0 r\o 2} +
\Bigg({3F_0^2\o 8} - {F_1\o 2}\Bigg)~r^2 + \Bigg({F_2\o 2} +
{5F_0^3\o 16} - {3F_0 F_1 \o 4}\Bigg)~r^3 ~+ \cr &~~~~~~~~ +
\Bigg({3F_1^2 \o 8} - {F_3\o 2} + {3F_0 F_2 \o 4} - {15 F_0^2 F_1
\o 16} + {35 F_0^4 \o 128}\Bigg)~r^4,}} where $F_i$ are defined
above as \fzfo\ and \newff. One can easily see that in the absence
of $b$ field $\tau_{(1)}$ is just the identity (up to the orders
that we consider), and in general this is given approximately by
\eqn\tauodef{\tau_{(1)} ~\approx~ {1\o \sqrt{\kappa_o}} ~ + ~
{\cal O}(r^5),} where $\kappa_o$ is given by the expansion \bswde.
The above relation is only approximate as we have worked only up
to ${\cal O}(r^4)$. For higher powers of $r$, or even large $r$,
we need more detailed analysis. It is easy to see even for
non-zero $b$ that \eqn\limta{{}^{\rm lim}_{r\to 0}~\tau_{(1)} ~ =
~ 1 ~ + ~ {\cal O}(r^5),} which gives a square ($x, \theta_1$)
torus at the far IR\foot{A more appropriate result would be to
consider $\tau_{(1)} = 1 + r {\tilde b}^2$ instead of just
$\tau_{(1)} = 1$ where ${\tilde b} = \sqrt{{a_1 b^2 \alpha_6 \o
{\cal F}_6(r_0)}}$ is the effective B-field. For small $r$ this
tells us how the $b$ field affects the complex structure of the
($x, \theta_1$) torus.}. For the other complex structure of the
($y, \theta_2$) torus we find \eqn\sectau{\eqalign{\tau_{(2)}^2 ~
= ~ & 1 + {\cal F}_6(r_0)^{-1}\Big[\Big( {\alpha_6 - G_0 {\cal
F}_6(r_0)}\Big)~ r + \Big({G_1 {\cal F}_6(r_0) - \alpha_6 G_0 +
\beta_6}\Big)~r^2 ~ + \cr & ~~~~ + \Big({\alpha_6 G_1 - G_2 {\cal
F}_6(r_0) - \gamma_6 G_0 + \gamma_6}\Big)~r^3 + \Big({\beta_6 G_1
- \alpha_6 G_2 - \gamma_6 G_0 + \delta_6}\Big)~r^4\Big],}} where
we have defined $G_i$ as \eqn\gijoe{\eqalign{&G_0 = {\alpha_5 \o
{\cal F}_5(r_0)}, ~~~~ G_1 = {\alpha^2_5 \o {\cal F}^2_5(r_0)} -
{\beta_5 \o {\cal F}_5(r_0)},\cr  & G_2 = {\gamma_5 \o {\cal
F}_5(r_0)} - {2\alpha_5\beta_5 \o {\cal F}^2_5(r_0)} + {\alpha^3_5
\o {\cal F}^3_5(r_0)}, \cr & G_3 = -{\delta_5 \o {\cal F}_5(r_0)} +
{\beta^2_5 \o {\cal F}^2_5(r_0)} + {2\alpha_5\beta_5 \o {\cal
F}^2_5(r_0)} - {3\alpha^2_5\beta_5 \o {\cal F}^3_5(r_0)} +
{\alpha^4_5 \o {\cal F}^4_5(r_0)},}} and all the variables in the
above relations have already been defined. We also see some
interesting consequences when we apply the second line of
\morerel\ to \sectau. The complex structure $\tau_{(2)}$ becomes
\eqn\setaun{\tau_{(2)} ~ = ~ 1 + {\cal O}(r^5)} even for small but
finite $r$. This means that the ($y, \theta_2$) torus is exactly
square at far IR, but the ($x, \theta_1$) torus is only
approximately square at far IR. At finite $r$, $\tau_{(2)}$
remains square, whereas $\tau_{(1)}$ receives $b$ dependent
corrections (see footnote above).

The conclusions about the complex structures that we gave above are not in any way unexpected. Combining
the relation \albethga\ with the second line of \morerel, we can reproduce both the complex structures
as evaluated above. What is interesting however, is that now we can write the two tori metric completely
in terms of the $\tau_{(i)}$ in \setaun\ and \tauodef\ as:
\eqn\mettoti{ds^2_{\rm tori} ~=~ {\rm x}_1~\vert dy + {\rm i} d\theta_2\vert^2 +
{\rm x}_2~\vert dx + {\rm i} \kappa_o^{-{1\o 2}}
d\theta_1\vert^2,}
where $\kappa_o$ is defined in \bswde\ and ${\rm x}_{1,2}$ are unknown functions that have to be determined
from the expansion above. In the far IR, $ds^2_{\rm tori}$ becomes the metric for two square tori with
some $r$ dependent coefficients. The $b$ field simply distorts one of the tori so that it scales in
some particular way as we move along the radial direction.

It is now easy to determine ${\rm x}_{1,2}$ from the expansion above. All we require is to represent them as
some series like:
\eqn\series{{\rm x}_1 = 1 + \sum_{i=1} {\rm x}_{(1i)} r^i, ~~~~~~ {\rm x}_2 = 1 + \sum_{i=1} {\rm x}_{(2i)} r^i,}
with the generic terms ${\rm x}_{(ai)} \ne 0$ for $a= 1,2; i= 1, 2, ...$. The set of steps required to get to the
final answer is to first evaluate the quantity ${\rm x}_2 \o \kappa_o$ and secondly compare these results to the
relations \albethga\ and \cdrelcha. For ${\rm x}_1$ we can compare the metrics \metmat, \mettoti\ with
\morerel. We will not show these analyses
here\foot{One would require the expansion $\kappa^{-1}_o = 1 + F_0 r + (F_0^2 - F_1) r^2 -
(F_2 + F_0^3 - 2 F_0 F_1) r^3
+ (F_1^2 - F_3 + F_0^4 + 2 F_0 F_2 - 3 F_0^2 F_1) r^4$ with $F_i$ defined in \fzfo\ and \newff\ to do the
analysis.},
but readers can easily verify the following relations:
\eqn\emrel{{\rm x}_1 - {\cal E} ~ = ~ {\cal O}(r^5), ~~~~ {\rm x}_2 - {\cal D} ~ = ~ {\cal O}(r^5),}
which specifies the metric \mettoti\ at least up to ${\cal O}(r^5)$ in the expansion above.

Once we have a relation like \emrel, the rest follows rather
straightforwardly. The structure that we are alluding to is almost
like \newstrt\ but is a little more complicated. For the present
case we have: \eqn\miko{\eqalign{&{\cal A} - {\cal D}\cdot {\cal
E} ~ = ~ {\cal O}(r^5),\cr & {\cal B} - {\kappa_o^{-1} \cdot {\cal
D}^{-1} \cdot {\cal F}^{-1} \o \big(1 + b_0~{\rm cot}~\langle
\theta_1\rangle\big)\big(1 + c_0~{\rm cot}~\langle
\theta_2\rangle\big)} ~ = ~ {\cal O}(r^5),}} which differ from
\newstrt\ in a crucial way. The above relations could be
simplified further to take into account the complex structures of
the two tori. Assuming $b_0, c_0$ to be very small, an obvious
simplification occurs when in \miko\ we have the second relation
modified to \eqn\mlll{{\cal B} - \kappa_o^{-1} \cdot {\cal D}^{-1}
\cdot {\cal E}^{-1} ~ = ~ {\cal O}(r^5),} where $\kappa_o^{-1}$
expansion was given earlier as a footnote. We also see that the
combination of \miko\ and \mlll\ is close to the structure that we
had earlier. This is good, because it means that our earlier
choice of metric still survives possible dipole deformations. Of
course at the far IR we shouldn't detect any observable effects of
the dipoles, so this is not too surprising. In fact at the IR
there could be further simplification coming from the fact that we
are at small $r$. One such simplification is to look for the
behavior of: \eqn\xonextwo{\vert {\rm x}_1 - {\rm x}_2\vert ~ = ~
\sum_{i} ~ \vert {\rm x}_{(1i)} - {\rm x}_{(2i)}\vert ~r^i} which
clearly is very small at $r \to 0$. What happens for finite $r$?
In the absence of a $b$ field, every term on the RHS of \xonextwo\
for $i \le 5$ vanishes. We will assume that this continues to hold
even in the presence of the $b$ field because dipole deformations
will change results only in the far UV and not in IR. This would
naturally then imply \eqn\xoxt{\vert {\rm x}_1 - {\rm x}_2\vert ~
= ~ {\cal O}(r^5)} up to the order that we made our analysis so
far. This simplifies \mettoti. But this is not all. A few more
simplifications follow immediately (we give only a partial
analysis): \eqn\simplif{\eqalign{& {\alpha_1 \o {\cal F}_1(r_0)} ~
= ~ {2\alpha_6 \o {\cal F}_6(r_0)}, ~~~ {\beta_1 \o {\cal
F}_1(r_0)} ~ = ~ {\alpha^2_6 \o {\cal F}^2_6(r_0)} + {2\beta_6 \o
{\cal F}_6(r_0)}, ~~~ {\alpha_2 \o {\cal F}_2(r_0)}~=~ {(a_1 b^2
-2) \alpha_6 \o {\cal F}_6(r_0)}, \cr &{\beta_2 \o {\cal
F}_2(r_0)}~ =~ {(3-a_1^2b^4-2a_1b^2)\alpha^2_6 \o {\cal
F}^2_6(r_0)} - {(a_2b^4 + a_1b^2)\alpha_6 \o {\cal
F}_6(r_0)}-{(2+a_1b^2)\beta_6 \o {\cal F}_6(r_0)}, \cr & {\alpha_3
\o {\cal F}_3(r_0)}~=~{(1+a_1b^2)\alpha_6 \o {\cal F}_6(r_0)}, ~~~
{\beta_3 \o {\cal F}_2(r_0)} = {(1+a_1b^2)\beta_6 \o {\cal
F}_6(r_0)} + {a_1^2 b^4 \alpha^2_6 \o {\cal F}^2_6(r_0)} - {a_2
b^4 \alpha_6 \o {\cal F}_6(r_0)},}} and more involved relations
for ($\gamma_i, \delta_i$) $i=1,2,3$ in terms of ($\alpha_6,
\beta_6, \gamma_6, \delta_6$). Taking the complex coordinates for
the two base tori to be \eqn\bastor{dz_1 ~= ~ dx + {{\rm i} \o
\sqrt{\kappa_o}}~d\theta_1, ~~~~~~~ dz_2 ~ = ~ dy + {\rm i}~
d\theta_2,} the final metric is a simple modification of
\metmetnow\ that we had earlier:
\eqn\metmatnow{\eqalign{ds^2_{\cal M}~ = ~& {\cal F}(r)^2~dr^2 +
\kappa_o^{-1}{\cal F}(r)^{-2}~\Big(dz + f_1(r, \theta_1)~ dx +
f_2(r, \theta_2)~dy\Big)^2~ + \cr & ~~~~~~~~~~ + {\cal F}(r)~\vert
dz_1\vert^2 + {\cal F}(r)~\vert dz_2\vert^2,}} which is as good as
it gets because this is very close to \metmetnow; at far IR we
expect this to coincide with \metmetnow. Whether this could be
possible needs to be worked out now. We therefore require a way to
evaluate

\noindent $\bullet$ The warp factor ${\cal F}(r)$, and

\noindent $\bullet$ The functions $f_1(r, \theta_1)$ and $f_2(r, \theta_2)$

\noindent to complete this side of the story. The crucial difference between the present metric and \metmetnow, other
than the appearance of $\kappa_o$, is that both $f_{1,2}$ could be functions of $r$ also. The base tori, as we
observed earlier, are almost square and are deformed a little bit by the $b$ field.

\subsec{Dipole deformations and decoupling of the KK states}

The second part of the story is to introduce the seven-branes to
our background \metmatnow. These will back-react on the metric
\metmatnow\ to change the geometry. We will show that this is not
difficult to work out. In addition to that, we will find that
because of the seven-branes there will now be a non-trivial
axion-dilaton switched on.

The third and the final part of the story is to bring back the D5
branes. We have already discussed a consistent way to do this:
construct the D5 branes as bound states on a single D7 brane! This
will guarantee a fully supersymmetric background with non-trivial
fluxes on a non-trivial metric. Of course this is not the {\it
only} way to have a supersymmetric background with seven-branes,
$D5$ branes and fluxes. From F-theory, discussed earlier and in
\gtone, \realm, \gttwo\ we know that using F-theory we can have a
fully consistent background with separated D5s and D7s (along with
primitive fluxes). So the background that we construct in this
section is clearly not the most generic; although it is simple
enough to illustrate all the important ingredients of our
analysis.

Before we move ahead to determine the warp factor, $f_i$ and the branes,
we want to make some comments on the field theory
interpretation. One of the main difficulties in dealing with supergravity
duals to confining field theories is to decouple
the KK masses from the scale of the SUSY field theory. This is a crucial thing because we have wrapped D5 branes
on a non-trivial ${\bf P}^1$ globally or on the ($y, \theta_2$) torus locally. These wrapped branes will
generically have  KK modes from dimensional reduction along the compact ($y, \theta_2$) direction.
There have recently been proposals on decoupling the KK modes for
dipole deformed theories. The works of Lunin-Maldacena and
Gursoy-Nunez \nifuli\  have discussed the field theory living on
D-branes when there is an NS flux with one leg on the brane and one leg orthogonal to the brane.

The conclusion for the solutions of \nifuli\ was that, by turning on the NS flux, the masses of the KK modes
grow because the sizes of the $S^2$ or $S^3$ cycles
decrease.
Of course for our case there are no
 non-trivial three-cycles in the geometry.
We have an explicit two-cycle and we know the local geometry
around this cycle. A non-trivial three-cycle should appear after
the transition; the geometry after the transition is not covered
by the discussion in this paper. The only consistency condition
which can be checked is that the size of the resolved two-cycle is
indeed zero in the IR, as will be seen by the corresponding values of
volumes with and without $B$-fields.

In our language, the proposal of \nifuli\ tells us that, by
turning the NS field in the $(y, \theta_1)$ direction, the KK
modes are the only sector charged under $U(1)_{y} \times
U(1)_{\theta_2}$ where the gluons and the gluinos are chosen to
not be charged under $U(1)_x$. The dipole deformation then appears
only in the KK spectrum and does not change the four-dimensional
field theory.

Now we can ask how  the masses of the KK modes are changed in our
case. In the (near)--local solution, the masses of the KK modes
are inversely proportional to the area of the (quasi)--torus $(y,
\theta_2)$. For constant values of $y$ and $\theta_2$, the area of
the torus depends on the value of $\tau_{(2)}$ as well as the $dz$
fibration. The results we want to compare are:

\noindent (a) The nonzero corrections to $\tau_{(2)}$ starting from
${\cal O}(r^3)$ for the non-deformed case\foot{It could even be ${\cal O}(r^2)$ as all the warp factors are
taken to be linear in $r$.}.

\noindent (b) The non-zero corrections to $\tau_{(2)}$ starting from
${\cal O}(r^5)$ for the dipole deformed case.

\noindent (c) The non-zero corrections to the whole metric due to the underlying $b$ field.

\noindent In the discussion
below -- to be presented soon -- we will argue from this simple analysis that,
near the local solution limit,
the volume of the two-cycle on which we have wrapped $D5$ branes {\it decreases} in the dipole deformed theory.
This would imply that the KK masses are indeed bigger in the dipole deformed theory and they
can be decoupled from the QCD scale.

Coming back to the issues of warp factors and other things we now
have to determine the functional forms of $f_i(r, \theta_i)$ for
our case. As expected, the equations of motion for $f_i$ are more
complicated than \fcrela\ that we had earlier. We haven't been
able to work out the full details, but an approximate equation can
be given for $f_i$ that relate the $\theta_i$--variation of $f_i$
to the warp factors that we had before, in the following way:
\eqn\fcrelanow{{1\o \sqrt{1-b^2}}{\del f_i\o \del \theta_i} +
{\alpha_6 \o {\cal F}_6(r_0)} + {f_i~{\rm cot}~\theta_i \o
\sqrt{1-b^2}} ~ = ~ -{2\beta_6 \o {\cal F}_6(r_0)}~r + {\cal
O}(r^2)} where $i = 1,2$ and the $\sqrt{1-b^2}$ dependence above
is only approximate and is valid near the point radially away from
the chosen point \choipoi.

The solution to the above equation is not difficult to find, and
it is given by the following functional form that is a slight
modification of what we had earlier in \finow\foot{Again for 
$\theta_i \ne 0$. For $\theta_i =0$ the fibration becomes a total 
derivative.}: 
\eqn\finownow{f_i ~
= ~ \sqrt{1-b^2} \Bigg[{\alpha_6 \o {\cal F}_6(r_0)} + {2\beta_6
\o {\cal F}_6(r_0)}~r + {\cal O}(r^2)\Bigg] ~{\rm cot}~\theta_i}
This is again encouraging because the modification from \finow\ is
very small. In fact in the far IR the metric fibration in
\metmatnow\ will be exactly the same as in \metmetnow\ if we
replace the $Q$ in \calcnow\ by \eqn\qnew{Q ~ = ~ {\alpha_6
\sqrt{1-b^2}\o {\cal F}_6(r_0)}.} The above value of $Q$ is not
exact, as we have made some simplifying assumptions to get to
this.
 How far this value of $Q$ is away from the exact answer
will be determined later in the paper. Our naive expectation would
be to extrapolate the value of $Q$ in \finow\ to the present case.
This will tell us that we might be off by a quantity $\delta Q$
where $\delta Q$ is given by the following expression:
\eqn\delqu{\delta Q ~ = ~ {\Big(1+a_1 b^2 -
\sqrt{1-b^2}\Big)\alpha_6 \o {\cal F}_6(r_0)},} where $a_1$ is
still an undetermined constant. Such a change will no doubt have
an effect on the original equation \fcrelanow\ that determines the
$f_i$ as we shall discuss later, but we have reasons to believe
that the $r$ and $\theta_i$ dependence of $f_i(r, \theta_i)$ may
still survive the correction proposed in \delqu\ or its correct
generalisation thereof to be presented later. This would mean that
the metric with a $b$ field deformation may take the final form
\eqn\desjardins{\eqalign{ds^2_{\cal M}~ = ~& {\cal F}(r)^2~dr^2 +
\kappa_o^{-1}{\cal F}(r)^{-2}~\Big(dz + \Delta_1(b)~ {\rm
cot}~\theta_1~ dx + \Delta_2(b)~ {\rm cot}~\theta_2~dy + {\cal
O}(r)\Big)^2~ + \cr & ~~~~~~~~~~ + {\cal F}(r)~\vert dz_1\vert^2 +
{\cal F}(r)~\vert dz_2\vert^2,}} where we have put in all the $b$
field corrections in $\Delta_i(b)$ and $r$ dependent corrections
as ${\cal O}(r)$ in the $dz$--fibration. Observe also that we
haven't yet determined the coefficients in the warp factor ${\cal
F}$. These coefficients will eventually be related to some
conserved charges in the system, but an exact determination of
these -- along the lines of \shiu\ -- will be postponed to future
publications.

At this point we should put in the seven-branes. The seven-branes
are not sources of $B_{NS}$ fields, so we would expect the $b$
field to remain unaffected. However string coupling would
definitely be altered by the seven-branes. Without them
 the string coupling $g_s$ is affected only by the
background $b$ field in the following way: \eqn\gstring{g^2_s ~ =
~ {(g_s^o)^2\o 1-{a_1 b^2 \alpha_6 \o {\cal F}_6(r_0)}~r +
\Big[{a_2 b^4 \alpha_6 \o {\cal F}_6(r_0)} - {a_1 b^2 \beta_6 \o
{\cal F}_6(r_0)} - {a_1 b^2 \alpha_5\alpha_6 \o {\cal
F}_5(r_0){\cal F}_6(r_0)}\Big]~r^2 + {\cal O}(r^3)},} where
$g_s^o$ is the string coupling in the absence of $b$ field. We now
need to see how the seven branes would modify the result. First,
of course we have to figure out their back-reaction  on the
geometry \desjardins. The generic ansatz for the metric of a
seven-brane oriented along spacetime directions $x^{0,1,2,3}$ and
at a point on the ($x, \theta_1$) directions is given by
\eqn\dseve{ds^2_{\rm D7} ~ = ~ h^{-1} ~ds^2_{\Vert} ~ + ~
h~ds^2_{x \theta_1},} where $h$ is the so-called harmonic
function. This would mean that the final metric with $b$ fluxes,
and seven-branes will be to modify \desjardins\ by the warp factor
$h$ in the following way: \eqn\sudesjardins{\eqalign{ds^2_{\cal
M}~ = ~& h(r)^{-1} \Big[ds^2_{0123} + {\cal F}(r)^2~dr^2 + {\cal
F}(r)~\vert dz_2\vert^2\Big] + h(r)~{\cal F}(r)~\vert
dz_1\vert^2~+ \cr & ~ + h(r)^{-1} \kappa_o^{-1}{\cal
F}(r)^{-2}~\Big(dz + \Delta_1(b)~ {\rm cot}~\theta_1~ dx +
\Delta_2(b)~ {\rm cot}~\theta_2~dy + {\cal O}(r)\Big)^2.}} This is
almost the final metric that we want for our case because we
expect switching on gauge fluxes on the D7 brane to get bound
states of $D5$ branes will alter the above metric only by an
additional warp factor. The string coupling ${\tilde g}_s$ before
switching on the gauge fluxes is easy to work out and is given by
\eqn\gsnew{{\tilde g}_s ~ = ~ {g_s \o h^4}} where $g_s$ is given
by \gstring\ above. All we  need now to complete the story is to
add gauge fluxes. Our initial analysis told us that the gauge
fluxes giving rise to D5 branes charges can be evaluated as
\qfive\ assuming that the integral is over a finite sphere ${\bf
P}^1$. Unfortunately this may not always be true, because our
initial choice of metric given as \rzmet\ is {\it not} realised in
the present set-up. In fact the metric on the D7 brane can be
calculated precisely from \sudesjardins\ and is given at a point
$x = x_0, \theta_1 = \theta_{10}$ by the following metric:
\eqn\medsetn{ds^2_s = h^{-1} \Bigg[ds^2_{0123} + {\cal F}^2~dr^2 +
{\cal F}~\vert dz_2\vert^2 + {1\o \kappa_o {\cal F}^{2}}~\Big(dz +
\Delta_2(b)~ {\rm cot}~\theta_2~dy\Big)^2 \Bigg].} A careful look
at the metric suggests something very interesting: the metric
resembles closely  a locally deformed Taub-NUT space! Therefore
all the nice properties of a Taub-NUT space could presumably be
applied here (with of course certain modifications to take into
account the deformations\foot{The warp factors ${\cal F}, h$ are
not quite that of a Taub-NUT space, because we are viewing
everything locally. But a slice of the metric does have a strong
resemblance to a KK monopole.}). In particular the metric
\medsetn\ can give an answer to the puzzle that we raised above,
namely, the non existence of a finite sized two-cycle. The metric
\medsetn\ can in fact support {\it normalisable} anti-selfdual
harmonic forms and therefore we can identify the gauge fluxes with
these forms. A study of such harmonic forms has been done earlier
in \imamura\ and recently in \robbins\ by considering various
possible deformations of the Taub-NUT metric. However all the
analysis done before were for globally deformed Taub-NUT. Here we
have only the local version so to apply our techniques we have to
assume that the ($y, \theta_2$) torus will eventually become a
sphere (or a squashed sphere) globally. This would mean that
${\cal F}^{-1} \vert dz_2\vert^2 ~ \to ~ r^2 d\Omega_2^2$ with
$\Omega_2$ being the metric of a (squashed) two-sphere globally.
We now define a one-form on this space, \eqn\onefo{\zeta ~ = ~
g(r) \Big(dz + \Delta_2(b)~ {\rm cot}~\theta_2~dy\Big)} with one
assumption: $\Delta_2$ is only a function of the $b$ field. The
function $g(r)$ can be explicitly determined by considering the
fact that the two form $\omega$ constructed out of this is
anti-selfdual, and is given by \eqn\geer{g(r) ~ = ~ {\rm exp}
\Bigg[-\Delta_2 \int {dr\o r^2 {\cal F}^2 \sqrt{\kappa_o}}\Bigg]}
where the integral should be from any point $r$ to $r \to \infty$
in the full global geometry. In the absence of a global picture,
all we can confirm here is that $g(r)$ is a normalisable function.
This would then mean that we can define our gauge fluxes -- which
we will switch on the D7 brane -- as \eqn\fdeff{F ~ \equiv ~N
\omega ~= ~ N d\zeta} with $\omega$ normalisable but not globally
defined; and $N$ is an integer specifying the number of $D5$
branes. Once such fluxes are specified, the background RR field
will be determined. Alternatively, we can assume that the warp
factor ${\cal F}$ sufficiently curves the $dz$ fibration to allow
non-trivial two-cycles {\it a l\`a} the two-cycles in the metric
\rzmet. In either case, bound states of D5 branes wrapped
on two-cycles ($y, \theta_2$) will be created. With this, the
final metric with D5s, D7 and fluxes, is very close to the one
presented above in \sudesjardins\ and is modified only by
appropriate warp factor\foot{There is a little more to it. Indeed the 
metric will pick up extra warp factors, but the background fluxes will 
also change. One can see this from an earlier analysis of \sltwoz\ 
done for topologically trivial background geometry.}. 
As we can clearly see now,  near the
point $r = r_0$ the final local metric is exactly the one that we
had presented earlier in \gtone, \realm\ and \gttwo! This
therefore serves as a very strong confirmation of our result.

One last step still remains: we need to determine the size of the
two-cycle on which we have wrapped D5 branes. Now that we have an
almost complete description of the local geometry, the metric of
the two-cycle will not be too difficult to determine. For the
constant value of ($r, z, x, \theta_1$), the metric  is diagonal
and is given by \eqn\twcy{ds^2_{\rm two-cycle} ~ = ~ h^{-1} dy^2
\Bigg[{\cal F}(r_0) + {\Delta_2^2 ~{\rm cot}^2~\theta_2 \o
\kappa_o~{\cal F}^2(r_0)}\Bigg] + h^{-1}~{\cal
F}(r_0)~d\theta_2^2} where we see that the $dz$ fibration also
contributes to the metric of the two-cycle as one would have
expected. Note also that the metric of the two-cycle depends on
the $b$ field from two different sources: $\Delta_2$ and
$\kappa_o$. Thus the volume of the two-cycle is given by
\eqn\voltwo{{\rm Vol}_{\rm b} ~ = ~ {{\cal F}(r_0)\o h}\sqrt{1 +
{\Delta_2^2~{\rm cot}^2~\theta_2 \o \kappa_o~ {\cal F}^3(r_0)}},}
where ${\rm Vol}_{\rm b}$ denotes the volume calculated with
back-reactions from the $b$ field. In the absence of the $b$ field
we know that ${\rm Vol}_0 ~ = ~ {{\cal F}(r_0)\o h}\sqrt{1 + {{\rm
cot}^2~\theta_2 \o {\cal F}(r_0)^3}}$, and therefore the change in
the volume is given by \eqn\chvol{\delta V ~ = ~ {\rm Vol}_0 -{\rm
Vol}_{\rm b} ~ = ~ {{\rm cot}^2~\theta_2 \o 2h~{\cal F}^2(r_0)}
\Bigg(1- {\Delta_2^2\o \kappa_o}\Bigg).} To see which volume is
bigger we need the behavior of ${\Delta_2^2 \o \kappa_o}$ in terms
of the order by order expansion that we have been using. Our
earlier determination of \delqu\ may simply suggest
\eqn\kdel{\kappa_o ~< ~ 1 ~~~~~ {\rm and}~~~~~ \Delta_2 ~< ~ 1}
because ${\cal O}(r^{2n+1})$ terms for $\kappa_o$ are dominant and
contribute negatively to the sum in the series. Similarly,
$\Delta_2$ terms also seem to be suppressed as can be seen from
\qnew. Unfortunately, these considerations do not help us to see
the behavior of the ratio ${\Delta_2^2 \o \kappa_o}$ and therefore
we need a more detailed analysis to figure this out.

Our first step would be to determine which of $\kappa_o$ and
$\Delta_2^2$ is bigger. In our earlier order-by-order expansion,
if we keep the expansion only to the first order in $r$, then one
can show with some effort \eqn\valofD{\Delta_2^2 ~ = ~ 1 - a_1 b^2
\Bigg(1 + {\alpha_6 \o {\cal F}_6(r_0)}\Bigg) r_0} where we have
taken a point $r = r_0$ to do the analysis. From this, and using
the expansion of $\kappa_o$, it is easy to show that
\eqn\kDsq{\kappa_o ~-~\Delta_2^2 ~ = ~ a_1 b^2} implying $\kappa_o
> \Delta_2^2$, at least to the order that we have considered. This
also means that \eqn\volco{{\rm Vol}_{\rm b} ~~ < ~~ {\rm
Vol}_{0}} and therefore the volume of the two-cycle decreases when
we switch on a dipole deformation, again up to the orders that we
have considered in our expansion. The question now is whether the
result remains unchanged if we take higher order terms in our
expansion series. For this we need to find a closed form for
$\Delta_2^2$. We therefore make the following observations:

\noindent $\bullet$ $\Delta_2^2 = 1$ when $\kappa_o =1$. Furthermore
$\Delta_2^2$ can only be functions of $\kappa_o$ and the warp factors ${\cal F}, h$ because these are the only
unknown variables in our system.

\noindent $\bullet$ In the dual F-theory side $\Delta_2 {\rm cot}~\theta_i$ with $i = 1, 2$ appear explicitly
as four-form G-fluxes. In M-theory these G-fluxes now couple not to the co-dimension four surfaces {\bf A} and
{\bf B} (in fig 1), but to oriented co-dimension surfaces.

\noindent $\bullet$ The $B$ field component that gave rise to the
dipole deformation $b$ picks up other additional
components\foot{This would explain why one of the complex
structure \tuone\ in \cstnno\ is affected by the $b$-field. All
the components of the $b$ field have one leg along the $\theta_1$
direction. Therefore the ($x, \theta_1$) torus should definitely be
affected. On the other hand due to (a) the fact that the other components of the
$b$ field are along all the isometry directions, and (b) the
inherent gauge invariance of the $b$ field, the ($y, \theta_2$) torus
is not affected but only the $U(1)$ fibration is, as apparent from
\desjardins.} -- now parallel to the $x$ and $z$ directions -- that
are also proportional to $\Delta_2$. Although this is a generic
result, the new components are suppressed compared to  the results
that one would get without $B$ fields.

\noindent The last observation is particularly useful to pin-point the precise value of $\Delta_2$. In this paper
we will not go through the analysis, as this could be derived easily from the inherent F-theory picture. The final
result in compact form can be written as
\eqn\delcompa{\Delta_2^2 ~ = ~ {\kappa_o {\cal F} - 1\o {\cal F} -1}}
which can now be easily shown to reproduce \kDsq\ and hence would finally explain the decoupling of the KK states.

Before finishing this section, we should evaluate the background
$H$--fluxes. Since the dipole $b$--field cannot be gauged away,
there must exist the corresponding $H_{NS}$. Furthermore this
should correspond to our earlier expected ansatz \bnsbrr. Now
that we know most of the details regarding our background we
should be able to verify this. Taking the metric factors
correctly, we can show that locally there is indeed a $H_{NS}$
given by \eqn\hnsnow{{\cal H} ~ = ~ {\cal H}_0 ~{\rm
cosec}^2~\theta_2 ~dy \wedge d\theta_2 \wedge d\theta_1,} which is
exactly of the form \bnsbrr\ as one might have expected\foot{The above equation for 
${\cal H}$ \hnsnow\ is fine as long as we study far IR values. Due to the 
subtleties mentioned in \sltwoz, there will be other $r$ dependent components. A full
analysis of this will require more inputs than what we have mentioned here. These
and other issues will be addressed in \gwyn.}. 
The
constant ${\cal H}_0$ could also be determined for our case once
we put in the value of $\Delta_2$ in \delcompa. For the metric
\desjardins, ${\cal H}_0$ turns out to be \eqn\hzerodj{{\cal H}_0
~ = ~ {\sqrt{{\cal F}(1-\kappa_o)(\kappa_o {\cal F} -1)}\o {\cal
F} -1}} where ${\cal F}$ and $\kappa_o$ are all measured at $r =
r_0$ and therefore ${\cal H}_0$ is a constant. Once we know
$H_{NS}$, the $H_{RR}$ -- that forms the D5 sources -- is easily
determined from primitivity as shown earlier in \bnsbrr. Along
 with the metric \sudesjardins, string-coupling $g_s$
\gsnew\ and the axion (which we leave for the readers to derive)
the full background satisfying all the equations of motion can be
completely determined.

\newsec{Conclusions and future directions}

In this paper we have addressed several issues concerning the generic realm of
gauge-gravity dualities. Our first starting point was to clarify the fibration structure 
of the local metric that we have presented earlier in \gtone, \realm\ and \gttwo. A naive
analysis would have led to a constant fibration of the form \metformj. That this is not the 
full story can only become apparent if we carefully consider the warp factors. With these 
considerations, our first result is the
\medskip

$\bullet$ {Metric given by \metnownow\ with non-trivial $U(1)$ fibration.}
\medskip

\noindent The metric clearly tells us how one should view the local geometry. It is interesting to note that 
there may exist a family of such solutions, all coming from different possible realisations of global geometries. 
The local solutions are supersymmetric, but their naive global extensions may not be supersymmetric. In fact 
in this paper we haven't been able to find a globally defined metric that forms the gravity dual of ${\cal N} =1$
gauge theory with fundamental (and possibly bi-fundamental) flavors. Part of the reason lies in the UV picture 
which, for our case, is complicated by the presence of local and non-local seven branes, $D5$ branes and 
$D3$ branes. One thing is of course clear: 

\medskip
$\bullet$ {The full background is conformally K\"ahler with fluxes and branes,}
\medskip

\noindent although it could be made non-K\"ahler using the underlying F-theory picture. We give example of all these
cases by solving equations of motion order by order in powers of $r$. For a given patch, exemplified by fig. 3,
the metric is well defined and the full global geometry would be to add up all the patches. From F-theory we know 
that the manifold has at least one ${\bf P}^1$ on which we have wrapped $D5$ branes. The local seven branes i.e the 
$D7$ branes when brought near the $D5$s would also wrap the ${\bf P}^1$. On an isolated $D7$ brane, the 
$D5$ branes could exist as bound states of $D5-D7$. This is naturally supersymmetric and susy is only broken 
down to ${\cal N} =1$ by the background geometry. 
 
The story in the heterotic theory is more interesting. The background is not dual to the type IIB background,
and therefore not constrained by the type IIB structure. Global metrics in heterotic theory do exist, and
one example of this was already presented in \gttwo. In our language the global metric that we determined in 
\gttwo\ is what we called the metric after geometric-transition (GT). Our result therein was that 

\medskip
$\bullet$ The heterotic metric after GT was a warped version of a MN-type \mn\ metric.
\medskip

\noindent This is of course for minimal susy. For ${\cal N} =2$ the metric is given by a variant of 
the background of type \kimn. On the other hand {\it before} GT the situation is more intriguing. The 
complete local background can be found for this case, and our results are
 
\medskip
$\bullet$ The metric is given by \gulukola\ (or as \hettori\ in a simplified form).
\medskip
$\bullet$ The torsion is computed in terms of the torsion classes ${\cal W}_i$ in sec. 3.2.
\medskip
$\bullet$ The complete mathematical structure of this manifold is given in sec. 3.3
\medskip
\noindent For the mathematical parts, we have found a family of solutions given by 
holomorphic ${\bf C}^*$ fibrations arising from topologically nontrivial fibrations over the Kodaira surface ${\bf S}$.
These manifolds are generically non-K\"ahler as can be seen from the non-zero values of ${\cal W}_{3,4,5}$. They are
also complex because ${\cal W}_1 = {\cal W}_2 = 0$ can be imposed on the solutions. So we find
\medskip
$\bullet$ New non-compact, non-K\"ahler complex manifolds in heterotic theory
\medskip
\noindent whose local metric can be easily determined. The global story is another issue which we will dwell on 
in future works. The story, however, is not complete unless we figure out the vector bundles. In \gttwo\ we 
showed how vector bundles could be pulled through a conifold transition. A similar analysis should be done here
because our manifold is the one {\it before} GT\foot{The local geometry that we analysed here has no holomorphic
${\bf P}^1$, only holomorphic $T^2$. 
This is of course similar to our earlier conclusion for the type IIB case. On the other hand the 
geometry of \gttwo\ is global and has non-trivial ${\bf S}^3$ on which we did the conifold transition.}. 
Additionally, it is also interesting to 
ask whether topologically non-trivial holomorphic ${\bf C}^*$
fibrations exist for arbitrary ${\bf S}$. In fact looking at the
mathematics, it would seem that these {rarely} exist except for the
cases where the Kodaira surface ${\bf S}$ has torsion\foot{A brief explanation of how 
the torsion in $H^2({\bf S,Z})$ arises is as follows: 
Consider ${\bf S}$ as a $T^2$ fibration over ${\bf B} = T^2$.  Each of the two $S^1$
fibrations has a chern class, which can be identified with an integer via
$H^2({\bf B,Z})={\bf Z}$.  If we call these integers $r$ and $s$, then if $r$ and $s$ are
relatively prime, then $H^2({\bf S,Z}) = {\bf Z}^4$.  But if $r$ and $s$ have greatest
common divisor $m > 1$, then $H^2({\bf S,Z}) = {\bf Z}^4 + {\bf Z}_m$.}
in $H^2$. The
physical implication of this result is not clear to us at this stage. 
Maybe this is restricting the choice of the intrinsic torsion 
in our framework, but we have no concrete conclusion on this right now.

For the type IIB theory we also have additional results. Once we know that the fibrations in the metric 
can be non-trivial, we can ask whether there could be other effects. One new effect could in principle 
come from the back-reactions of the $B$ fields. For our case, due to the presence of branes and orientifold
planes at the orientifold corner of the moduli space, the $B$ fields backreact as {\bf dipole} deformations
in the field theory. Although the background geometry is complicated, we have been able to evaluate the 
local metric. Our results are 

\medskip
$\bullet$ The metric is given by \sudesjardins.
\medskip
$\bullet$ The three-form $NS$ flux is given by \hnsnow.
\medskip
$\bullet$ The three-from $RR$ flux is given by the Hodge dual of \hnsnow, i.e \bnsbrr.
\medskip
$\bullet$ The string coupling is given by \gsnew.
\medskip
\noindent The five-form fluxes and the axion can be evaluated from above. Of course all these results would be
further influenced by the subtleties mentioned in \sltwoz. But these additional corrections would {not} be
visible in the IR of the gauge theory {\it except} for one thing:
\medskip
$\bullet$ Volume of the two-cycle {\it shrinks} due to the dipole deformation,
\medskip
\noindent resulting in the KK modes -- from the dimensional reduction on the two-cycle of the wrapped branes
-- becoming heavier. Therefore they can be integrated out from the gauge theory, giving rise to pure 
${\cal N} = 1$ YM theory at the IR.  Hence for all IR purposes our solutions for the background will
be robust.  

\vskip.2in

\centerline{\bf Acknowledgements}

\noindent We would like to thank Melanie Becker, Robert Brandenberger, Josh Guffin, Karl Landsteiner, 
Carlos Nunez and Diana Vaman for helpful conversations.  
KD, MG and RG are supported in part by NSERC grants. 
The research
of SK is supported in part by NSF grants DMS-02-44412 and DMS-05-55678.
The research of AK is supported in part by DFG--the German Science Foundation, DAAD--the German Academic 
Exchange Service, the European RTN Program MRTN-CT-2004-503369 and the University of Maryland.
RT is supported by PPARC; and 
thanks the Galileo Galielei Institute for
Theoretical
Physics for the hospitality and the INFN for partial support during the
completion of this work.

\noindent {\bf Preprint Numbers:} ILL-(TH)-06-5, LTH--704


\listrefs

\bye